\documentclass[hidelinks,12pt]{article}
\usepackage{graphicx}
\usepackage{latexsym,amsmath,amsfonts,amssymb}
\usepackage{bbm}
\usepackage{slashed}
 \usepackage[all]{xy}
\usepackage{cite}
\usepackage[toc]{appendix}
\usepackage[linktoc=page]{hyperref}
\renewenvironment{thebibliography}[1]{%
\begin{oldthebibliography}{#1}%
\setlength\itemsep{-1mm}%
}%
{%
\end{oldthebibliography}%
}
\newcommand{\dvol}{d\mathrm{vol}}
\newcommand{\vol}{\mathrm{vol}}

\newcommand{\wt}{\widetilde}

\newcommand{\ie}{\textit{i.e.}}

\numberwithin{equation}{section}

\newcommand{\nn}{\nonumber}
\newcommand{\mat}[1]{\begin{pmatrix} #1 \end{pmatrix}}

\newcommand{\be}{\begin{equation}} \newcommand{\ee}{\end{equation}}
\newcommand{\bea}{\begin{equation} \begin{aligned}} \newcommand{\eea}{\end{aligned} \end{equation}}

\newcommand{\cF}{\mathcal{F}}

\newcommand{\cM}{\mathcal{M}}
\newcommand{\cN}{\mathcal{N}}
\newcommand{\cO}{\mathcal{O}}
\newcommand{\cP}{\mathcal{P}}
\newcommand{\cQ}{\mathcal{Q}}

\newcommand{\cT}{\mathcal{T}}

\newcommand{\bC}{\mathbb{C}}

\newcommand{\bP}{\mathbb{P}}

\newcommand{\bR}{\mathbb{R}}
\newcommand{\bZ}{\mathbb{Z}}
\newcommand{\fg}{\mathfrak{g}}

\def\coeff#1#2{\relax{\textstyle {#1 \over #2}}\displaystyle}

\DeclareMathOperator{\Tr}{Tr}

\numberwithin{equation}{section} % Equations numbered as <section>.#

\newcommand\equ[1] {\begin{equation}#1\end{equation}}

\newcommand\eqs[1] {\begin{align}#1\end{align}}
\newcommand\eqsn[1] {\begin{align*}#1\end{align*}}

\renewcommand\( {\left(}
\renewcommand\) {\right)}

\newcommand\N {{\cal N}}

\begin{document}

%%%%%%% title page %%%%%%%%%

\begin{titlepage}

\begin{flushright}
Imperial/TP/2015/FB/04\\
ITP-UU-15-16
\end{flushright}

\vspace*{0.5cm}

\begin{center}
{\LARGE \bf Two-dimensional SCFTs from D3-branes} \\

\vspace*{2cm}

{\bf Francesco Benini${}^{(1,2)}$, Nikolay Bobev${}^{(3)}$ and P. Marcos Crichigno${}^{(4)}$}
\bigskip
\bigskip

{\small
${}^{(1)}$
Blackett Laboratory, Imperial College London\\
South Kensington Campus, London SW7 2AZ, United Kingdom
\vskip 2mm
${}^{(2)}$
International School for Advanced Studies (SISSA) \\
Via Bonomea 265, 34136 Trieste, Italy
\vskip 2mm
${}^{(3)}$
Instituut voor Theoretische Fysica, KU Leuven \\
Celestijnenlaan 200D, B-3001 Leuven, Belgium
\vskip 2mm
${}^{(4)}$
Institute for Theoretical Physics and Spinoza Institute, Utrecht University\\
Leuvenlaan 4, 3854 CE Utrecht, The Netherlands
\vskip 2mm
}

\bigskip
\texttt{f.benini@imperial.ac.uk,~nikolay@itf.fys.kuleuven.be,~p.m.crichigno@uu.nl} \\

\end{center}

\vspace*{0.8cm}

\begin{abstract}

\noindent 

\end{abstract}

\noindent We find a large class of two-dimensional $\cN=(0,2)$ SCFTs obtained by compactifying four-dimensional $\mathcal{N}=1$ quiver gauge theories on a Riemann surface. We study these theories using anomalies and $c$-extremization. The gravitational duals to these fixed points are new AdS$_3$ solutions of IIB supergravity which we exhibit explicitly. Along the way we uncover a universal relation between the conformal anomaly coefficients of four-dimensional and two-dimensional SCFTs connected by an RG flow across dimensions. We also observe an interesting novel phenomenon in which the superconformal R-symmetry mixes with baryonic symmetries along the RG flow.

\end{titlepage}

%%%%%%%%%%%%%%%%%%%%%%%%%%%%%%%%%%%%%

{\small
\setlength\parskip{-0.5mm}
\setcounter{tocdepth}{2}
\tableofcontents
}

%%%%%%%%%%%%%%%%%%%%%%%%%%%%%%
\section{Introduction}
%%%%%%%%%%%%%%%%%%%%%%%%%%%%%%

Understanding the space of consistent conformal field theories (CFTs) is of great importance since this would provide insight into a classification of the possible phases of quantum field theories. One can hope that this hard problem becomes more manageable if one introduces additional symmetries, such as supersymmetry or conformal symmetry, to restrict the class of possible theories. In two spacetime dimensions there is a further simplification since the conformal group is infinite-dimensional. Despite this favorable circumstance, the classification of two-dimensional superconformal field theories (SCFTs) is far from complete. Therefore it is important to understand the space of consistent two-dimensional SCFTs and to sharpen our tools to study such theories. The goal of this work is to provide evidence for the existence of a novel class of 2d SCFTs with $\cN=(0,2)$ supersymmetry which arise from the twisted compactification of 4d SCFTs on a Riemann surface, and to employ a variety of techniques to understand their physics. 

Two-dimensional CFTs are also very interesting for a different reason. Gravity in three-dimensional asymptotically AdS space is
% supposed to be
one of the simplest toy models for
% understanding
quantum gravity---see for example \cite{Witten:2007kt}. Thus constructing and classifying possible AdS$_3$ solutions of string theory, and understanding their holographic duals, is of great importance to uncover the structure of quantum gravity in three dimensions. Besides, gravitational theories in AdS$_3$ also provide good laboratories to test and explore the AdS/CFT correspondence in detail---in fact such a setup was the precursor of holography \cite{Brown:1986nw}. These two alternative vantage points provide
% a different perspective and
further motivation for the work presented here.

Our goal is to study four-dimensional superconformal field theories (SCFTs) with $\mathcal{N}=1$ supersymmetry compactified on a Riemann surface with a partial topological twist. The main tools we use are anomalies, $c$-extremization, and holography. The basic idea is simple and dates back to the work of Witten \cite{Witten:1988xj}. On a general curved manifold supersymmetry is generically broken because there are no covariantly-constant spinors. If however the supersymmetric QFT at hand has a continuous R-symmetry, one can turn on a background field for it which cancels the spin connection on the curved manifold. This procedure of preserving supersymmetry on curved spaces is called the ``topological twist.'' We will be interested in studying 4d $\mathcal{N}=1$ theories on $\mathbb{R}^2\times \Sigma_{\mathfrak{g}}$ where $\Sigma_{\mathfrak{g}}$ is a smooth Riemann surface of genus $\mathfrak{g}$. Since the 4d theory has a $U(1)_R$ R-symmetry and the structure group of $\Sigma_{\mathfrak{g}}$ is $SO(2)$, we can generically preserve $\cN=(0,2)$ supersymmetry on $\Bbb R^2$ and thus, at energies below the scale set by the size of the Riemann surface, we have a 2d supersymmetric field theory. These 2d theories are the main subject of our work. In particular, we will argue that generically they will be superconformal and, by using the anomalies of the 4d theory, we will be able to calculate the anomalies of its 2d ``offsprings." An interesting generalization is possible if the 4d theory has continuous flavor symmetries. Then supersymmetry is preserved even when one turns on background magnetic flux on the Riemann surface for these symmetries. In this way from a single 4d SCFT one can obtain a multi-parameter family of candidate 2d theories labeled by the genus of the Riemann surface and the choice of background magnetic flavor fluxes. Since the magnetic flux on a compact Riemann surface must be appropriately quantized, this leads to a discrete family of theories. While anomalies provide a powerful calculational tool, they are not always well-suited to answering dynamical questions, thus in general it is hard to rigorously argue that the 2d SCFTs in question actually exist. One possible approach to remedy this situation is to employ holography and construct explicit AdS$_3$ vacua which are holographic duals to the SCFTs of interest. This is often possible if the parent 4d theory has itself a holographic dual description as we demonstrate explicitly.

These general ideas were made very concrete in \cite{Bershadsky:1995vm, Bershadsky:1995qy, Maldacena:2000mw, Benini:2012cz, Benini:2013cda} where they were applied to the case of 4d $\mathcal{N}=4$ SYM theory.\footnote{See also \cite{Johansen:2003hw} and \cite{Kapustin:2006hi} for related work on four-dimensional $\mathcal{N}=1$ and $\mathcal{N}=2$ theories, respectively.}  Here we argue that the setup is much more general and provide evidence for this claim by analyzing in detail the $Y^{p,q}$ family of superconformal quiver gauge theories \cite{Benvenuti:2004dy}. Using the knowledge of the 't~Hooft anomalies for these theories, we calculate the central charges of the 2d theories obtained from them upon twisted compactification on $\mathbb{R}^2 \times \Sigma_{\mathfrak{g}}$. An important role in this analysis is played by $c$-extremization \cite{Benini:2012cz, Benini:2013cda}, which is a tool that allows us to unambiguously determine the superconformal R-symmetry in two dimensions and thus the correct conformal anomalies. The reason we choose this class of theories is that they have explicit AdS$_5$ holographic duals, constructed in \cite{Gauntlett:2004yd}. This provides us with the reasonable expectation that the 2d SCFTs will also have weakly-coupled duals in type IIB supergravity. This expectation indeed bears fruit and we are able to construct new explicit warped AdS$_{3}\times_{w} \mathcal M_{7}$ solutions of IIB supergravity which are dual to the 2d SCFTs of interest.

A novel phenomenon that arises from the study of this class of field theories is that the R-symmetry generically mixes along the RG flow not only with usual mesonic flavor symmetries, but also with the baryonic flavor symmetry available in all $Y^{p,q}$ quivers. This is rather surprising from the supergravity perspective because, unlike mesonic symmetries, the baryonic symmetry does not arise from isometries of the metric, but rather from the RR 4-form potential on a topological three-cycle.

Finally, we should point out that the AdS$_3$ solutions we construct can be thought of as the near-horizon limit of BPS black strings in five dimensions. The entropy density of these black strings is related to the central charge of the dual 2d CFT and thus our successful match of the supergravity and field theory central charges can also be viewed as a microscopic counting of the degrees of freedom of the black strings.

The ideas and techniques discussed in this paper are similar to the ones employed by Maldacena-N\'u\~nez in \cite{Maldacena:2000mw} as well as in the more recent literature \cite{Bah:2011vv, Bah:2012dg, Benini:2012cz, Benini:2013cda, Bobev:2014jva}, see also \cite{Kutasov:2013ffl, Kutasov:2014hha} for relevant recent work. The supersymmetric AdS$_3$ solutions of IIB supergravity we find have only 5-form flux turned on. These backgrounds fall under the classification of \cite{Kim:2005ez} and indeed some of our solutions have been studied previously in \cite{Gauntlett:2006af, Gauntlett:2006qw, Gauntlett:2006ns, Donos:2008ug, Donos:2008hd, Gauntlett:2007ts, Kim:2012ek}.%
\footnote{Many of our solutions are actually ``T-dual'' to M-theory solutions in \cite{Gauntlett:2006qw}.}
More recently, AdS$_3$ solutions arising from string and M-theory have also been analyzed in \cite{Bah:2015nva, Bea:2015fja, Donos:2014eua, Karndumri:2013dca, Karndumri:2013iqa, Almuhairi:2011ws} (see also \cite{deBoer:2011zt} for related work). On the field theory side there have been interesting constructions of 2d $\cN=(0,2)$ SCFTs and dualities between them by employing compactifications of a higher-dimensional SCFT in \cite{Gadde:2013sca, Gadde:2013lxa, Gadde:2014ppa, Putrov:2015jpa, Franco:2015tna, Gadde:2015wta}.

We begin our exploration in the next section with a brief review of the $Y^{p,q}$ quiver gauge theories and we then proceed to compactify these theories on a Riemann surface and study the system at low energies. We also discuss a universal feature of RG flows connecting 4d $\mathcal{N}=1$ and 2d $\cN=(0,2)$ SCFTs. As an illustration of this general result, in Section \ref{subsec:D3dP} we consider the 4d $\mathcal{N}=1$ SCFTs arising from D3-branes at del Pezzo singularities. In Section \ref{Supergravity Solutions} we switch gears and discuss the construction of explicit AdS$_3$ solutions of IIB supergravity, which are holographic duals to the 2d SCFTs of interest.  We conclude in Section \ref{sec:conclusions} with a short summary and a number of directions for future work. In the various appendices we present technical details which pertain to the construction and analysis of the supergravity solutions discussed in Section \ref{Supergravity Solutions}.

%%%%%%%%%%%%%%%%%%%%%%%%%%%%%%
\section{Field theory}
\label{Field theory}
%%%%%%%%%%%%%%%%%%%%%%%%%%%%%%

%%%%%%%%%%%%%%%%%%%%%%%%%%%%%%
\subsection{$Y^{p,q}$ quivers}
%%%%%%%%%%%%%%%%%%%%%%%%%%%%%%

Let us first summarize some of the salient features of the $Y^{p,q}$ family of four-dimensional $\cN=1$ superconformal field theories. We will follow the notation and conventions of \cite{Benvenuti:2004dy} and take the coprime integers $p,q$ to satisfy $p>0$ and $0\leq q\leq p$. The theories are quiver gauge theories, with $2p$ nodes each representing an $SU(N)$ gauge group. The matter fields are in chiral multiplets and transform in bifundamental representations of pairs of gauge groups, as dictated by the quiver diagram. The theories have an $SU(2)_1 \times U(1)_2 \times U(1)_B \times U(1)_R$ continuous global symmetry, where $SU(2)_1 \times U(1)_2$ is a mesonic flavor symmetry (and we denote the Cartan of $SU(2)_1$ with $U(1)_1$), $U(1)_B$ is a baryonic symmetry and $U(1)_R$ is the superconformal R-symmetry. The matter fields can be organized into four groups, dubbed $\{Y,Z,U^{\alpha},V^{\alpha}\}$ with $\alpha=1,2$, according to their charges under the global symmetry as we summarize in the following table:
\be
\label{4Dcharges}
\begin{array}{c|c|c|c|c|c}
\text{Fields} & \text{multiplicity} & U(1)_1 & U(1)_2 & U(1)_R & U(1)_B \\
\hline
Y            & p+q & 0	& -1	& R_Y	& p-q   \\
Z            & p-q  & 0	& 1	& R_Z	& p+q  \\
U^1        & p     & 1	& 0	& R_U	& -p     \\
U^2        & p     & -1	& 0	& R_U	& -p     \\
V^1        & q     & 1	& 1	& R_V	& q      \\
V^2        & q     & -1	& 1	& R_V	& q      \\
\lambda & 2p   & 0	& 0   & 1 		& 0
\end{array}
\ee
By $\lambda$ we denoted the gaugini in vector multiplets, transforming in the adjoint representation of the gauge groups. The R-charges of the matter chiral multiplets are
\bea
\label{Rdef}
R_Y &= \frac{(2p-q) w +2pq - w^2}{3q^2} \;,\qquad\qquad &
R_U &= \frac{4p^2 -2p w }{3q^2} \;, \\
R_Z &= \frac{(2p+q) w - 2pq - w^2}{3q^2} \;,\qquad\qquad &
R_V &= \frac{3q-2p + w }{3q} \;,
\eea
where we have defined
\be
\label{wdef}
w \equiv \sqrt{4p^2-3q^2} \;.
\ee
One should keep in mind that the fermions in chiral multiplets have R-charge $1$ less than that of the multiplet.
When $w\in\mathbb{Z}$, the central charges of the 4d theory are rational.

The conformal anomaly coefficients, or central charges, $a$ and $c$ of the $Y^{p,q}$ theories, can be computed using the well-known relation \cite{Anselmi:1997am} between conformal and R-symmetry 't Hooft anomalies in $\cN=1$ SCFTs:
\be
a = \frac{9}{32} \Tr(R^3) - \frac{3}{32} \Tr(R) \;, \qquad\qquad c = \frac{9}{32} \Tr(R^3) - \frac{5}{32} \Tr(R) \;.
\ee
Using the charges in \eqref{4Dcharges} and \eqref{Rdef}, one finds
\be
\label{acYpq}
a(Y^{p,q}) + \frac{3p}8 = c(Y^{p,q}) + \frac p4 = \dfrac{3p^2(3q^2-2p^2+pw)}{4q^2(2p+w)} \, N^2 \;.
\ee
This is obtained%
\footnote{If some chiral multiplet is in the adjoint rather than in the bifundamental representation, the implicit multiplicity is $N^2-1$ and the $\cO(1)$ terms are different. This only happens for $Y^{1,1} \cong \bC^2/\bZ_2 \times \bC$. One obtains $a(Y^{1,1}) = \frac12 N^2 - \frac{5}{12}$ and $c(Y^{1,1}) = \frac12 N^2 - \frac13$.
\label{foo: adjoint}}
by noticing that the bifundamentals have implicit multiplicity $N^2$, while the gaugini have multiplicity $N^2-1$.
At leading order in $N$, the two central charges are equal because for this class of quiver gauge theories and at that order, the linear R-symmetry 't~Hooft anomaly vanishes: $\Tr R= \cO(1)$.

There are some cases of special interest. The theory $Y^{p,0}$ is a $\mathbb{Z}_p$ orbifold of the Klebanov-Witten (KW) theory \cite{Klebanov:1998hh} and has central charges
\be
\label{a,c 4D Ypq}
a(Y^{p,0}) \,\simeq\, c(Y^{p,0}) \,\simeq \frac{27 p}{64}\, N^2\;,
\ee
at leading order in $N$.
The theory $Y^{p,p}$ is a $\mathbb{Z}_p$ orbifold of the $\mathcal{N}=2$ quiver theory which itself is obtained by a $\mathbb{Z}_2$ orbifold of $\mathcal{N}=4$ SYM. The central charges for this theory are
\be
a(Y^{p,p}) \simeq c(Y^{p,p}) \simeq \frac{p}{2} \, N^2 \simeq 2p \, a_{\cN=4} \;,
\ee
where in the last equality we have emphasized the relation to the central charge of $\mathcal{N}=4$ SYM at leading order. 

It is worth collecting here the explicit expressions for the linear and cubic 't Hooft anomalies for the quiver gauge theories of interest. After a straightforward algebraic calculation one finds that the 20 independent cubic 't Hooft anomalies are:
\bea
k_{111} &= k_{222} = k_{122} = k_{12B} = k_{12R} = k_{1BB} = k_{1BR} = k_{1RR} = k_{2RR} = k_{BBB}=k_{BRR}=0 \\
k_{112} &= 2q N^2 \;,\qquad k_{11B}= 2(q^2-p^2)N^2 \;,\qquad k_{11R} = \frac{2}{3q^2} \big( pw^2+(q^2-2p^2)w-2pq^2 \big) N^2 \\
k_{2BB} &= 2p^2 q N^2 \;,\qquad k_{2BR} =\frac{2p^2}{3q}(w-2p) N^2 \;,\qquad k_{22R} = \frac{2}{3q^2}(2p^2+pq+q^2)(w-2p) N^2 \\
k_{22B} &= 2p^2 N^2 \;,\qquad k_{BBR} = -\frac{2p^2}{3}(p+w)N^2 \;,\qquad k_{RRR}=\frac{8p^2}{9q^4}(w^3+9pq^2-8p^3) N^2 -2p \;.
\eea
The linear 't Hooft anomalies are
\be
k_{1} = k_2 = k_B = 0 \;, \qquad\qquad k_R = -2p\;.
\ee
The identity $9k_{JRR} = k_{J}$, valid for any flavor (non-R) symmetry $J$ in a 4d $\mathcal{N}=1$ SCFT is clearly obeyed \cite{Intriligator:2003jj}. As pointed out in \cite{Benvenuti:2004dy}, baryonic symmetries are such that $k_{BBB}=k_{B}=0$. For general flavor symmetries this is not necessary, although for the $Y^{p,q}$ theories we also have $k_{111}=k_{222}=k_1=k_2=0$.

%%%%%%%%%%%%%%%%%%%%%%%%%%%%%%
\subsection{2d central charges}
\label{2D central charges}
%%%%%%%%%%%%%%%%%%%%%%%%%%%%%%

In this section we consider compactifications of generic four-dimensional $\cN=1$ field theories on compact (\ie{} with no punctures) Riemann surfaces $\Sigma_\fg$ of genus $\fg$, performing a partial topological twist so as to preserve $\cN=(0,2)$ supersymmetry in two dimensions. Under the assumption that the theories flow to interacting SCFTs (which could be tested holographically, for instance), we would like to compute their central charges. To do this, we exploit the fact that in two-dimensional $\cN=(0,2)$ SCFTs the R-symmetry can be identified by a $c$-extremization principle \cite{Benini:2012cz,Benini:2013cda}, and then the central charges are related to its 't~Hooft anomalies. We begin by providing explicit examples in the case of  $Y^{p,q}$ quivers and then discuss an approach for generic four-dimensional $\cN=1$ field theories.

The calculation proceeds as in \cite{Benini:2013cda}. To perform the partial topological twist, we turn on a background gauge field along the generator 
\be
\label{background generator}
T = b_1 T_1 + b_2 T_2 + B T_B+ \frac{\kappa}{2} T_R \;,
\ee
where $T_{1,2}$, $T_B$ are the generators of $U(1)_{1,2}$ and $U(1)_B$, respectively, while $T_R$ is the generator of the $U(1)_R$ {\it superconformal} R-symmetry. We have defined $\kappa$ as the normalized curvature of the Riemann surface: $\kappa=1$ for $\mathfrak{g}=0$, $\kappa=0$ for $\mathfrak{g}=1$, and $\kappa=-1$ for $\mathfrak{g}>1$. When the flavor flux $b_{1}$ is nonzero, the $SU(2)_{1}$ flavor symmetry of the system is broken to $U(1)_{1}$. For $b_1=0$ the $SU(2)_{1}$ symmetry is intact.

An important point is that the background flux \eqref{background
generator} must be properly and carefully quantized. We turn on an
external flux
\be
F = T \, \dvol_{\Sigma_\fg} \;,
\ee
where the volume form is normalized $\int \dvol_{\Sigma_\fg} = 2\pi
\eta_\Sigma$ and $\eta_\Sigma = 2|\mathfrak{g}-1|$ for $\mathfrak{g}\neq1$, $\eta_\Sigma =1$ for $\mathfrak{g}=1$. Then for every gauge-invariant operator
$\cO$, the effective number $n$ of flux units felt by the associated
particles and defined by
\be
\label{quantization condition}
\frac1{2\pi} \int_{\Sigma_\fg} F \cdot \cO = \eta_\Sigma \, T \cdot \cO \equiv n\, \cO \;,
\ee
should be an integer: $n \in \bZ$. This is the standard Dirac quantization condition. Since we have fixed the origin of the flavor flux around the 4d superconformal R-symmetry, which in the case of $Y^{p,q}$ quivers typically assigns irrational charges, one generically needs an irrational flavor flux to balance it. In particular, zero flavor flux is generically not allowed unless the superconformal R-charges are rational. When a twist by the pure superconformal R-symmetry is in fact possible, we refer to it as the ``universal twist,'' for reasons that will become clear below.

Next, we define the trial 2d R-symmetry to be a general linear combination of the 4d R-symmetry and the Abelian flavor symmetries, \ie 
\be
\label{eq:Ttrdef}
T_\text{tr} = \epsilon_1 T_1 + \epsilon_2 T_2+\epsilon_B T_B +T_R \;,
\ee
where the real parameters $\epsilon_{i}$'s are unfixed at the moment and we construct the trial central charge 
\be
\label{crtrgen} 
c_r^\text{tr} = -3 \eta_\Sigma \sum_\sigma m_\sigma t_{\sigma} (q_R^{(\sigma)})^2 \;.
\ee
The sum above is over the 4d fermionic fields labelled by $\sigma$, $m_\sigma$ is their  multiplicity, $q_R^{(\sigma)}$ is the charge under the trial R-symmetry in \eqref{eq:Ttrdef}, and $t_\sigma$ is the charge under the background gauge field in \eqref{background generator}. Here we have used the relation $c_r^\text{tr} =3k_{RR}$ (see  \cite{Benini:2012cz,Benini:2013cda} for details)  and that  the net number of right-moving minus left-moving 2d chiral massless fermions is computed by the index theorem:
\be
n_r^{(\sigma)}-n_{\ell}^{(\sigma)} = -t_{\sigma}\eta_{\Sigma} \;.
\ee
For the case of $Y^{p,q}$ quivers, the various parameters are summarized in the following table:
\be
\label{tqcharges}
\begin{array}{c|c|c|c}
\text{Fields} & m_\sigma & t_\sigma & q_R^{(\sigma)} \\
\hline
Y            & (p+q)N^2 &  \frac{\kappa}{2} (R_Y-1) - b_2  +B(p-q)            & R_Y -1 - \epsilon_2 +\epsilon_B(p-q)                     \\
Z            & (p-q)N^2  &  \frac{\kappa}{2}  (R_Z-1) +b_2 +B(p-q)        & R_Z -1 + \epsilon_2     +\epsilon_B(p-q)               \\
U^1        & pN^2     &  \frac{\kappa}{2} (R_U-1) +b_1  -Bp           & R_U -1 + \epsilon_1 -\epsilon_Bp                    \\
U^2        & pN^2     &  \frac{\kappa}{2}  (R_U -1)-b_1 -Bp            & R_U -1 - \epsilon_1 -\epsilon_Bp                     \\
V^1        & qN^2     &  \frac{\kappa}{2}  (R_V-1) +b_2 + b_1 +Bq  & R_V -1+\epsilon_2 + \epsilon_1+\epsilon_Bq  \\
V^2        & qN^2     &  \frac{\kappa}{2}  (R_V-1) +b_2 - b_1 +Bq   & R_V -1 +\epsilon_2 - \epsilon_1 +\epsilon_Bq  \\
\lambda & 2p(N^2-1)   &  \frac{\kappa}{2}                 & 1   
\end{array}
\ee
We recall that for $Y^{1,1}$ the multiplicities are different, see footnote \ref{foo: adjoint}.

At this point we invoke the principle of $c$-extremization, stating that the 2d superconformal R-symmetry is the one extremizing the trial central charge \eqref{crtrgen}, whose value at the extremum is the actual right-moving central charge $c_{r}$ of the 2d SCFT. With the ingredients given above, these can be calculated for any $Y^{p,q}$ quiver, Riemann surface, and background fluxes. In full generality the result is lengthy, so in the following subsections we discuss some cases of particular interest. When carrying out the extremization procedure, one must often treat the cases $\kappa=0$  ($\fg=1$) and $\kappa\neq 0$  ($\fg \neq 1$)  separately, as we do below.

%%%%%%%%%%%%%%%%%%%%%%%%%%%%%%
\subsubsection{$Y^{p,0}$ on $\Sigma_{\fg\neq 1}$}
\label{Yp0 on Sigma}
%%%%%%%%%%%%%%%%%%%%%%%%%%%%%%

We begin with the special case $q=0$. For $p=1$ this corresponds to the KW theory, while for general values of $p$ we have a $\bZ_p$ orbifold of it that preserves $\cN=1$ supersymmetry. Assuming $\kappa \neq 0$  (and thus  $\kappa^2=1$)  the trial central charge is extremized at
\bea
\epsilon_1 &= \frac{2b_1\kappa \big(16 b_2^2 -(4B p-\kappa)^2 \big)}{1-8(b_1^2+b_2^2+2B^2p^2)+32 B p \kappa(b_1^2-b_2^2)} \;, \\
\epsilon_2 &= \frac{2b_2\kappa \big(16 b_1^2 -(4B p+\kappa)^2 \big)}{1-8(b_1^2+b_2^2+2B^2p^2)+32 B p \kappa(b_1^2-b_2^2)} \;, \\
\epsilon_B &= \frac{2 \big( 2(b_1^2-b_2^2)-B p \kappa (1+8b_1^2+8b_2^2-16B^2p^2) \big)}{p \big(1-8(b_1^2+b_2^2+2B^2p^2)+32 B p \kappa(b_1^2-b_2^2) \big)} \;.
\eea
We note, rather surprisingly, that even when the background baryonic flux $B$ vanishes, we have $\epsilon_{B} \neq 0$ and thus the two-dimensional superconformal R-symmetry is mixed with the baryonic symmetry. Only when the flavor fluxes $b_{1,2}$ are also set to zero there is no mixing and the 2d and 4d R-symmetries coincide. This is a generic feature of all the examples we will discuss below. 

Evaluating the trial central charge at the extremum we find
\be
\label{crYp0}
c_r = -3 p \kappa \eta_{\Sigma} \left[ \frac{3-16(b_1^2+b_2^2+2B^2p^2)-256 \big( b_1^2b_2^2+B^4p^4-B^2p^2(b_1^2+b_2^2) \big) }{ 4 \big( 1-8(b_1^2+b_2^2+2B^2p^2)+32 B p \kappa(b_1^2-b_2^2) \big) }N^2-1 \right] \;.
\ee
An interesting case is obtained by setting the mesonic flavor fluxes to zero, \ie{} $b_1=b_2=0$:
\be
\label{conifolduni}
c_r = -3 p \kappa \eta_{\Sigma} \left[\frac14(3+16 B^2p^2)N^2-1\right] \qquad\qquad \text{for $b_1 = b_2 = 0$} \;.
\ee
This can be positive only for $\kappa=-1$. Another useful specialization is obtained by setting $B=0$:
\be
\label{conifolduni1}
c_r = -3 p \kappa \eta_{\Sigma} \left[ \frac{3-16(b_1^2+b_2^2+16 b_1^2b_2^2)}{4 \big( 1-8( b_1^2+b_2^2) \big) }N^2-1\right] \qquad \text{for $B=0$} \;.
\ee
Interestingly, both for $\kappa=1$ and $\kappa=-1$ there are regions in the $(b_1,b_2)$-plane where $c_r$ is positive. Finally, we note that setting $B=b_1=b_2=0$ (\ie{} when the  twist is purely along the superconformal R-symmetry in the UV) which requires $\kappa=-1$, one has 
\be
c_r = (\fg-1) \left[ \frac{32}{3} \, a(Y^{p,0}) - 2p \right] \qquad\qquad\quad \text{for $b_1 = b_2 = B = 0$} \;,
\ee
where $a(Y^{p,0})$ is the 4d central charge of the $Y^{p,0}$ theory given in \eqref{a,c 4D Ypq}. We will see that the leading order of this simple relation between the 2d central charge and the 4d anomaly coefficient $a$, is a {\it universal} feature that holds for a large class of theories justifying the name ``universal twist".

Before moving to other examples, let us analyze the $Y^{p,0}$ theory on a Riemann surface with $\kappa=-1$ in more detail, since this is one of the examples that we will revisit holographically in Section~\ref{Supergravity Solutions}. Specifically, we set $b_{1}=b_{2}=0$, but admit a nonzero baryonic flux $B$. The R-charges of the fields $(Y,Z,U)$ are $(\tfrac{1}{2}, \tfrac{1}{2}, \tfrac{1}{2})$ and the baryonic charges are $(p,p,-p)$, respectively.  It is easy to see that the quantization condition \eqref{quantization condition} for the background R-flux \eqref{background generator} imposes
\be
pNB=\frac{1}{4(\fg-1)} n_B \;,
\ee
where $n_B$ is an even (odd) integer  if $N(\fg-1)$ is even (odd).\footnote{To see this, consider for instance a baryonic operator made out of $N$ fields $Y$, with gauge indices appropriately contracted. The total baryonic charge is $pN$ and the R-charge is $N/2$. Thus, the quantization condition \eqref{quantization condition} in this background imposes
$2(\mathfrak g-1)N(pB-1/4)=n$, where $n\in \Bbb Z$.  Equivalently, we may write this as  $pNB=\frac{1}{4(\mathfrak g-1)}n_{B}$, where we defined $n_{B}\equiv(2n+N(\mathfrak g-1))$. We note that  $n_{B}$ is an even (odd) integer if $N(\mathfrak g-1)$ is even (odd).} Using this, equation \eqref{conifolduni} can be written as
\be
\label{conifolduni3}
c_r = \frac{32}{3}(\mathfrak g-1) \, a(Y^{p,0})+\frac{3p \, n_{B}^{2}}{2(\mathfrak g-1)}-2p(\mathfrak{g}-1)\,,
\ee
with $a(Y^{p,0})$ given in \eqref{a,c 4D Ypq}. We will reproduce this result holographically to leading order in $N$ in Section~\ref{Supergravity Solutions}.

%%%%%%%%%%%%%%%%%%%%%%%%%%%%%%
\subsubsection{$Y^{p,0}$ on $T^{2}$}
%%%%%%%%%%%%%%%%%%%%%%%%%%%%%%

\begin{figure}[t]
\centering
\includegraphics[width=4.3cm]{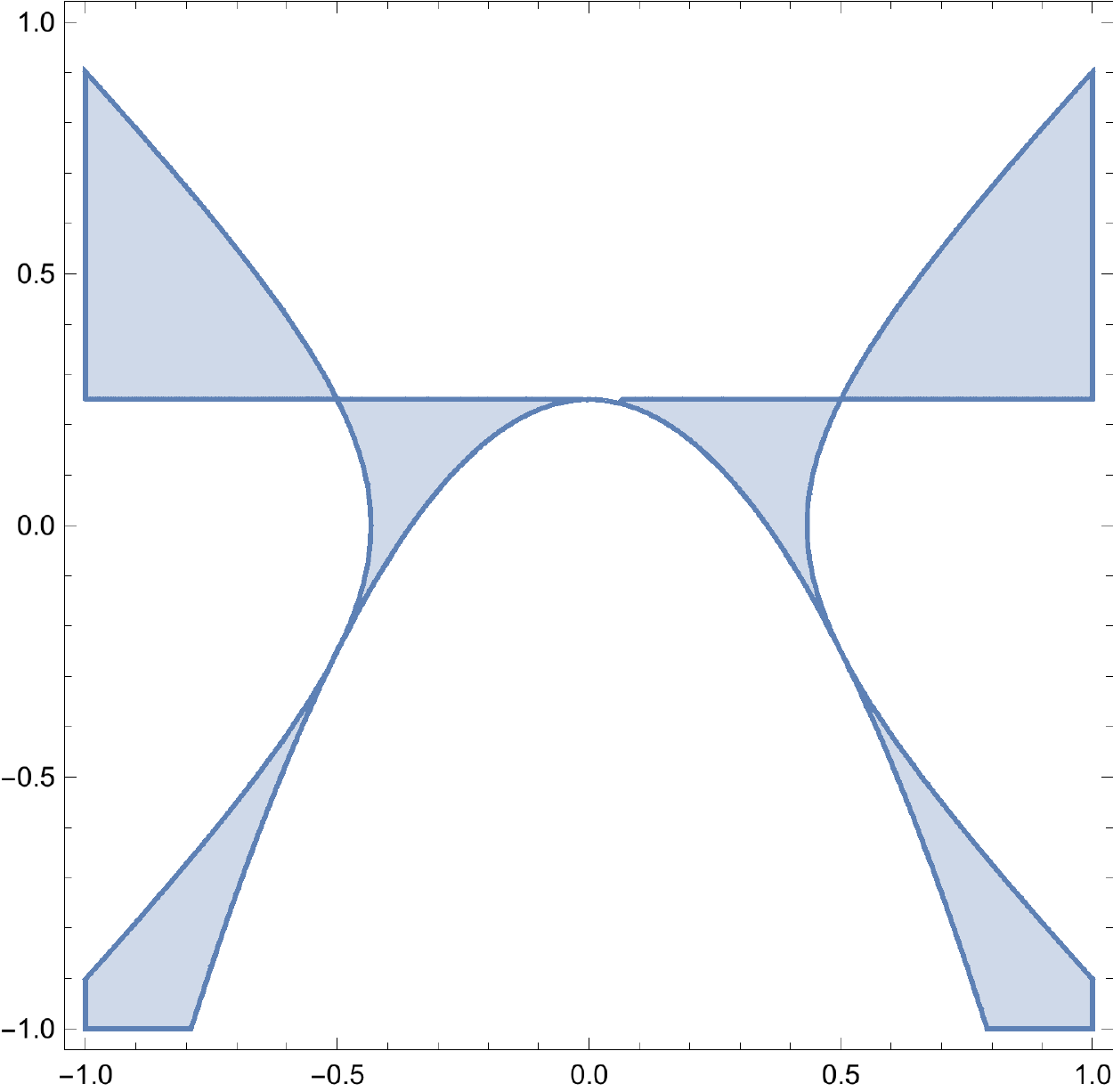}
\qquad
\includegraphics[width=4.3cm]{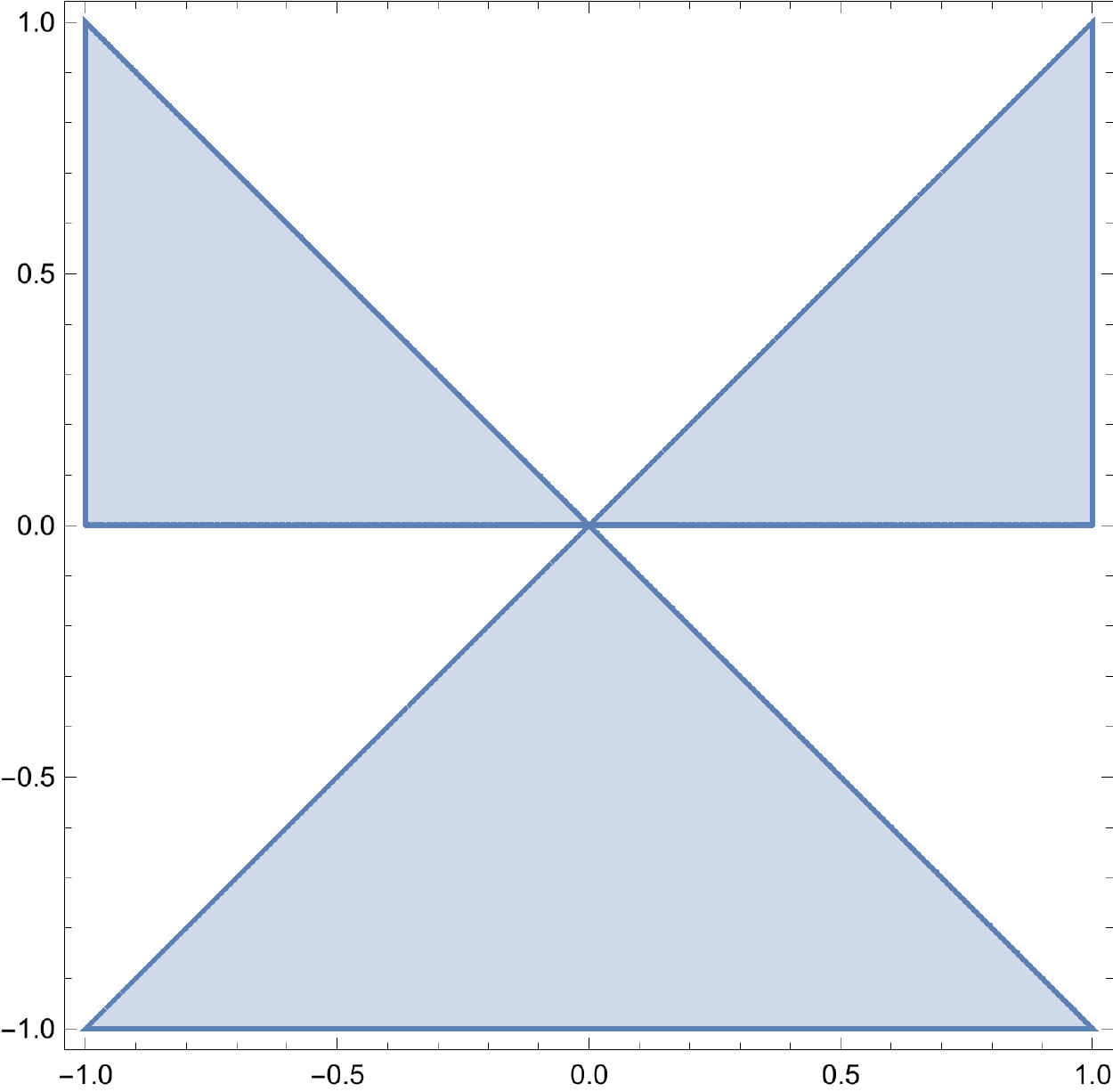}
\qquad
\includegraphics[width=4.3cm]{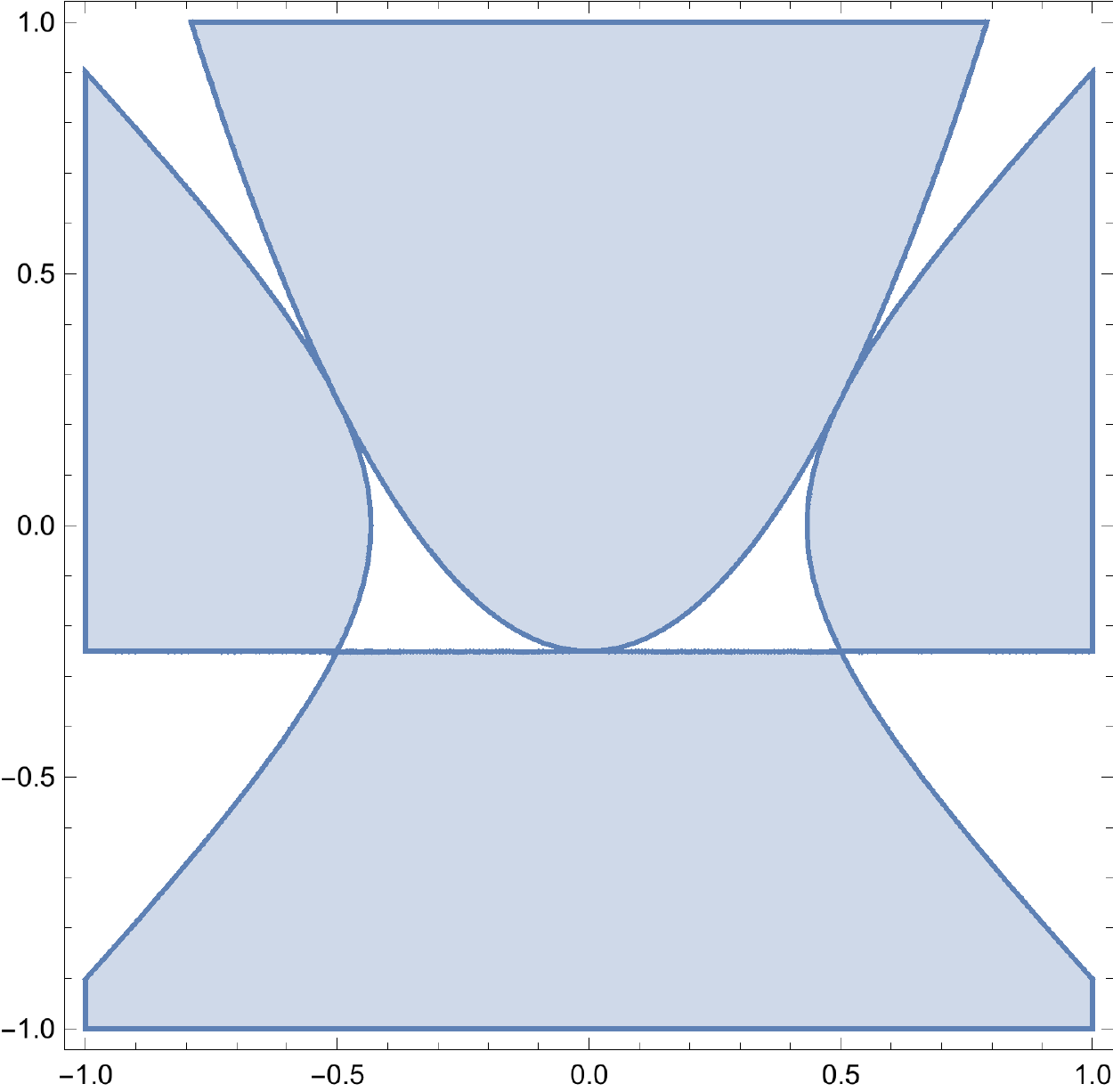}
\caption{In blue, the regions in the $(b_2,pB)$-plane where the central charge for $Y^{p,0}$ with $b_1=0$ is positive. The cases $\kappa=\{1,0,-1\}$ are presented from left to right and the horizontal and vertical axes represent $b_2$ and $pB$ respectively.
\label{Yp0plot}}
\end{figure}

Setting $\kappa=0$ and in the presence of generic background fluxes one finds
\be
\epsilon_1 = \frac{b_1(b_2^2 -B^2p^2)}{Bp(b_1^2-b_2^2)} \;,\qquad
\epsilon_2 = \frac{b_2(b_1^2 -B^2p^2)}{Bp(b_1^2-b_2^2)}  \;,\qquad
\epsilon_B = \frac{2B^2p^2-(b_1^2+b_2^2)}{2p(b_1^2-b_2^2)} \;,
\ee
which leads to the central charge
\be
\label{crYp0}
c_r = 6 \eta_{\Sigma}\frac{(b_1^2-B^2p^2)(b_2^2-B^2p^2)}{B(b_1^2-b_2^2)}N^2 \;.
\ee
When $B=0$ (or $b_{1}=b_{2}=0$) the trial central charge is linear in the parameters $\epsilon_{1,2}$ (or $\epsilon_{B}$) and one cannot apply $c$-extremization directly. When $B\neq0$ but one of the fluxes $b_{1,2}$ vanishes, one finds
\bea
\epsilon_1 &= 0 \;, & \epsilon_2 &= \frac{B p}{b_2} \;, & \epsilon_B &= \frac{1-2B^2p^2}{2pb_2^2} \;, & c_r &= 6\eta_{\Sigma}\frac{Bp^2(b_2^2-B^2p^2)}{b_2^2}N^2 & &\text{for } b_1=0 \;, \\
\epsilon_1 &= -\frac{B p}{b_1} \;, & \epsilon_2 &= 0 \;, & \epsilon_B &= \frac{-1+2B^2p^2}{2pb_1^2} \;, & c_r &= 6\eta_{\Sigma}\frac{Bp^2(-b_1^2+B^2p^2)}{b_1^2}N^2 & & \text{for } b_2=0 \;.
\eea
The case $b_{1}=0$ is special because the $SU(2)_{1}$ factor in the flavor symmetry is restored, and the analysis of Section~\ref{Supergravity Solutions} will focus on this case. As one can check from the expressions above, there are always regions in the $(b_2, B)$-parameter space where the central charge is positive, for $\kappa=0,\pm1$. This is illustrated in Figure~\ref{Yp0plot}, where we have limited the analysis to leading order in $N$ for simplicity.

\paragraph{Special twists.} Finally, let us comment on the boundaries of the colored regions in Figure~\ref{Yp0plot}. At generic points on these boundaries the central charge either diverges or becomes zero and such points are therefore excluded. Exceptions to this rule may appear at points where two such contours intersect. To obtain the value of the central charge at such points, and determine whether such a twist leads to a candidate unitary CFT in the IR, one should insert the value of the fluxes into the trial central charge \eqref{crtrgen} first, and then extremize it. Carrying this out for $\kappa=0$ and $\kappa=1$ one finds that all the boundaries in Figure~\ref{Yp0plot} are completely excluded, to leading order in $N$. For  $\kappa=-1$ the situation is more interesting and one finds that there are three special points that lead to a finite and positive central charge in the IR, namely the points $A=(1/2,1/4), \, B=(-1/2,1/4)$ and $C=(0,-1/4)$. Let us discuss these special twists for $\kappa=-1$ in more detail. The trial central charges at large $N$ read:
\eqs{\nonumber
A:&\qquad c_{tr}=\frac32 pN^{2}(\mathfrak g-1)(4-(-1+2\epsilon_{2}+2p\epsilon_{B})^{2})\,,\\ \label{ctr special Yp0}
B:&\qquad c_{tr}=\frac32 pN^{2}(\mathfrak g-1)(4-(1+2\epsilon_{2}-2p\epsilon_{B})^{2})\,,\\ \nonumber
C:&\qquad c_{tr}=\frac32 pN^{2}(\mathfrak g-1)(-4\epsilon_{1}^{2}+(1-2p\epsilon_{B})(3+2p\epsilon_{B}))\,.
}
We note that for each twist $c_{tr}$ does not depend on certain mixing parameters $\epsilon_{i}$. For the $A$ and $B$ twists it does not depend on $\epsilon_{1}$ and depends only a particular combination of $\epsilon_{2}$ and $\epsilon_{B}$ while for the $C$ twist it is independent of $\epsilon_{2}$. This implies that there are no mixed anomalies between the corresponding flavor symmetry
and the R-symmetry and thus mixing with it is irrelevant. Thus the corresponding flavor symmetry does not act at low energies and simply decouples.

Extremizing the trial central charges \eqref{ctr special Yp0} one finds that the central charges in the IR coincide and are given by
\equ{\label{ctr special Yp0explicit}
c_{r}(A)=c_{r}(B)=c_{r}(C) =6p(\mathfrak g-1)N^{2}\,.
}
It would be interesting to study these twists, and the putative CTFs they lead to, in more detail.

%%%%%%%%%%%%%%%%%%%%%%%%%%%%%%
\subsubsection{$Y^{p,p}$ on $\Sigma_{\fg\neq 1}$}
\label{subsec:YppQFT}
%%%%%%%%%%%%%%%%%%%%%%%%%%%%%%

Another special case of interest is $q=p$. For $p=1$ one has a quiver with two nodes with $\mathcal{N}=2$ supersymmetry which is a $\mathbb{Z}_2$ orbifold of $\mathcal{N}=4$ SYM \cite{Kachru:1998ys}. In this case the chiral field $Y$ is in the adjoint. For all other values of $p$ we have a $\mathbb{Z}_p$ orbifold of this $\mathcal{N}=2$ theory which preserves only $\mathcal{N}=1$ supersymmetry. 

Assuming  $\kappa\neq 0$, the trial central charge \eqref{crtrgen} is extremized for
\bea
\epsilon_1 &= - \frac{2b_1\kappa(6b_2+\kappa)}{1-12(b_1^2+b_2^2+B^2p^2+b_2Bp)} \;, \\
\epsilon_2 &=   \frac{12b_2^2-2b_2\kappa-8(b_1^2+b_2^2+B^2p^2+b_2Bp)}{1-12(b_1^2+b_2^2+B^2p^2+b_2Bp)} \;, \\
\epsilon_B &= \frac{4(b_1^2+b_2^2+B^2p^2+b_2Bp)-(12b_2^2+12 b_2 B p+2 B p \kappa)}{p \big( 1-12(b_1^2+b_2^2+B^2p^2+b_2Bp) \big) } \;,
\eea
and the right-moving central charge  reads
\be
\label{crYpp}
c_r = 3 p \kappa \eta_{\Sigma}\left[\frac{72(1-3\kappa b_2)(b_1^2+B^2p^2+b_2Bp)-8(1-9b_2^2)}{9 \big( 1-12(b_1^2+b_2^2+B^2p^2+b_2Bp) \big)}N^2+1 \right] \;.
\ee
For $b_1=b_2=0$ this simplifies to
\be
c_r = -3 p \kappa \eta_{\Sigma}\left[\frac{8(1-9B^2p^2)}{9(1-12B^2p^2)}N^2-1\right] \;.
\ee
On the other hand for $B=0$ one finds
\be
c_r = -3 p \kappa \eta_{\Sigma}\left[\frac{8 \big( 1-9(b_1^2+b_2^2)+27 \kappa b_1^2b_2 \big)}{9 \big( 1-12(b_1^2+b_2^2) \big)}N^2-1\right] \;.
\ee

%%%%%%%%%%%%%%%%%%%%%%%%%%%%%%
\subsubsection{$Y^{p,p}$ on $T^{2}$}
%%%%%%%%%%%%%%%%%%%%%%%%%%%%%%

\begin{figure}[t]
\centering
\includegraphics[width=4.3cm]{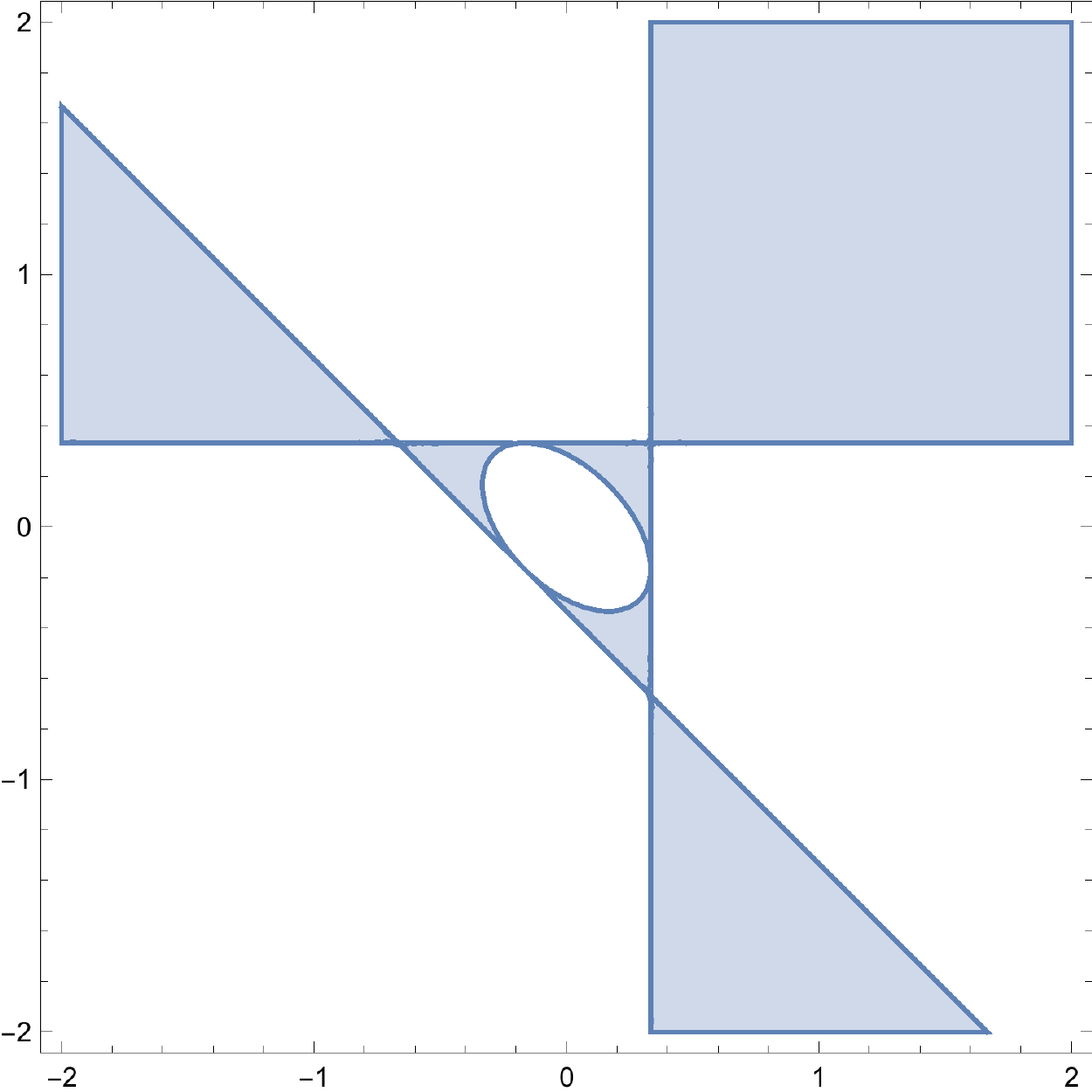}
\qquad
\includegraphics[width=4.3cm]{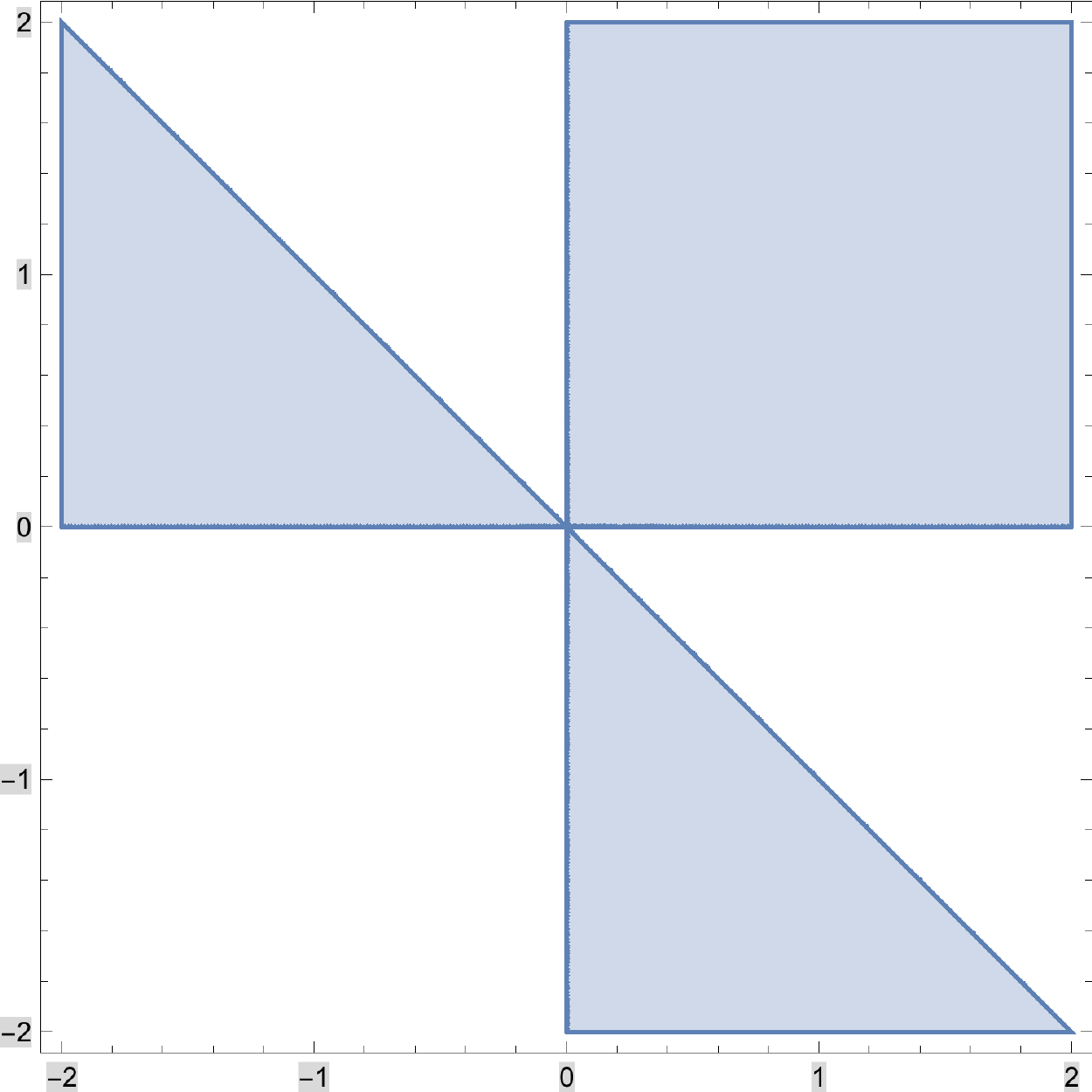}
\qquad
\includegraphics[width=4.3cm]{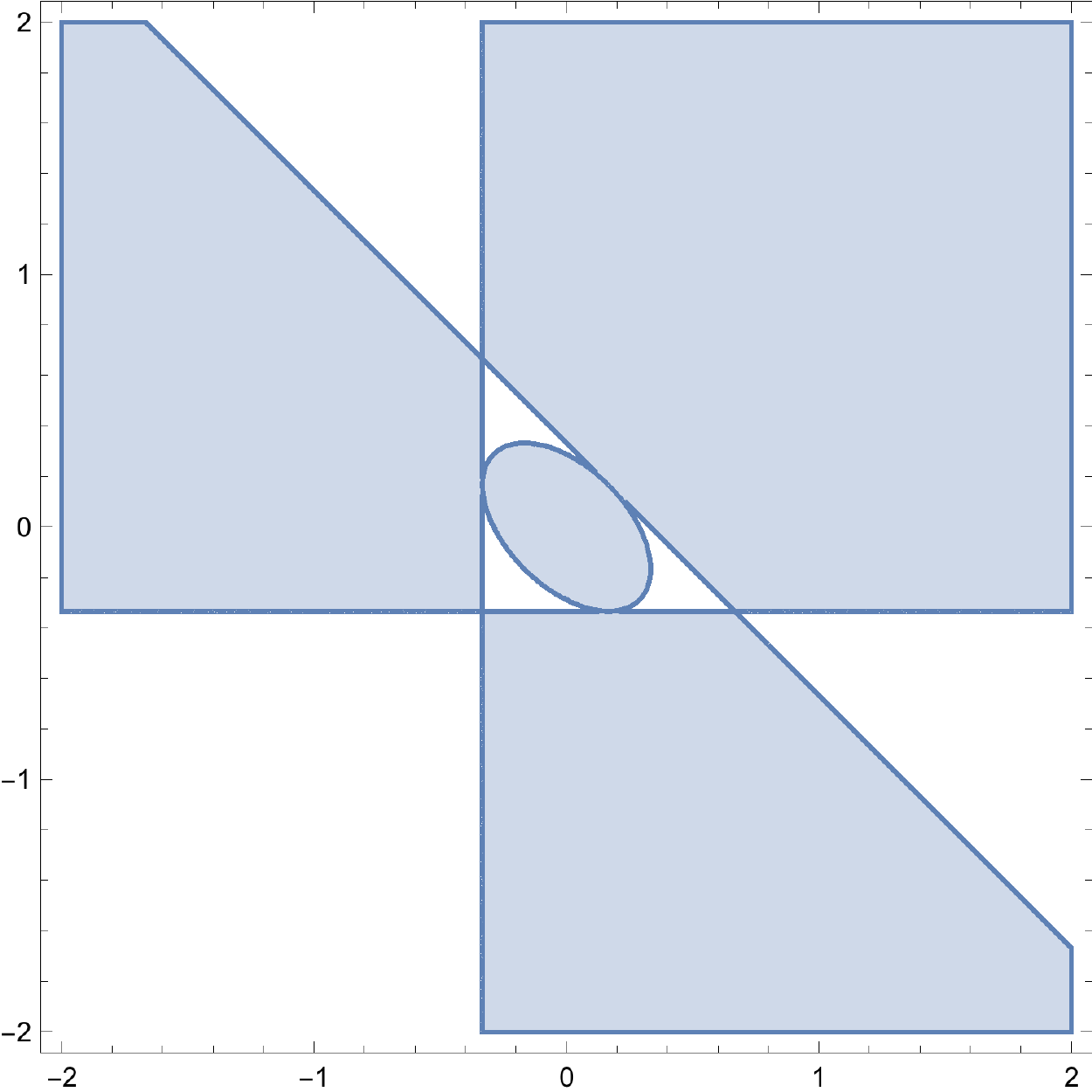}
\caption{In blue, the regions in the $(b_2,pB)$-plane where the central charge for $Y^{p,p}$ with $b_{1}=0$ is positive. The cases $\kappa=\{1,0,-1\}$ are presented from left to right.
\label{Yppplot}}
\end{figure}

Setting $\kappa=0$ one finds
\be
\epsilon_1 = \frac{b_1b_2}{b_1^2+b_2^2+B^2p^2+b_2Bp} \;, \qquad
\epsilon_2 = \frac{2}{3} - \frac{b_2}{b_1} \epsilon_1 \;, \qquad
\epsilon_B = \frac{-b_1^2+2b_2^2-B^2p^2+2b_2Bp}{3p(b_1^2+b_2^2+B^2p^2+b_2Bp)} \;,
\ee
which leads to the central charge
\be
\label{crYp0}
c_r = 6 \eta_{\Sigma}\frac{b_2p(b_1^2+B^2p^2+b_2Bp)}{b_1^2+b_2^2+B^2p^2+b_2Bp}N^2 \;.
\ee
If $B=b_1=0$ the $c$-extremization procedure seems to be applicable but one finds $c_r=0$ and thus does not lead to a candidate unitary CFT. When $B=b_2=0$ or $b_1=b_2=0$, the trial central charge is linear in the parameters $\epsilon_i$ so one cannot apply $c$-extremization directly. We will thus take at least two of the background fluxes to be non-trivial. We summarize some of the results for the $Y^{p,p}$ quivers with $b_{1}=0$ in Figure~\ref{Yppplot}. 

\paragraph{Special twists.}

As in the $Y^{p,0}$ case discussed above, generic points on the boundaries of the colored regions in Figure~\ref{Yppplot} are excluded as they lead to either a vanishing or diverging central charge. The only exceptions are found for $\kappa=-1$, at the intersection points between the ellipse and the straight lines, namely $A'=(-\frac13,\frac16),B'=(\frac16,-\frac13)$ and $C'=(\frac16,\frac16)$. The  trial central charges at large $N$ read:
\eqs{ \nonumber
A':&\qquad c_{tr}=\frac23 pN^{2}(\mathfrak g-1)(8-6\epsilon_{2}-9\epsilon_{2}^{2})\,,\\ \label{ctr special Ypp}
B':&\qquad c_{tr}=\frac23 pN^{2}(\mathfrak g-1)(-9\epsilon_{1}^{2}+(2-3p\epsilon_{B})(4+3p\epsilon_{B}))\,,\\ \nonumber
C':&\qquad c_{tr}=\frac23 pN^{2}(\mathfrak g-1)(-9\epsilon_{1}^{2}+(4-3\epsilon_{2}-3p\epsilon_{B})(2+3\epsilon_{2}+3p\epsilon_{B}))\,.
}
As seen from these expressions, and discussed below equation \eqref{ctr special Yp0}, the three twists lead to certain flavor symmetries decoupling in the IR (this is manifested by the mixing parameter $\epsilon_{i}$ not appearing in the trial central charge). The corresponding central charges of the candidate CFTs in the IR read again:
\equ{
c_{r}(A')=c_{r}(B')=c_{r}(C') =6p(\mathfrak g-1)N^{2}\,.
}
It is curious to note that these values are the same as the ones presented in \eqref{ctr special Yp0explicit}. It will be interesting to investigate further whether there is a relation between these classes of two-dimensional CFTs.

\bigskip

Finally, we comment that one might have naively expected that the central charges in \eqref{crYpp} with $B=0$ can be compared to the ones derived in \cite{Benini:2013cda}, since the theories considered here arise as the IR fixed points of $\mathbb{Z}_2\times \mathbb{Z}_p$ orbifolds of $\mathcal{N}=4$ SYM further placed on a Riemann surface, while the theories in \cite{Benini:2013cda} came from pure $\mathcal{N}=4$ SYM on a Riemann surface. However this is not the case and the central charges in \eqref{crYpp} differ from the ones in \cite{Benini:2013cda}. This suggests that the RG flow from four to two dimensions does not commute with the orbifold action. From the field theory point of view, one of the reasons is the role played by the $U(1)_B$ symmetry which is absent in $\mathcal{N}=4$ SYM (and therefore in the setup of \cite{Benini:2013cda}), but clearly plays a crucial role in the present construction since it mixes along the RG flow with the $U(1)_R$ symmetry.

%%%%%%%%%%%%%%%%%%%%%%%%%%%%%%
\subsubsection{$Y^{p,q}$ on $T^{2}$}
%%%%%%%%%%%%%%%%%%%%%%%%%%%%%%

Let us now take $\kappa=0$ and keep $p$ and $q$ general. For general values of the flavor and baryonic fluxes the central charges are lengthy and we will refrain from presenting them here. When we set $b_1=0$ we get an enhanced $SU(2)$ flavor symmetry and this will be the case of interest in the supergravity analysis. Let us focus on this choice of background flux. The trial central charge \eqref{crtrgen} is extremized for
\be
\epsilon_1 = 0 \;,\qquad
\epsilon_2 = \frac{p+w}{3q} -\frac{3pb_2^2}{3q(b_2^2+b_2Bq+B^2q^2)}\;,\qquad
\epsilon_B = \frac{4p-2w}{3q^2} -\frac{pB^2}{b_2^2+b_2Bq+B^2q^2} \;.
\ee
The right-moving central charge is particularly simple:
% (recall that for $\kappa=0$ we have $\eta_{\Sigma}=1$)
\be
\label{crYpqkappa0}
c_r = 6 p^2B\left[1-\frac{B^2p^2}{b_2^2+b_2Bq+B^2q^2}\right] N^2 \;.
\ee
If  in addition we set $b_2=0$ the result is
\begin{equation}
\label{DGKus}
c_r = -\frac{6Bp^2(p^2-q^2)}{q^2}N^2 \;,
\end{equation}
which is positive only for $B<0$. This result looks very similar to the central charges found in supergravity in Section 4.1 of \cite{Donos:2008ug}. Indeed after the redefinition $p=q_\text{\tiny DGK}$, $q=p_\text{\tiny DGK} + q_\text{\tiny DGK}$ and $BN^2 = N_{\text{\tiny DGK}} M_{\text{\tiny DGK}}$, the central charge in \eqref{DGKus} becomes
\be
c_r = \frac{6 p_\text{\tiny DGK} q_\text{\tiny DGK}^2 ( p_\text{\tiny DGK} + 2 q_\text{\tiny DGK})}{(p_\text{\tiny DGK} + q_\text{\tiny DGK})^2} N_{\text{\tiny DGK}} M_{\text{\tiny DGK}} \;,
\ee
which is identical to equation (4.18) in \cite{Donos:2008ug}.%
\footnote{It is clear from the analysis of \cite{Donos:2008ug} that there is an allowed range for the parameters in which $p_{\text{\tiny DGK}}\leq0$, $q_{\text{\tiny DGK}}>0$ and $q_{\text{\tiny DGK}}\geq|p_{\text{\tiny DGK}}|$. This is the range compatible with the values of the parameters $p$ and $q$ in $Y^{p,q}$, \ie{} with $p>0$ and $p \geq q\geq 0$.}

%%%%%%%%%%%%%%%%%%%%%%%%%%%%%%
\subsubsection{$Y^{p,q}$ on $\Sigma_{\fg\neq 1}$}
%%%%%%%%%%%%%%%%%%%%%%%%%%%%%%

Finally, we discuss the generic case of $Y^{p,q}$ on a Riemann surface with $\kappa\neq0$. For general values of $p$ and $q$ and general background fluxes it is straightforward to apply the general $c$-extremization procedure as outlined above, but the results are too unwieldy to present explicitly. Therefore we will restrict ourselves to a few special values of the background fluxes while keeping $p$ and $q$ general.

For $b_1=b_2=0$ the expression for the central charge is relatively complicated and takes the form
\bea
c_r = & -3\eta_{\Sigma} \kappa \bigg[ \frac{16}{9}a(Y^{p,q}) \\
&+\frac{4p^2q^2 B^2 [w^2(2p^3-p^2w+3q^2w) - 36 \kappa Bq^2(p^2-q^2)(pw+w^2+6q^2B^2(q^2-pw-2p^2))]}{3(2p^2+pw-q^2)(2pw-w^2+6\kappa B q^2(p-w)-12 B^2 q^4)^2}N^2\\
&+\frac{48p^2q^4 B^4(p+w) [2p^4-p(p^2+q^2)w+(p^2-4q^2)w^2+pw^3]}{3(2p^2+pw-q^2)(2pw-w^2+6\kappa B q^2(p-w)-12 B^2 q^4)^2}N^2 -\frac{p}{3} \bigg] \;.
\eea
If instead we set $b_1=B=0$ we find
\be
c_r = -3\eta_{\Sigma} \kappa \left[ \frac{16}{9}a(Y^{p,q}) +\frac{8b_2^2p^2 w}{w(2p+w)-4b_2^2(2p+w)^2}N^2 -\frac{p}{3}\right] \;.
\ee
Finally, we note that by setting the remaining flux $b_2=0$ (which requires $\kappa=-1$) one finds to leading order in $N$ again the relation
\be
\label{conifolduni2}
c_r = \frac{32}{3}(\mathfrak{g}-1) \, a(Y^{p,q})+\mathcal O(1) \;.
\ee
As we explain in the next section, this relation holds not only for $Y^{p,q}$ quivers at large $N$, but quite generally for a large class of 4d $\mathcal{N}=1$ SCFTs on Riemann surfaces, twisted by the four-dimensional superconformal R-symmetry (when this is possible).

%%%%%%%%%%%%%%%%%%%%%%%%%%%%%%
\subsection{A universal RG flow across dimensions}
\label{subsec:universal}
%%%%%%%%%%%%%%%%%%%%%%%%%%%%%%

Here we would like to show that when a four-dimensional $\mathcal{N}=1$ SCFT is placed on a Riemann surface with a partial topological twist, there is a universal relation between the conformal anomalies in two and four-dimensions. Our result is valid under the assumption that the 2d theory in the IR is indeed a SCFT with normalizable vacuum, and that there are no accidental IR symmetries. Whether this is true or not is a dynamical question which we will not be able to address in general. However if the four-dimensional theory has a gravitational dual we will establish the existence of the two-dimensional SCFT holographically.

Suppose that we have a 4d $\cN=1$ supersymmetric theory (not necessarily conformal) with global symmetry $U(1)_R \times U(1)_F \times G_F$ where $U(1)_R$ is an R-symmetry, $U(1)_F$ is a flavor symmetry, and $G_F$ is some additional non-Abelian global symmetry.\footnote{The results below generalize easily to the case where there is more than one Abelian factor in the flavor group. We refrain from discussing the general case to avoid clutter in the formulae.} The 't~Hooft anomalies of this theory are encoded in the following 6-form anomaly polynomial:
\bea
\label{4D anomaly form}
I_{6} &= \frac{k_{RRR}}{6} c_{1}(\cF_{R})^{3}+\frac{k_{FFF}}{6} c_{1}(\cF_{F})^{3}+\frac{k_{RRF}}{2} c_{1}(\cF_{R})^{2}c_{1}(\cF_{F})+\frac{k_{RFF}}{2} c_{1}(\cF_{R})\, c_{1}(\cF_{F})^{2} \\
&\quad -\frac{k_{R}}{24} c_{1}(\cF_{R})\,p_{1}(\cT_{4})-\frac{k_{F}}{24} c_{1}(\cF_{F})\,p_{1}(\cT_{4}) \;.
\eea
Here $k_{ABC}$ and $k_A$ are the cubic and linear 't Hooft anomalies, $c_1(\cF)$ is the Chern class of the bundle with curvature $\cF$,  $p_1(\cT_4)$ is the Pontryagin class of the four-manifold on which the theory is placed, and the powers of all characteristic classes are with respect to the wedge product. When the theory has a Lagrangian description, one can easily compute the anomalies as $k_{ABC}=\Tr(ABC)$ and $k_A=\Tr(A)$ where the trace is over all chiral fermions in the theory.%
\footnote{One should represent all fermions with right-moving chiral fields. Otherwise, the correct formulae should be $k_{ABC} = \Tr \gamma_5 ABC$ and $k_A = \Tr \gamma_5 A$, where $\gamma_5$ is the 4d chirality matrix.}

In a similar fashion one can encode the anomalies of a 2d theory with $\cN=(0,2)$ supersymmetry in the 4-form anomaly polynomial%
\footnote{For simplicity we again assume that the 2d theory has only a single Abelian factor in the flavor group.}
\be
\label{2D anomaly form}
I_4 = \frac{k_{RR}}{2} c_{1}(\cF_{R})^{2}+\frac{k_{FF}}{2} c_{1}(\cF_{F})^{2}+k_{RF}\,c_{1}(\cF_{R}) \, c_{1}(\cF_{F})-\frac{k}{24} p_{1}(\cT_{2}) \;,
\ee
where all the Chern and Pontryagin classes are the ones in 2d. The coefficients $k_{AB}$ are the quadratic 't~Hooft anomalies, while $k$ is the gravitational anomaly. In a theory with Lagrangian description they are given by the formula $k_{AB} = \Tr \gamma_3 AB$ and $k=\Tr \gamma_3$, where the trace is over all complex chiral fermions in the theory and $\gamma_3$ is the 2d chirality matrix (positive on right-movers).

If the theories are actually superconformal and $R$ is the superconformal R-symmetry, the relations between conformal and 't~Hooft anomalies in 4d and 2d take the following form:
\be
a = \frac{9}{32} k_{RRR} - \frac{3}{32} k_R \;,\qquad c = \frac{9}{32} k_{RRR} - \frac{5}{32}k_R \;, \qquad c_r = 3k_{RR} \;,\qquad c_r - c_l = k \;.
\ee
Here $c_{l,r}$ are the 2d left- and right-moving central charges. Superconformal symmetry also enforces $9k_{RRF} = k_F$ in 4d \cite{Intriligator:2003jj} and $k_{RF}=0$ in 2d \cite{Benini:2012cz}.

We place the 4d theory on a compact Riemann surface and implement a partial topological twist which preserves $\cN=(0,2)$ supersymmetry in the remaining two dimensions. At the level of R-symmetry and flavor symmetry line bundles, this topological twist amounts to the following replacement:
\be
\label{relation classes 4d 2d}
\cF_{R}^{(4D)} \to \cF_{R}^{(2D)} - \frac{\kappa}{2} \, t_\fg \;,\qquad\qquad \cF_{F}^{(4D)} \to \cF_{F}^{(2D)} + \epsilon \, \cF_{R}^{(2D)} + \mathfrak{b} \, t_\fg \;.
\ee
Here $t_\fg$ is the Chern class of the tangent bundle to the Riemann surface normalized in such a way that $\int_{\Sigma_\fg} t_\fg = \eta_\Sigma$. The R-symmetry background is fixed by supersymmetry. The parameter $\mathfrak{b}$, instead, represents the freedom to turn on a background magnetic flux through the Riemann surface for the $U(1)_F$ symmetry---such a parameter should be properly quantized as in (\ref{quantization condition}). We are interested in flows that lead to 2d fixed points. We have introduced the parameter $\epsilon$ because by $\cF^{(2D)}_R$ we now mean the 2d \emph{superconformal} R-symmetry, which in general is a mix between some R-symmetry derived from four dimensions and the Abelian flavor symmetries. As in Section~\ref{2D central charges}, the value of $\epsilon$ at the 2d fixed point is fixed by $c$-extremization.

To calculate the anomalies of the IR 2d SCFT, we plug the background \eqref{relation classes 4d 2d} into the 6-form \eqref{4D anomaly form}, integrate the result over $\Sigma_\fg$ (notice that $t_\fg^2=0$) and then read off the $I_4$ anomaly polynomial of the 2d theory. Extremizing the trial value of $k_{RR}(\epsilon)$ with respect to $\epsilon$ we find
\be
\label{epsilonuni}
\epsilon = -\frac{\kappa \, k_{RRF} - 2 \mathfrak{b}  \, k_{RFF}}{ \kappa \, k_{RFF} - 2 \mathfrak{b} \, k_{FFF}} \;,
\ee
and the right-moving central charge is
\be
\label{2d cR anomalies 4D}
c_r = \frac{3\eta_\Sigma}2 \bigg[ {-\kappa} \, k_{RRR} + \frac{(\kappa \, k_{RRF} - 2 \mathfrak{b}  \, k_{RFF})^2 + 2 \mathfrak{b}  \, k_{RRF} (\kappa \, k_{RFF} - 2 \mathfrak{b}  \, k_{FFF}) }{ \kappa \, k_{RFF} - 2 \mathfrak{b}  \, k_{FFF}} \bigg] \;.
\ee
The values of the other 2d anomalies are
\be
k_{FF} = \frac{\eta_\Sigma}2 \big( 2 \mathfrak{b} \, k_{FFF} - \kappa \, k_{RFF} \big) \;,\qquad c_r - c_l = \frac{\eta_\Sigma}2 \big( 2 \mathfrak{b} \, k_F - \kappa \, k_R \big) \;,\qquad k_{RF} = 0 \;.
\ee
The relation $k_{RF}=0$ precisely corresponds to the fact that we have extremized $c_r$.

Consider now the case of a 4d SCFT with $k_F=0$, and perform the partial topological twist using the exact 4d superconformal R-symmetry, \ie{} $R$ is the 4d superconformal  R-symmetry and take $ \mathfrak{b} =0$ (in cases where the R-symmetry flux on $\Sigma_\fg$ is properly quantized). Since $9k_{RRF} = k_F = 0$, from (\ref{epsilonuni}) it follows that $\epsilon=0$. This means that the IR 2d superconformal R-symmetry coincides with the UV 4d one, and no mixing with $U(1)_F$ occurs along the RG flow. For such an RG flow across dimensions, which is unitary only for $\kappa=-1$, we obtain a universal relation
\be
\label{universalmat}
\mat{ c_r \\ c_l } = \frac{16}3 (\fg-1) \mat{ 5 & -3 \\ 2 & 0} \mat{a \\ c} \;.
\ee
This result is reminiscent of the universal RG flow between four-dimensional $\mathcal{N}=2$ and $\mathcal{N}=1$ SCFTs discussed in \cite{Tachikawa:2009tt}. In our case, the RG flow is between four-dimensional $\mathcal{N}=1$ SCFTs and two-dimensional $\cN=(0,2)$ SCFTs. 

We note that for $c_{r}$ to be positive, the four-dimensional theory should satisfy
\be
\label{bound a/c}
\frac35 < \frac ac \;,
\ee
or $k_{RRR}>0$. This lower bound is compatible with the Hofman-Maldacena (HM) \cite{Hofman:2008ar} window $\frac12\leq \frac{a}{c} \leq\frac32$ for $\N=1$ SCFTs, but it places a restriction on the class of theories for which this RG flow can lead to unitary 2d SCFTs with a normalizable vacuum in the IR. On the other hand, the upper bound of the HM window implies that the 2d SCFTs at hand obey the bound $c_{r}/c_{l}\leq 9/4$. We emphasize that these inequalities hold only for the universal twist of four-dimensional theories with $k_{F}=0$.

Finally, we note that for theories with $a=c$ (\ie{} with $k_R = 0$), one has $c_r = c_l$ and
\be
\label{universalnum}
c_r = \frac{32}3 (\fg-1)\, a \;.
\ee
Notice that if $k_F \simeq 0$ or $k_R \simeq 0$ only at leading order in $N$, the statements above are still true at leading order. Since the $Y^{p,q}$ quivers have $k_F \simeq k_R \simeq 0$, the result in \eqref{universalnum} is an explanation of the universal result observed at large $N$ in many of the examples seen above, as in \eqref{conifolduni}. For CFTs with weakly coupled supergravity duals we have $a \simeq c$ and thus the universal relation in \eqref{universalnum} holds. In Section \ref{subsec:univsugra} and Appendix \ref{General formulas for the central charge} we indeed show how this comes about on the supergravity side. 

Since the $Y^{p,q}$ quiver gauge theories generically have {\it two} Abelian flavor symmetries for generic $p,q$ (one mesonic and one baryonic), the formula \eqref{2d cR anomalies 4D} is not applicable. Of course, the approach taken above is valid and one can repeat the analysis  in the case of several Abelian flavor symmetries, but we do not provide the results here. The case of $Y^{1,0}$ with purely baryonic flux $\mathfrak{b}=B$ is an exception; in this case the full flavor symmetry is enhanced to $SU(2)_{1}\times SU(2)_{2}\times U(1)_{B}$ and there can be no mixing of the R-symmetry with any flavor symmetry other than the baryonic one. In this case using the values for the anomaly coefficients $k_{***}$, one can verify that \eqref{2d cR anomalies 4D}  reproduces \eqref{conifolduni}.

Some interesting examples of $\mathcal{N}=1$ SCFTs with holographic duals for which we have $k_R\neq 0$ are discussed in \cite{Benini:2009mz, Bah:2011vv, Bah:2012dg}. For these theories one should use the general formula in \eqref{universalmat}.

%%%%%%%%%%%%%%%%%%%%%%%%%%%%%%
\subsection{D3-branes at del Pezzo singularities}
\label{subsec:D3dP}
%%%%%%%%%%%%%%%%%%%%%%%%%%%%%%

To illustrate the utility of our general result in \eqref{universalmat} let us consider the $\mathcal{N}=1$ SCFT describing the low-energy dynamics of D3-branes at the tip of Calabi-Yau three-folds which are complex cones over del Pezzo singularities, $dP_k$, $k=0,1,\ldots, 8$. These theories were originally introduced in \cite{Morrison:1998cs} and admit a dual holographic description. For $k=3,\dots,8$ the theories do not have any flavor symmetry, have rational R-charges and thus should provide an ideal testing ground for our universal formula. In the case of $dP_0 \cong \bP^2$, the theory has $SU(3)$ flavor symmetry which however cannot mix with the R-symmetry. We can also add $\bP^1 \times \bP^1$ to the list, which has $SU(2)^2$ flavor symmetry not mixable with the R-symmetry. Finally, we can also consider D3-branes in flat spacetime---of which the $\bP^2$ case is a $\bZ_3$ orbifold---giving $\cN=4$ SYM at low energies. The cases of $dP_1 \cong Y^{2,1}$ and $dP_2$, instead, are different because they have an Abelian flavor symmetry that can mix with the R-symmetry, and in fact the 4d R-charges are irrational: these theories cannot be placed on a Riemann surface in the ``universal way'' (although they can if we allow for flavor fluxes).

The conformal anomalies for the quiver gauge theories arising from the $dP_{k=3,\dots,8}$  singularities were computed for example in \cite{Intriligator:2003wr}. At leading order in $N$---or formally for gauge group $U(N)$---they are given by
\be
a_{dP_k} = c_{dP_k} = \frac{27}{4(9-k)} N^2 \;.
\ee 
The conformal anomalies of $\cN=4$ SYM are
\be
a_{\cN=4} = c_{\cN=4} = \frac14 N^2 \;.
\ee
The case of $dP_0 \cong \bP^2$ gives a $\bZ_3$ orbifold of $\cN=4$ SYM with conformal anomalies
\be
a_{\bP^2} = c_{\bP^2} = \frac34 N^2 \;.
\ee
Finally, the line bundle over $\bP^1 \times \bP^1$ gives the Klebanov-Witten theory:
\be
a_\text{KW} = c_\text{KW} = \frac{27}{64} N^2 \;.
\ee
From the universal formula \eqref{universalnum} we find the central charges of the two-dimensional SCFTs that arise from the compactification of the 4d SCFTs on a Riemann surface with $U(1)_R$ twist:
\be
\label{c2ddPk}
c^{(2d)}_{\cN=4} = \frac{8(\fg-1)}3 N^2 \;,\qquad c^{(2d)}_\text{KW} = \frac{9(\fg-1)}2 N^2 \;,\qquad c^{(2d)}_{dP_k} = \frac{72(\fg-1)}{(9-k)} N^2 \;.
\ee
These field theory results nicely reproduce a dual supergravity calculation presented in \cite{Gauntlett:2006qw} as we now show.

In Section 6.1 of \cite{Gauntlett:2006qw} the authors found a class of $AdS_3$ solutions of type IIB supergravity based on the six del Pezzo surfaces $dP_{k=3,\dots,8}$, on $\bP^2$ and on $\bP^1 \times \bP^1$. The internal seven-dimensional manifold is topologically a Sasaki-Einstein 5d manifold fibered over a closed Riemann surface of genus $\mathfrak{g}>1$. The Sasaki-Einstein manifold is in turn a $U(1)$ bundle over the four-dimensional K\"ahler-Einstein base. One can think of these solutions as arising from the backreaction of D3-branes transverse to the 5-manifold which wrap the Riemann surface. The central charges of the $\cN=(0,2)$ SCFTs dual to these solutions were computed in \cite{Gauntlett:2006qw}:
\be
\label{cJerome}
c_\text{sugra} = \frac{36 M |\chi|}{m^2 h^2 l} n^2 \;.
\ee
Here $\chi = 2-2\mathfrak{g}$ is the Euler number of the Riemann surface, $l = \text{gcd} \big\{ m,|\chi| \big\}$, $h = \text{gcd} \big\{ \frac{M}{m},|\chi| \big\}$. The numbers $(M,m)$ are as follows: for $\mathbb{CP}^2$ we have $(M,m) = (9,3)$, for $\mathbb{P}^1\times \mathbb{P}^1$ we have $(M,m) = (8,2)$, and for $dP_k$ with $k=3,\ldots,8$ we have  $(M,m) = (9-k,1)$. Finally, the number $n$ is expressed in terms of $N$ through
\be
N = - \frac{M}{m h }n \;.
\ee
The integer $N$ is the quantized 5-form flux through the five-cycle transverse to the Riemann surface wrapped by the D3-branes, and should then be identified with the rank of the gauge group in the dual field theory. We can rewrite the supergravity central charge as
\be
\label{cJeromeN0}
c_\text{sugra} = \frac{72 |\fg-1|}{M l} N^2 \;.
\ee
Using the values of $M$ and $l$ given above, we find perfect agreement with the field theory result in \eqref{c2ddPk}. For $dP_{3\dots8}$ and $\bP^1 \times \bP^1$ we have immediate matching. For the circle bundle over $\bP^2$ that gives $S^5$, \ie{} for $\cN=4$ SYM, one notices that the adjoints have R-charge $\frac23$ and so there are gauge-invariant mesonic operators of fractional R-charge: the twist is only possible on surfaces whose $\fg-1$ is a multiple of 3, then $l=3$ and the central charges match. Alternatively, for the line bundle over $\bP^2$ which leads to $\bC^3/\bZ_3$, the field theory has bi-fundamentals of R-charge $\frac23$ but the gauge-invariant mesons have integer R-charge $2$, and the twist is possible for any genus; then $l=1$ and the central charges match.

This agreement between field theory and gravitational calculations provides strong evidence that the supergravity solutions found in Section 6.2 of \cite{Gauntlett:2006qw} are dual to the 2d $\cN=(0,2)$ SCFTs which arise from a twisted compactification on $\Sigma_\fg$ of the 4d $\cN=1$ $dP_k$ SCFTs.%
\footnote{We were informed by Jerome Gauntlett that he has arrived at the same conclusion by an independent field theory calculation of the two-dimensional central charges.} In fact, we will show in the next section that this matching  holds for twisted compactifications on $\Sigma_\fg$ of a general class of 4d $\cN=1$ SCFTs with gravity duals, whenever twisting by the pure 4d superconformal R-symmetry is possible. We will also provide new examples of gravity duals to field theories twisted by baryonic flux and match their central charges. A generalization to include flavor flux is also possible, and although we provide the local backgrounds explicitly,  we leave a global analysis of these solutions and a matching of their central charge for future work.

%%%%%%%%%%%%%%%%%%%%%%%%%%%%%%
%%%%%%%%%%%%%%%%%%%%%%%%%%%%%%
\section{Supergravity solutions}
\label{Supergravity Solutions}
%%%%%%%%%%%%%%%%%%%%%%%%%%%%%%
%%%%%%%%%%%%%%%%%%%%%%%%%%%%%%

We are interested in constructing type IIB supergravity solutions of the warped-product form AdS$_3\times_{w} \mathcal M_{7}$ preserving $\cN=(0,2)$ supersymmetry. The concrete four-dimensional $\mathcal{N}=1$ SCFTs discussed above arise from D-branes at the tip of conical Calabi-Yau manifolds. This suggests that  the only non-vanishing flux in the supergravity solutions of interest is the self-dual 5-form. Thus, we search for solutions of the form
\bea
\label{eq:SGAnsatz}
ds^2_{10} &= L^2 \, \big( e^{2\lambda}  ds^2_{\text{AdS}_3} + ds^2_{\cM_7} \big) \;, \\
g_s F_{(5)} &= L^4 \,(1+\ast_{10}) \, \dvol_{\text{AdS}_3} \wedge F_{(2)} \;,
\eea
where $F_{(2)}$ is a 2-form on $\mathcal M_{7}$. The most general solution with these properties was analyzed in \cite{Kim:2005ez,Gauntlett:2007ts}, where it was shown that the internal manifold $\mathcal M_{7}$ must  locally be a $U(1)$ bundle over a six-dimensional K\"ahler manifold, whose K\"ahler potential satisfies a fourth-order nonlinear partial differential equation. Explicit solutions were further studied in \cite{Gauntlett:2006ns,Donos:2008ug}. Here, rather than searching for explicit solutions for the six-dimensional base, motivated by the field theory analysis we assume that $\mathcal M_{7}$ is a five-dimensional fibration over a Riemann surface with $SU(2)\times U(1)\times U(1)$ isometry and derive a set of BPS and Bianchi equations for this Ansatz. Of course, the final solution can be written in the form derived in \cite{Kim:2005ez,Gauntlett:2007ts}, as we have checked.

When only the metric and 5-form flux are turned on (\ie{} without any non-trivial axio-dilaton or 3-form flux), the supersymmetry variations of the spin-$\frac12$ fermions in type IIB supergravity vanish identically and the gravitino variation is given by%
\footnote{We follow the conventions of \cite{Gauntlett:2005ww}.}
\be
\label{Gravitino variations IIB}
\delta\psi_\mu = \partial_{\mu} \epsilon+ \dfrac{1}{4}\omega_{\mu ab}\Gamma^{ab}\epsilon + \dfrac{i}{192}F_{\mu \nu_1\nu_2\nu_3\nu_4}\Gamma^{\nu_1\nu_2\nu_3\nu_4}\epsilon = 0 \;,
\ee
where $\epsilon$ is a complex ten-dimensional spinor satisfying the chirality condition $\Gamma^{12345678910} \epsilon= \epsilon$.\footnote{In our notation, we denote the time direction by $x^{1}$, rather than the more conventional $x^{0}$.} The self-dual 5-form $F_{(5)} = (1+\ast_{10}) G_{(5)}$ must satisfy the Bianchi identity $dF_{(5)}=0$, and we will make a choice for $G_{(5)}$ such that
\be
\label{Bianchi FG text}
d \, G_{(5)}=d \ast_{10} G_{(5)}=0 \;.
\ee 
In principle, solutions to \eqref{Gravitino variations IIB} and \eqref{Bianchi FG text} are not necessarily solutions to the equations of motion. In our setup, however, we have checked that solving \eqref{Gravitino variations IIB} and \eqref{Bianchi FG text} for the Ansatz in \eqref{eq:SGAnsatz} leads to solutions to the equations of motion.

Once we have constructed a globally well-defined supergravity solution of the form \eqref{eq:SGAnsatz}, the central charge of the dual CFT is given by the Brown-Henneaux formula \cite{Brown:1986nw}
\begin{equation}
c_{\text{sugra}} = \dfrac{3L}{2G_{N}^{(3)}}\;,
\end{equation}
where $G_{N}^{(3)}$ is the 3d Newton constant (see Appendix~\ref{General formulas for the central charge} for conventions and explicit formulas).

\subsection*{The Ansatz}

The field theory setup suggests that we should be looking for solutions in which $\mathcal M_{7}$ is a five-dimensional fibration over a Riemann surface $\Sigma_{\mathfrak g}$:
\begin{displaymath}
    \xymatrix{ \mathcal M_{5}\ar[r] & \mathcal M _{7} \ar[d] \\
                 & \Sigma_{\mathfrak g} }
\end{displaymath}
In addition we require $\cM_5$---and therefore also $\cM_7$---to have $SU(2)\times U(1) \times U(1)$ isometry, corresponding to the flavor and R-symmetry of the dual field theory. We denote the coordinates of AdS$_3$ by $\{t,z,r\}$, the coordinates of $\Sigma_{\mathfrak g}$ by $\{x_{1},x_{2}\}$, and the remaining coordinates by $\{y,\theta, \phi, \beta, \psi\}$. The most general Ansatz compatible with these requirements is\footnote{Here we omit the overall scale factors of $L$ and $g_{s}$ from \eqref{eq:SGAnsatz}. These must be reinstated when computing the central charge.} 
\bea
\label{MetricAnsatz}
ds^2_{10} &= f_1(y)^2ds^{2}_{\text{AdS}_3} + f_2(y)^2 ds^{2}_{\Sigma_{\mathfrak g}}+ f_3(y)^2 ds^{2}_{S^{2}} + f_4(y)^2dy^2+f_5(y)^2 D\beta^2 \\
&\qquad + f_6(y)^2 \big( D\psi+f_7(y)D\beta \big)^2 \;, \\[.5em]
G_{(5)} &= \dvol_{\text{AdS}_3} \wedge \Big[ G_1(y) \, \dvol_{\Sigma_{\mathfrak g}} +G_2(y) \, \dvol_{S^{2}}+G_3(y) \, dy\wedge D\beta \\
&\qquad +G_4(y) \, dy\wedge \big( D\psi +f_7(y)\,D\beta \big) \Big] \;,
\eea
where 
\bea\nn
& ds^2_{\text{AdS}_3} = \dfrac{-dt^2+dz^2+dr^2}{r^2} \;, \quad ds^{2}_{\Sigma_{\mathfrak g}} = e^{2h(x_1,x_2)}\left(dx_1^2+dx_2^2\right) \;, \quad ds^{2}_{S^{2}} = d\theta^2+\sin^2\theta d\phi^2 \;,\\
& D\beta = d\beta + c \cos\theta\, d\phi + a_2A_{\fg}(x_1,x_2) \;, \quad D\psi = d\psi +b\cos\theta\, d\phi + a_3 A_\fg(x_1,x_2) \;,\quad dA_\fg = \dvol_{\Sigma_\fg}
\eea
and $\dvol_{X}$ is the volume form on $X$ (see Appendix~\ref{BPS equations and Bianchi identities} for details). The real parameters $\{b,c,a_2,a_3\}$ are for the moment free but will be constrained by the BPS equations. We choose $0\leq \theta \leq \pi$ and $0\leq \phi\leq2\pi$ so that $ds^{2}_{S^{2}}$ is the metric on the round $S^{2}$. The ranges of the other coordinates will be determined by requiring that the metric is globally compact and smooth and they depend on the details of the particular solutions. We will discuss this in more detail for some concrete examples below.  The parameters $a_2$ and $a_3$ specify the fibration of the five-manifold over $\Sigma_{\mathfrak{g}}$ and thus we expect them to be related to the flavor flux $b_2$ and the R-symmetry flux fixed to $\kappa/2$ by supersymmetry \eqref{background generator}. Since we impose an $SU(2)$ isometry, our solutions will capture supergravity duals of the field theory setup in Section \ref{Field theory} with vanishing flavor flux $b_1$ in \eqref{background generator}.

 In principle, a term of the form $G_{5}(y)D\beta \wedge D\psi$ in the flux is allowed, but it is easy to show that $G_{5}=0$ follows from $\delta \psi_{5}=0$ (see Appendix~\ref{BPS equations and Bianchi identities}). The function $h(x_1,x_2)$ encodes the constant curvature metric on the genus $\mathfrak{g}$ Riemann surface and is given by
\begin{equation}\label{hdef}
h(x_1,x_2) = \begin{cases} -\log \frac{1 + x_1^2 +x_2^2}2 & \text{for } \fg = 0 \\ \frac12 \log 2\pi & \text{for } \fg = 1 \\ - \log x_2 & \text{for } \fg >1\;.\end{cases}
\end{equation}
We define the normalized curvature  $\kappa = 1$ for $\mathfrak{g}=0$, $\kappa = 0$ for $\mathfrak{g}=1$, and $\kappa = -1$ for $\mathfrak{g}>1$. The symmetries of the problem suggest that we impose the following projectors on $\epsilon$:
\begin{equation}
\Gamma^{12}\epsilon = - \epsilon\;, \qquad \Gamma^{45}\epsilon = i \epsilon\;, \qquad \Gamma^{67}\epsilon = i \epsilon\;, \qquad \Gamma^{89}\epsilon = i \epsilon\;.
\end{equation}
As shown in Appendix~\ref{BPS equations and Bianchi identities}, the BPS equations impose $f_{6}(y)=\alpha_{2} f_{1}(y)$, with $\alpha_{2}$ a non-vanishing constant. The function $f_4(y)$ in \eqref{MetricAnsatz} can be freely adjusted by choosing an appropriate coordinate $y$. It is convenient to make a choice such that
\equ{f_5(y)= \frac{1}{f_1^2(y) f_ 4(y)}\,.
}
To simplify the BPS equations for the remaining functions $f_{1},f_{2}, f_{3}, f_{4},f_{7}$ it is instructive to rewrite them in terms of  the functions $\mathcal P_{1},\mathcal P_{2},\mathcal P_{3},\mathcal{Q},\mathcal P_{7}$, defined by:
\eqs{\nonumber
f_1^2(y)=\,& \sqrt{\frac{\mathcal P_2(y) \mathcal P_3(y)}{\mathcal P_1(y)}}\,, \quad f_2^2(y)= \sqrt{\frac{\mathcal P_2(y) \mathcal P_1(y)}{\mathcal P_3(y)}}\,, \quad f_3^2(y)= \sqrt{\frac{\mathcal P_3(y) \mathcal P_1(y)}{\mathcal P_2(y)}}\,,\\\label{nice form ansatz}
f_4^{2}(y)=\,&  \frac{\sqrt{\mathcal P_{1}(y)\mathcal P_{2}(y) \mathcal P_{3}(y) }}{\mathcal Q(y)}\,,
\qquad f_{7}(y)=\frac{\mathcal P_{7}(y)}{\mathcal P_{2}(y)\mathcal P_{3}(y)}\,.
}
This form of the Ansatz combined with reality and positivity of the metric requires that
\equ{
\text{sign} \big( \mathcal P_{1}(y)\mathcal P_{2}(y)\mathcal P_{3}(y) \big)=+ \;, \qquad \text{sign}\,\mathcal Q(y)=+\,.
}
The range of $y$ will be restricted by the  zeros of the function $\mathcal Q(y)$, between which $\mathcal Q(y)>0$. We shall assume that $y$ takes values in the finite range $[y_{1},y_{2}]$ between two such zeros and that in this range\footnote{Another option is that two of the three functions $\mathcal P_{1,2,3}$ are negative and the remaining one is positive. However, by simple redefinitions one can choose them all to be positive. For instance, if $\mathcal P_{1,3}<0$  and $\mathcal P_{2}>0$, one may redefine $(\mathcal P_{1}, \mathcal P_{3},\mathcal P_{7})\to (-\mathcal P_{1}, -\mathcal P_{3},-\mathcal P_{7})$, which leaves the Ansatz invariant.}
\equ{
\cP_{1}(y)>0\,, \qquad \cP_{2}(y)>0\,, \qquad \cP_{3}(y)>0\,, \qquad y\in [y_{1},y_{2}] \,.
}
In what follows we will often omit the argument $y$ in all the functions.

\subsection{General solution}

As shown in Appendix~\ref{BPS equations and Bianchi identities}, the BPS equations imply that the functions $\cP_{2,3}$ are linear in $y$:
\eqs{
\mathcal P_{2} =a_{2} \, y+C_{2}\,,\qquad \mathcal P_{3}= - c \, y +C_{3}\,,
}
 where $C_{2,3}$ are two integration constants. The functions $\cP_{1,7}$ are fixed in terms of $\cP_{2,3}$ and $\mathcal Q$ by
\equ{\label{solutions P1, P7}
\mathcal P_{1}= \frac{\alpha_{2}\left( a_{3}  \mathcal P_{3}-b \mathcal P_{2}+\mathcal P_{7}'\right)}{2 }\,,\qquad 
\mathcal P_{7}= \frac{4  \alpha_{1}\mathcal P_{2}\mathcal P_{3}-\mathcal Q'}{4  \alpha_{2}}\,,
}
where $\alpha_{1}$ is another integration constant\footnote{As discussed in Appendix~\ref{Linear transformations on beta psi} it is always possible to set  $\alpha_1=0$ by a coordinate transformation.} and  prime denotes derivative with respect to $y$. For consistency of the BPS equations one must impose the constraints
\equ{ \label{relations alphas}
 c\, \alpha_1 + b \, \alpha_2=-\frac{1}{2}\,, \qquad\qquad a_2 \, \alpha_1 + a_3\, \alpha_2 = \frac{\kappa}{2}\,.
}
Similarly, the 5-form flux  in the Ansatz \eqref{MetricAnsatz} is determined by:
\bea
\label{flux general nice}
G_{1} &= \frac{8  \, \cP_{1}\cP_{2}-2  \kappa \, \cP_{2} \cP_{3}+a_{2}\mathcal Q'}{4  \, \cP_{1}} \;, \qquad & G_{2} &= \frac{8  \, \cP_{1}\cP_{3}-2  \,  \cP_{2} \cP_{3}-c\, \mathcal Q'}{4  \, \cP_{1}} \;, \\
G_{3} &=\frac{2 \, \cP_{2}\cP_{3}(\cP_{2}+\kappa \, \cP_{3})+(c\, \cP_{2}-a_{2}\, \cP_{3})\mathcal Q'}{4 \cP_{1}\cP_{2}\cP_{3}}\;, \qquad & G_{4} &= -\alpha_{2}\(\frac{\cP_{2}\cP_{3}}{\cP_{1}}\)' \;.
\eea
Thus, the metric and the 5-form are completely determined in terms of the integration constants and the single (yet unknown) function $\mathcal Q$. The final constraint is the Bianchi identity (\ref{Bianchi FG text}), which implies a fourth-order ODE for the function $\mathcal Q$. Remarkably this ODE can be integrated twice into the following second-order ODE:
\equ{\label{diff equation for Q}
 \mathcal Q'^{2}-2 \mathcal Q \big( \mathcal Q'' -2 (\mathcal P_{2}+\kappa \mathcal P_{3}) \big) +\mathcal P_{2}\mathcal P_{3}\(- 4  y^{2}\kappa+\delta_{1}+\delta_{2} y\)=0 \;,
}
where $\delta_{1,2}$ are new integration constants. Thus, the supergravity backgrounds we are after are completely characterized by solutions to (\ref{diff equation for Q}). Although we have not found the most general solution to this equation, it is easy to see that the most general {\it polynomial} solution is at most a cubic: 
\equ{\label{polynomial ansatz Q}
\mathcal Q = q_3\,y^3 + q_2\,y^2 + q_1\,y + q_0\;.
}
In this case the functions $\cP_{1},\cP_{7}$ in \eqref{solutions P1, P7}  become linear and quadratic in $y$, respectively. Specifically, we have 
\eqs{\label{solutions Ps text}
\mathcal P_{1}&= C_{1}y+C_{0}\,,\qquad \mathcal P_{2} =a_{2}  \, y+C_{2}\,,\qquad \mathcal P_{3}= - c  \, y +C_{3}\,, \qquad 
\mathcal P_{7}= \frac{4  \alpha_{1}\mathcal P_{2}\mathcal P_{3}-\mathcal Q'}{4 \alpha_{2}}\,,
}
where $C_{0}\equiv\frac{1}{4}(C_{2}+C_{3}\kappa-q_{2})$ and $C_{1}\equiv\frac{1}{4}(a_{2}-c \kappa -3 q_{3})$. 

The solutions seem to depend on the parameters $\{\alpha_{1}, \alpha_{2}, a_{2},a_{3},b,c,\kappa,q_{i}, C_{2},C_{3},\delta_{1},\delta_{2}\}$. However, these are not all independent. The parameters $\alpha_{1,2}$ can be set to a convenient value by a choice of coordinates (see Appendix~\ref{Linear transformations on beta psi}), and we consider $a_{3},b$  fixed in terms of other parameters by \eqref{relations alphas}. Finally, plugging the expressions for $\mathcal Q, \cP_{2},\cP_{3}$ into the Bianchi identity  \eqref{diff equation for Q} leads to a number of nonlinear constraints among the remaining  parameters $\{a_{2},c,\kappa,q_{i}, C_{2},C_{3},\delta_{1},\delta_{2}\}$, with many different branches of solutions, depending on the values of $a_{2},c, \kappa$. All the explicit supergravity solutions that we discuss below and in Appendix~\ref{General  solution app} arise from different solutions to these constraints. 

For the purpose of comparison with the field theory analysis of Section~\ref{Field theory}, we are interested in solutions describing  $Y^{p,q}$ manifolds fibered over the Riemann surface. Of course, our general Ansatz captures not only those solutions, but all  solutions with (at least) the same isometry, including for instance AdS$_3\times S^{3}\times T^{4}$. In this section we will focus on the solutions relevant to the field theory analysis. Before we do this however it is important to understand whether $\mathcal{M}_7$ can have any conical singularities prior to specifying any particular solution.

\subsubsection{Absence of conical singularities}
\label{Absence of conical singularities}

One may worry that at zeros of the function $\mathcal{Q}$ the metric might be singular.  It is easy to see, however, that such potential singularities are at most conical, and in fact can always be removed by an appropriate choice of coordinate periodicity. This follows from the form of the Ansatz and the BPS equations and thus holds for any solution in this class. We first make a linear change of variables $\beta = w_1\tilde{\beta} + w_2 \tilde{\psi}$ and $\psi = w_3 \tilde{\beta} + w_4 \tilde{\psi}$ in \eqref{MetricAnsatz}, where $w_{1,2,3,4}$ are real constants, and study the metric in the $(y,\tilde \beta)$ subspace. As shown in Appendix~\ref{Linear transformations on beta psi}:
\begin{equation}
ds^2_{2} = \frac{\sqrt \mathcal P}{\mathcal Q} \left[dy^2 + w^{2} \dfrac{\alpha_2^2 \mathcal Q^{2}}{w_2^2\mathcal{P}_1\mathcal Q+\alpha_2^2(w_4\mathcal P_2\mathcal P_3+w_2\mathcal P_7)^2} \, \big( d\tilde \beta + \ldots \big)^2 \right]\;,
\end{equation}
where we defined $\mathcal P\equiv \mathcal{P}_1\mathcal{P}_2\mathcal{P}_3$ and $w^{2}\equiv(w_1w_4-w_2w_3)^2$.
Near a zero of $\mathcal Q$ at $y=y_i$ we expand $\mathcal Q(y) \approx \mathcal Q'(y_i)(y-y_i)$ and, defining the new radial variable
\begin{equation}
r^2 = 2|y-y_i|\;,
\end{equation}
we have
\begin{equation}
ds^2_{2} \,\approx\, \dfrac{2\sqrt{\mathcal P(y_{i})}}{|\mathcal Q'(y_i)|} \left[ dr^2 + \dfrac{w^{2}}{4} \dfrac{ \mathcal Q'(y_i)^2}{ \big( w_4\cP_{2}(y_{i})\cP_{3}(y_{i})+w_2\cP_7(y_i) \big)^2} \, r^2 \, \big( d\tilde \beta+\ldots \big)^2 \right]\;.
\end{equation}
To avoid conical singularities one has to ensure that the coefficient of $r^2(d\tilde \beta+\ldots)^2$ is the same at {\it all} zeros $y_{i}$ of $\mathcal Q$, and choose the periodicity of $\tilde \beta$ accordingly. The functional identity relating $\cP_{7}$ to $\mathcal Q'$ in \eqref{solutions P1, P7} ensures that this is indeed the case; choosing
\begin{equation}
w_4=-\alpha_1\;, \qquad\qquad w_2=\alpha_2\;,
\end{equation}
and using \eqref{solutions P1, P7}, the 2d metric near a zero of $\mathcal Q$ becomes
\begin{equation}
ds^2_{2}  \approx \dfrac{2\sqrt{\mathcal P(y_{i})}}{|\mathcal Q'(y_i)|} \left[ dr^2 + 4(\alpha_{1}w_1+\alpha_{2}w_3)^2r^2(d\tilde \beta+\ldots)^2 \right]\;.
\end{equation}
All conical singularities are avoided by fixing the periodicity of $\tilde \beta$ to be $2\pi$ and choosing, say, $w_{3}$ such that
\begin{equation}
\alpha_{1}w_1+\alpha_{2}w_3=\pm\frac{1}{2}\,.
\end{equation}

%%%%%%%%%%%%%%%%%%%%%%%%%%%%%%%%%%%%%%%%
\subsection{$Y^{p,q}$ on  $\Sigma_{\fg>1}$  with universal twist}
\label{subsec:univsugra}
%%%%%%%%%%%%%%%%%%%%%%%%%%%%%%%%%%%%%%%%

As we have seen on the field theory side, when the flavor flux $b_{2}$ vanishes, \ie{} for a twist performed using the UV superconformal R-symmetry, the RG flow is special and universal. It is natural to expect that $b_2=0$ is mapped to $a_2=0$ in our supergravity Ansatz. Indeed in this case the supergravity solutions considerably simplify as we now show. After setting $a_2=0$ there are still various branches of solutions depending on the values of $c$ and $\kappa$. Assuming $\kappa\neq 0$ and $c\neq 0$, one such solution is  (see Appendix~\ref{BPS equations and Bianchi identities}):
\bea
\label{solution universal Ypq fibered}
ds^2_{10} &=ds^{2}_{\text{AdS}_3} + \frac{3}{4} ds^{2}_{\Sigma_{\mathfrak g>1}} + \frac{3|1-c{y}|}{8}  (d\theta^2+\sin^2\theta \, d\phi^2) \\
&\qquad + \frac{9}{8}\frac{|1-c{y}|}{(2c{y}^3-3{y}^2+{a})}d{y}^2 +\frac{1}{8} \frac{(2c{y}^3-3{y}^2+{a})}{|1-c{y}|}\left(d\beta + c \cos\theta \, d\phi\right)^2 \\
&\qquad + \frac{1}{4} \Big( d\psi -\cos\theta \, d\phi+{y}\left(d\beta + c \cos\theta \, d\phi\right)-\frac{dx_1}{x_2} \Big)^2 \;,\\
G_{(5)}&=\dvol_{\text{AdS}_3} \wedge \Big( 2 \dvol_{\Sigma_{\mathfrak g>1}}+ \frac{|1-c y|}{4}\dvol_{S^{2}}+\frac{1}{4}dy\wedge (d\beta+c \cos \theta \, d\phi) \Big) \;,
\eea
where $a$ is the only remaining integration constant. This solution exists only for $\kappa=-1$ (\ie{} $\mathfrak g>1$). The internal metric is precisely the metric on $Y^{p,q}$ written in canonical form as in \cite{Gauntlett:2004yd}, fibered over $\Sigma_{\mathfrak g>1}$ in such a way that the fibration is non-trivial only along the Reeb vector $\partial_{\psi}$. This is a consequence of setting $a_{2}=0$. As in the case of the standard $Y^{p,q}$, since we have assumed\footnote{It is in fact possible to set $c=0$ in this solution, corresponding to $Y^{p,0}$ fibered over $\Sigma_{\mathfrak g>1}$. However, in this case there is a more general solution which we discuss in Section~\ref{Conifold on Sigma with baryonic flux}. }  $c\neq 0$  it can be rescaled to $1$ and  $0<a<1$. In fact, this metric looks like those found in Section 6.1 of \cite{Gauntlett:2006qw}, namely
\be
\label{Metriccneq0kneq061}
ds^2_{10} = ds^2_{\text{AdS}_3} + \frac{3}{4} ds^2_{\Sigma_{\mathfrak g>1}} + \frac{9}{4} \wt{ds}^2_{\text{SE}_5} \;,
\end{equation}
with $\wt{ds}^2_{\text{SE}_5}$ a five-dimensional Sasaki-Einstein manifold fibered over $\Sigma_{\mathfrak g>1}$. In the case at hand, SE$_{5}=Y^{p,q}$. Using the general formulas for the supergravity central charge presented in Appendix~\ref{General formulas for the central charge}, the volume of the $Y^{p,q}$ manifolds computed in \cite{Gauntlett:2004yd}, and the AdS$_{5}$/CFT$_{4}$ relation $a_{4D}^\text{sugra}=\frac{\pi^{3}N^{2}}{4 \text{Vol}(\text{SE}_{5})}$, the central charge of the two-dimensional CFT dual to the AdS$_3$ solution in \eqref{Metriccneq0kneq061} can be written as
\equ{\label{3/32  sugra}
c_\text{sugra}=\frac{32}{3}(\mathfrak g-1)\, a_{4D}^\text{sugra}\,.
}
This is in perfect agreement with the universal field theory result (\ref{universalnum}) obtained by $c$-extremization. This is strong evidence that the background in \eqref{solution universal Ypq fibered} describes the IR fixed point of 4d $Y^{p,q}$ SCFTs with AdS$_{5}$ gravity duals, placed on $\Bbb R^{2}\times \Sigma_{\mathfrak g>1}$ with a partial-topological twist along the UV superconformal R-symmetry. In fact, in this case there exists a consistent truncation of type IIB supergravity to five-dimensional minimal supergravity \cite{Buchel:2006gb}.\footnote{Five-dimensional minimal gauged supergravity arises also from a consistent truncation of more general type IIB \cite{Gauntlett:2007ma} and M-theory \cite{Gauntlett:2006ai} compactifications. Our universal flow should therefore also exist in these constructions.} Within the five-dimensional theory it is possible to construct the entire RG flow connecting the AdS$_5$ and AdS$_3$ backgrounds at hand analytically \cite{Klemm:2000nj, Maldacena:2000mw, Bobev:2014jva}. We believe that these supergravity and field theory results amount to very strong evidence for the proposed duality.

\subsection{ $Y^{p,0}$ on $\Sigma_{\fg>1}$ with baryonic flux}
\label{Conifold on Sigma with baryonic flux}

As seen on the field theory side,  in the case of $Y^{p,q}$ quivers an  interesting generalization of the universal twist arises by turning on background baryonic and mesonic flavor fluxes. Here we identify the gravity dual to $Y^{p,0}$ with purely baryonic flux $B$ ($i.e.$, $a_{2}=0$) discussed in Section~\ref{Yp0 on Sigma}.

Setting $a_{2}=c=0$ in \eqref{diff equation for Q} and solving for the remaining parameters one finds that only $\kappa=-1$ is allowed.  After some coordinate redefinitions and an overall rescaling, the metric and 5-form  read
\bea
\label{conifold with v}
ds^{2}_{10} &= ds^{2}_{\text{AdS}_3} + \frac{v^{2}+v+1}{4v}ds^{2}_{\Sigma_{\mathfrak g > 1}} + \frac{v^{2}+v+1}{4(v+1)} \( d \theta^2+\sin^2 \theta \, d \phi^2+\frac{1}{v}\big( dw^{2}+ \sin^{2}w \, d\nu ^{2} \big) \) \\
&\qquad \qquad \qquad \qquad \qquad \qquad  + \frac{1}{4} \(d \psi-\cos \theta \, d \phi-\cos w \, d \nu- \frac{dx_{1}}{x_{2}}\)^2 \;,\\
G_{(5)} &= \dvol_{\text{AdS}_3} \wedge \( \frac{(v+1)^{2}}{2v} \dvol_{\Sigma_\fg} + \frac{1}{2(v+1)} \Big( v^{2} \, \dvol_{S^2_{\theta\phi}} + \frac{1}{v} \dvol_{S^2_{w \nu}} \Big) \) \;,
\eea
where $v>0$ is a parameter controlling the relative size of $\Sigma_{\fg}$ and the two $S^{2}$'s in the metric. As we show below, the parameter $v$ also controls the baryonic flux in the dual field theory.

Taking the standard periodicities for the $S^{2}$ coordinates $\theta, w\in [0, \pi]$ and $\phi, \nu \in [0,2\pi)$,  the geometry of $\cM_7$ is a $U(1)$ bundle over $\Sigma_\fg \times S^2 \times S^2$. Letting $\psi \cong \psi + 2\pi \ell_\psi$,  the first Chern classes are $2 (1-\fg)/\ell_\psi$, $2/\ell_\psi$, and $2/\ell_\psi$, respectively. Thus, quantization imposes
\be
\psi \cong \psi + \frac{4\pi}{m}\,,
\ee
with $m$ an integer. In the case of maximal length, $m=1$, this describes a fibration of the conifold $Y^{1,0} \cong T^{1,1}$ corresponding to the well-studied Klebanov-Witten (KW) theory \cite{Klebanov:1998hh}. Higher values of $m$ describe a $\Bbb Z_{m}$ orbifold of this theory along $\psi$, but it is well-known that only $m=2$ (which corresponds to a fibration of $\Bbb F_{0}$) preserves supersymmetry. 
To obtain $Y^{p,0}$ we proceed as follows. We start with the conifold ($m=1$) which is subjected to the following identifications:
\be
(\phi, \nu ,\psi) \,\cong\, (\phi + 2\pi, \nu, \psi + 2\pi) \,\cong\, (\phi, \nu + 2\pi, \psi + 2\pi) \,\cong\, (\phi, \nu, \psi + 4\pi) \;,
\ee
and perform a $\bZ_p$ orbifold along $\nu$:
\be
\big( \phi, \nu, \psi \big) \,\cong\, \Big( \phi, \nu + \frac{4\pi}p, \psi \Big) \;.
\ee
This orbifold does preserve supersymmetry: one can check in the original conifold geometry--- the CY$_3$ given by the cone over $T^{1,1}$---that the Killing spinor is invariant under $\partial_\nu$, as is the holomorphic $(3,0)$-form $\Omega$. Notice that for $p=2$ this is the same as a $\bZ_2$ orbifold along $\psi$, but not for higher values of $p$.

Using the formulas provided in Appendix~\ref{General formulas for the central charge} we find that the central charge corresponding to \eqref{conifold with v} is given by
\be
\label{c conifold v}
c_\text{sugra} = 6p(\mathfrak g-1)\frac{v^{2}+v+1}{(1+v)^{2}}N^2 \;.
\ee
We claim that \eqref{conifold with v} describes the IR fixed point of the $Y^{p,0}$ field theory placed on $\Sigma_{\fg>1}$ with baryonic flux, considered in Section~\ref{Yp0 on Sigma}. The parameter $v$, as we shall show below, is related to the baryonic flux $B$ in the field theory. To make contact with the field theory result \eqref{conifolduni3}  for the central charge, we need to discuss the topology of $\cM_7$ in more detail. The 2-forms $\dvol_\Sigma$, $\dvol_{\theta\phi}$ and $\dvol_{w\nu}$ are closed and potentially in cohomology.  However because of the existence of $d e_\psi$, one linear combination vanishes in cohomology. Correspondingly, there are three 5-cycles one can construct in the geometry: 1) the $Y^{p,0}$ fiber of $\cM_7$ at a fixed point on $\Sigma_\fg$; 2) fibrations of $S^3$ (coming from $Y^{p,0} \cong S^3 \times S^2$) over $\Sigma_\fg$, and two representatives of $S^3$ are at fixed $(\theta,\phi)$ or fixed $(w,\nu)$. By integrating $*G_5$ on those 5-cycles we obtain, respectively:
\be
N = 4 \pi^3 \Lambda \frac{v^2+v+1}v \;,\qquad\qquad N_1 = \frac{\fg-1}{v+1} N \;,\qquad\qquad N_2 = (\fg-1) \frac{v}{v+1} N \;,
\ee
where $\Lambda\equiv\frac{1}{(2\pi l_{s})^{4}}\frac{L^{4}}{p\,g_{s}}$. Notice that
\be
\label{cohomology relations}
N_1 + N_2 = (\fg-1) N \;,\qquad n_B \equiv N_2 - N_1 = (\fg-1) \frac{v-1}{v+1} N \;.
\ee
The first relation is precisely the relation in homology, and $\dim H^5(\cM_7, \bR) = 2$. We interpret $N$ as the number of D3-branes, while $n_B$ is proportional to the baryonic flux. We note that $n_B$ is an even (odd) integer if $N(\fg-1)$ is even (odd). We also note that under the replacement $v \to 1/v$ (which exchanges the two $S^{2}$'s in (\ref{conifold with v})), one has $N\to N$ while $n_B\to- n_{B}$. The solution with $v=1$, and hence $n_B =0$, corresponds to the universal twist. Solutions with non-trivial baryonic flux have other values of $v$, fixed by
\be
\label{quant cond v}
v = \frac{(\fg-1)N + n_B}{(\fg-1)N - n_B} \;.
\ee
Finally, using \eqref{quant cond v} in \eqref{c conifold v} gives
\be
c_\text{sugra} = \frac{32}{3}(\mathfrak g-1) \, a(Y^{p,0})+\frac{3p \, n_B^2}{2(\mathfrak g-1)} \;,
\ee
which matches precisely the field theory result \eqref{conifolduni3} at large $N$. This provides strong evidence that  \eqref{conifold with v}  is the gravity dual describing the IR limit of $Y^{p,0}$ quiver gauge theories placed on $\Sigma_{\fg>1}$, twisted by the superconformal R-symmetry and  baryonic flux.

%We recall that the KW theory with baryonic flux was considered in Section~\ref{Yp0 on Sigma}. To make contact with the field theory result \eqref{conifolduni3}, we note that (\ref{cohomology relations}) and the fact that $N, N_1, N_2 \in\bZ$ imply that if $N(\fg-1)$  is odd, so is $n_B$ and we identify $n_B = N_B$. On the other hand, if  $N(\fg-1)$ is even, so is $n_B$ and we identify  $n_B = 2 N_B$. Thus, we can write
%\be
%c_\text{sugra} = \frac{32}{3}(\mathfrak g-1) \, a(Y^{p,0}) + \frac{3ph^2 N_B^2}{2(\mathfrak g-1)}
%\ee
%where $h=\text{gcd} \big\{ 2,N(\fg-1) \big\}$. 
% The solution with $N_B=0$ corresponds to setting $v=1$, and corresponds to the universal twist of the $Y^{p,0}$ theory over $\Sigma_\fg$.

\subsection{$Y^{p,q}$ on $T^{2}$ with baryonic flux}
\label{subsec:YpqT2sugra}

Another interesting solution is found in the case $c\neq 0$ (which we rescale to $1$ here). 
For simplicity we discuss only the case $\kappa=0$ and, as in the previous subsection, we set the mesonic flavor flux $a_{2}$ to zero. Setting $c=1$ and $a_{2}=0$ in \eqref{diff equation for Q}, solving the constraints among the remaining parameters and performing some coordinate redefinitions, we find:
\bea
\label{GDK}
ds^2_{10} &= \frac{b}{\sqrt{\tilde y}}ds^2_{\text{AdS}_3} + \frac{\sqrt{\tilde y}}b ds^2_{T^2} + \frac{1}{4 b \sqrt{\tilde y}}ds^{2}_{S^{2}} + \frac{b}{4\tilde{y}^{5/2} \big( b^2 - (1-\tilde y)^2 \big)} d\tilde y^2 \\
&\quad + \frac{b^2 - (1-\tilde y)^2 }{ 4b\sqrt{\tilde y} (b^{2}-1+2\tilde y)}(d\tilde \beta + \cos \theta\, d\phi)^2 + \frac{b^{2}-1+2\tilde y}{4b\sqrt{\tilde y}} \Big( d\tilde \psi-\frac{\tilde y}{b^{2}-1+2\tilde y}(d\tilde \beta +\cos \theta\,  d \phi) \Big)^2 \;.
\eea
The 5-form  is given in the Appendix---see \eqref{GDK App} and discussion below it. This solution was found previously in \cite{Donos:2008ug} (Section 4.1), where its central charge was computed and  shown to depend on  four integers $p_{\text{\tiny DGK}},q_{\text{\tiny DGK}},M_{\text{\tiny DGK}},N_{\text{\tiny DGK}}$ and is given by 
\equ{
c_{\text{sugra}}= \frac{6p_{\text{\tiny DGK}}q_{\text{\tiny DGK}}^{2}(p_{\text{\tiny DGK}}+2q_{\text{\tiny DGK}})}{(p_{\text{\tiny DGK}}+q_{\text{\tiny DGK}})^{2}} M_{\text{\tiny DGK}}N_{\text{\tiny DGK}}\,.
}
By making the identifications $p=q_{\text{\tiny DGK}}, q=p_{\text{\tiny DGK}}+q_{\text{\tiny DGK}}$ and $BN^2=M_{\text{\tiny DGK}}N_{\text{\tiny DGK}}$, the central charge matches exactly  our field theory result \eqref{DGKus}. This is strong evidence that \eqref{GDK} is the gravity dual to the IR limit of  $Y^{p,q}$ quiver gauge theories on $T^2$, twisted by the superconformal R-symmetry and  baryonic flux.

\subsection{Solutions with  flavor flux}

Finally, we briefly comment on turning on the mesonic flavor flux, which is controlled by the parameter $a_{2}$. There are several branches of solutions to the Bianchi identity, which we  give in detail in  Appendix~\ref{Solutions with flavor flux: a2 not zero}. Here we give only the local form of the solution, leaving a careful analysis of global properties and computation of the central charge for future work. Assuming $\mathfrak g>1$ and $c\neq 0$, the metric is given by 
\bea
\label{metric general nice text}
ds^{2}_{10} &= \sqrt{\frac{y(3+4a_{2}y)}{3(1-a)a_{2}^{2}-4C_{1}y}} \, ds^{2}_{\text{AdS}_3} + \sqrt{\frac{(3+4a_{2}y) \big( 3(1-a)a_{2}^{2}-4C_{1}y \big) }{ 16y} } \, ds^{2}_{\Sigma_{\mathfrak g>1}} \\
&\quad + \sqrt{\frac{y \big( 3(1-a) a_2^2 - 4C_{1}y \big)}{3+4a_{2}y}} \, ds^2_{S^2} + \frac{\sqrt{y (3+4a_{2}y) \big( 3(1-a) a_{2}^{2}-4C_{1}y \big) } }{ 4\mathcal Q} \, dy^2 \\
&\quad + \frac{4\mathcal{Q} \sqrt{ y \big( 3(1-a)a_2^2 - 4C_1 y \big)} }{ y^{2}(3+4a_{2}y)^{3/2}} \, D\beta^2 + \frac{1}{4}\sqrt{\frac{y(3+4a_{2}y)}{3(1-a)a_{2}^{2}-4C_{1}y}}\(D\psi+\frac{2\mathcal{Q}'}{y(3+4a_{2}y)}D\beta \)^2 \;,
\eea
where $D\beta=d\beta+\cos \theta\, d\phi + a_{2} A_\fg$, $D\psi=d\psi - \cos \theta\, d\phi - A_\fg$,
\begin{multline}
\cQ = \frac{3 (1-a) \big( 7+a_{2}^{2}+8C_{1}-4a_{2}(1+C_{1}) \big) }{ 16(a_{2}-1)^{2}} + \frac{3a_2 (1-a) (a_{2}-4C_{1}-5)}{4(a_{2}-1)} \, y \\
+ \frac34 \big( 1-4(a-1)a_{2}^{2}\big) y^2 + \frac13(1+a_{2}-4C_{1})y^{3} \;,
\end{multline}
where $C_1 = -\frac14 \big( 1+a_{2}+2\sqrt{1-a_{2}+a_{2}^{2}} \big)$; here the parameter $a$ is the only remaining integration constant after solving the BPS equations and the Bianchi identity and $a_{2}\neq 1$ controls the flavor flux. The 5-form flux, which we do not write here, is determined by the formulas in Appendix~\ref{Case kappa not 0, c not 0}. In the special case $a_{2}=0$ the solution coincides with (\ref{solution universal Ypq fibered}) with $c=1$ (up to a simple change of coordinates). The case $a_{2}=1$ is a special branch (see Appendix~\ref{Solutions with flavor flux: a2 not zero}). For $\fg=1$ there are two other branches of solutions, given in Appendix~\ref{Case kappa=0, c not 0}.

 %%%%%%%%%%%%%%%%%%%%%%%%%%%%%%%%%%%%%
\section{Discussion}
\label{sec:conclusions}
%%%%%%%%%%%%%%%%%%%%%%%%%%%%%%%%%%%%%

In this paper we have argued for the existence of a vast landscape of two-dimensional conformal field theories with $\cN=(0,2)$ supersymmetry. These theories arise through twisted compactifications of four-dimensional $\mathcal{N}=1$ SCFTs on a smooth Riemann surface. If the four-dimensional theory has a weakly-coupled supergravity dual, we can construct the holographic RG flows which in many cases lead to the supergravity duals to the two-dimensional IR fixed points. We have illustrated in detail how these general ideas work for the case of $Y^{p,q}$ quiver gauge theories that arise from D3-branes probing toric Calabi-Yau singularities. We have also argued that there is a universal RG flow across dimensions connecting 4d $\mathcal{N}=1$ and 2d $\cN=(0,2)$ SCFTs. This flow bears a resemblance to the universal flow between 4d $\mathcal{N}=2$ and $\mathcal{N}=1$ SCFTs discussed in \cite{Tachikawa:2009tt}.

Our supergravity solutions for the general $Y^{p,q}$ theories suggest that the two-dimensional SCFTs have, in general, large conformal manifolds. Some of the exactly marginal deformations are easy to identify. For $\fg>1$ there are the $3\fg-3$ complex structure deformations of the Riemann surface\footnote{We conjecture that similarly to the analysis in
\cite{Anderson:2011cz} the K\"ahler moduli of the Riemann surface correspond to irrelevant deformations.} (in the case $\fg=1$ there is one complex structure deformation).
Besides, one can turn on a flat connection for the $SU(2)$ flavor symmetry group that does not receive magnetic flux. As discussed in \cite{Benini:2009mz}, the $SU(2)$ flavor group leads to $3\mathfrak{g}-3$ independent complex moduli (one for $\fg=1$). There might be other marginal deformations, for instance coming from flat connections for the remaining $U(1)$ mesonic and $U(1)_B$ baryonic flavor symmetry depending on the flux turned on, as well as other less manifest moduli. For $\fg=0$ there are no complex structure deformations nor flat connections, and it is plausible that the corresponding two-dimensional SCFTs are isolated. This is surely an issue that deserves further study. Let us also remark that, although we have not studied supergravity solutions in which the $SU(2)$ flavor symmetry is broken, the field theory analysis in Section \ref{Field theory} suggests that there are two-dimensional SCFTs with only $U(1)_1\times U(1)_2$ flavor and $U(1)_B$ baryonic symmetry for any $\mathfrak{g}$ in some range of the parameters $\{p,q,b_1,b_2,B\}$.

An interesting finding of our study is that the R-symmetry mixes along the flow not only with mesonic flavor symmetries, but also with the baryonic symmetry. Thus, the R-symmetry of the 2d CFTs is realized by an isometry of the background combined with a gauge transformation of the RR potential. It would be interesting to study whether there is a geometric construction to determine the precise combination of isometries and RR transformations corresponding to the dual superconformal R-symmetry.

It is certainly desirable to have a more direct understanding of the 2d SCFTs uncovered by our construction. One way of thinking about these two-dimensional systems is to start from the four-dimensional theory on $\mathbb{R}^2\times \Sigma_{\mathfrak{g}}$ with a partial topological twist on $\Sigma_{\mathfrak{g}}$ and write down the BPS equations following from the Lagrangian of the theory. Then the two-dimensional theory at low energies will be a nonlinear sigma model on the moduli space of solutions to these BPS equations. A similar analysis has been performed for four-dimensional $\mathcal{N}=2$ SCFTs in \cite{Bershadsky:1995vm, Kapustin:2006hi}. The difficulty in this approach stems from the fact that the BPS equations for these four-dimensional theories are some appropriate generalizations of the Hitchin equations on $\Sigma_{\mathfrak{g}}$ and the moduli space of solutions is not known. An alternative approach would be to find a suitable two-dimensional gauged linear sigma model which in the IR describes the SCFTs of interest. It would be interesting to explore also whether there is a connection with the recent work in \cite{Tong:2014yna, Franco:2015tna, Franco:2015tya}.

The current work as well as the construction in \cite{Benini:2013cda} leads to the natural question of whether one can establish a useful correspondence between 2d CFTs and some 2d TQFT on the compactification Riemann surface. This correspondence should be in the spirit of similar proposals that relate four- and three-dimensional SCFTs with two and three-dimensional TQFTs, respectively \cite{Alday:2009aq, Gadde:2009kb, Dimofte:2011ju}. In the same spirit it is natural to extend our construction to Riemann surfaces with punctures.

It would also be interesting to compute the $T^2\times S^2$ partition function \cite{Benini:2015noa} for our theories and see whether we can match the result with the supergravity calculation. The field theory analysis should be accessible through the techniques developed recently in \cite{Benini:2015noa, Closset:2013sxa, Gadde:2015wta}.

The manifolds we constructed provide infinite-dimensional families of explicit 7d metrics of the type studied in \cite{Kim:2005ez,Gauntlett:2007ts}. These manifolds seem to provide a natural generalization to Sasaki-Einstein geometry and it would be very interesting to understand their geometry further.
% The principle of $c$-extremization in the dual SCFT suggests that these type of manifolds enjoy some type of volume-minimization principle analogous to \cite{Martelli:2005tp,Martelli:2006yb,Eager:2010yu}. The analysis however will probably be more complicated due to the important role played by the non-geometric baryonic $U(1)$ symmetries.
%%% Why giving this idea away?

We have restricted our supergravity analysis to AdS$_3$ solutions with $SU(2)\times U(1)\times U(1)$ isometry. It should be possible to relax this assumption and look for solutions with lower amount of symmetry. While this will be technically complicated since the BPS equations will reduce to PDEs, rather than ODEs, our field theory analysis suggests that these PDEs should have interesting solutions. Among them should be the solutions dual to $Y^{p,q}$ theories on $\Sigma_{\mathfrak{g}}$ with non-zero $b_1$ and $b_2$ flavor flux. In addition it is natural to expect that there is a generalization of our analysis to the $L^{p,q,r}$ quiver gauge theories which posses only $U(1)\times U(1) \times U(1)$ global symmetry \cite{Cvetic:2005ft,Cvetic:2005vk,Butti:2005sw}.

The field theory calculation of the central charges of the two-dimensional SCFTs performed in Section \ref{Field theory} is exact while the supergravity results are valid only to leading order in the rank of the gauge group, $N$. It will certainly be very interesting to understand how the $1/N^2$ corrections to the central charge arise on the supergravity side. This should amount to understanding higher-curvature corrections to our type IIB supergravity backgrounds along the lines of \cite{Baggio:2014hua}.

It would be nice to extend our analysis and find similar AdS$_2$ solutions of eleven-dimensional supergravity. These should fall in the classification of \cite{Kim:2006qu} (see also \cite{MacConamhna:2006nb}) and  be dual to M2-branes at the tip of a conical singularity, wrapping a compact Riemann surface. These solutions can be viewed as M2-brane black holes and the microscopic understanding of their entropy will be facilitated by the techniques recently developed in \cite{Benini:2015noa,Benini:2015eyy}.

\bigskip
\bigskip

%%%%%%%%%%%%%%%%%%%%%%%%%%%%%%%%%%%%%
\noindent \textbf{Acknowledgements }
%%%%%%%%%%%%%%%%%%%%%%%%%%%%%%%%%%%%%
\bigskip

\noindent We would like to thank Chris Beem, Jerome Gauntlett, Dario Martelli, Eoin \'O Colg\'ain, Phil Szepietowski,  Stefan Vandoren, and Wenbin Yan for useful discussions. FB is supported by the Royal Society as a Royal Society University
Research Fellowship holder, and by the MIUR-SIR grant RBSI1471GJ ``Quantum Field Theories at Strong Coupling: Exact Computations and Applications''. The work of NB is supported in part by the starting grant BOF/STG/14/032 from KU Leuven, by the COST Action MP1210 The String Theory Universe, and by the European Science Foundation Holograv Network. PMC is supported by the Netherlands Organization for Scientific Research (NWO) under the VICI Grant 680-47-603. This work is part of the D-ITP consortium, a program of the NWO that is funded by the Dutch Ministry of Education, Culture and Science (OCW). PMC would also like to thank the COST Action ``The String Theory Universe''  MP1210 for  travel support and the ITF at KU Leuven for hospitality.

%%%%%%%%%%%%%%%%%%%%%%%%%%%%%%%%%%%%%
\begin{appendices}
%%%%%%%%%%%%%%%%%%%%%%%%%%%%%%%%%%%%%

\section{BPS equations and Bianchi identities}
\label{BPS equations and Bianchi identities}

\renewcommand{\theequation}{A.\arabic{equation}}
\renewcommand{\thetable}{A.\arabic{table}}
\setcounter{equation}{0}
\label{appendixA}
%%%%%%%%%%%%%%%%%%%%%%%%%%%%%%%%%%%%%

Assuming that only the metric and 5-form flux are turned on, the type IIB  gravitino variations  read\footnote{We follow the conventions of \cite{Gauntlett:2005ww}.}
\begin{equation}
\label{spin32}
\delta\psi_\mu = \partial_{\mu} \epsilon+ \dfrac{1}{4}\omega_{\mu ab}\Gamma^{ab}\epsilon + \dfrac{i}{192}F_{\mu \nu_1\nu_2\nu_3\nu_4}\Gamma^{\nu_1\nu_2\nu_3\nu_4}\epsilon = 0\;.
\end{equation}
In our notation, we denote the ten dimensional coordinates by $x^{\mu}$, with $\mu=1,...,10$, and the signature is $(-,+,\cdots,+)$. The ten-dimensional spinor $\epsilon$ satisfies the chirality condition
\begin{equation}\label{chirality}
\Gamma^{12345678910} \epsilon= \epsilon\;.
\end{equation}
The supersymmetry variation of the spin-1/2 fermion in type IIB supergravity vanishes identically when there are no non-trivial dilaton-axion and 3-form fluxes turned on. In addition there is the Bianchi identity for the self-dual 5-form flux $F_{(5)} = (1+*_{10}) G_{(5)}$:
\equ{\label{Bianchi FG}
dF_{(5)}=0 \qquad  \Rightarrow \qquad d \, G_{(5)}=d\,  \ast_{10} G_{(5)}=0\,.
} 
The symmetries of the problem suggest that we impose the following projectors on the spinor $\epsilon$:
\begin{equation}\label{projectors}
\Gamma^{12}\epsilon = - \epsilon\;, \qquad \Gamma^{45}\epsilon = i \epsilon\;, \qquad \Gamma^{67}\epsilon = i \epsilon\;, \qquad \Gamma^{89}\epsilon = i \epsilon\;,
\end{equation}
which  together with the ten-dimensional chirality condition implies 
\begin{equation}
\Gamma^{310} \epsilon = -i \epsilon\;.
\end{equation}
The most general metric Ansatz compatible with our expectations is 
\bea
\label{MetricAnsatz app}
ds^2_{10} &= f_1(y)^2ds^{2}_{AdS_{3}} + f_2(y)^2 ds^{2}_{\Sigma_{\mathfrak g}}+ f_3(y)^2 ds^{2}_{S^{2}} + f_4(y)^2dy^2 + f_5(y)^2\left(D\beta \right)^2 \\
&\qquad + f_6(y)^2\left(D\psi+f_7(y)D\beta \right)^2 \;,\\  
G_{(5)} &= e^1\wedge e^2\wedge e^3\wedge \big[ g_1(y) e^4\wedge e^5+g_2(y) e^6\wedge e^7+g_3(y) e^8\wedge e^9+g_4(y) e^8\wedge e^{10}+g_5(y) e^9\wedge e^{10} \big]
\eea
where 
\eqsn{ds^{2}_{AdS_{3}} = \dfrac{-dt^2+dz^2+dr^2}{r^2}\,, \quad ds^{2}_{\Sigma_{\mathfrak g}}=e^{2h(x_1,x_2)}\left(dx_1^2+dx_2^2\right)\,, \quad ds^{2}_{S^{2}} = d\theta^2+\sin^2\theta d\phi^2\,,\\
D\beta=d\beta + c \cos\theta d\phi+ a_2A_{\fg}(x_1,x_2)\,, \quad D\psi=d\psi +b\cos\theta d\phi+a_3A_{\fg}(x_1,x_2)\,,\quad dA_{\fg}=\dvol{\Sigma_{\mathfrak g}}\;,
}
and we defined the vielbein
\begin{equation}
\begin{split}
e^{1} &= \dfrac{f_1}{r} \,dt\;, \qquad e^{2} = \dfrac{f_1}{r} \,dz\;, \qquad e^{3} = \dfrac{f_1}{r} \,dr\; \qquad e^{4} = f_2\,e^{h}dx_1\;, \qquad e^{5} = f_2\,e^{h}dx_2\,, \\
e^{6}&= f_3\,d\theta \;, ~~~ e^{7} = f_3\sin\theta \, d\phi \;, ~~~ e^{8} = f_4 \, dy\;, ~~~ e^{9}=f_5\,D\beta\;,\qquad e^{10} =f_6\left(D\psi +f_7\,D\beta \right)\;.
\end{split}
\end{equation}
The function $h(x_1,x_2)$ is defined in (\ref{hdef}). We denote the volume forms by 
\be
\dvol_{AdS_{3}}=\frac{1}{r^{3}}dt\wedge dz\wedge dr\,, \quad \dvol_{\Sigma_{\mathfrak g}}=e^{2h(x_{1},x_{2})}dx_{1}\wedge dx_{2}\,, \quad \dvol_{S^{2}}=\sin \theta \, d\theta \wedge d\phi \;.
\ee
In our normalization
\be
\vol(\Sigma_{\mathfrak g}) = \int_{\Sigma_{\mathfrak g}} \dvol_{\Sigma_{\mathfrak g}} = \begin{cases}
    4\pi |\mathfrak g-1|,& \quad  \mathfrak g \neq 1\\
    2\pi,              & \quad \mathfrak g=1 \;.
\end{cases}
\ee

Now we can use the gravitino variation in \eqref{spin32} along with the projectors in \eqref{chirality}, \eqref{projectors} to derive a set of differential equations for the unknown functions $f_{i},g_{i}$ appearing in the Ansatz (\ref{MetricAnsatz app}).  Before writing out all the equations we show that $g_5=0$. This follows from the $\delta\psi_5$ component of the gravitino variation\footnote{One can also argue that $g_5=0$ using the equations of motion.}
\begin{multline}
\delta\psi_5 = \dfrac{x_2}{f_2}\partial_{x_2}\epsilon + \frac{1}{8}\left[ 4\dfrac{f_2'}{f_2f_4} -2a_2\frac{ f_5}{f_2^2} - g_4 - i  g_5\right]\Gamma^{49}\epsilon \\
- \frac{i}{8}\left[ 2 \dfrac{f_6(a_3+a_2f_7)}{f_2^2} +g_1 - g_2-g_3\right]\Gamma^{34}\epsilon = 0\;.
\end{multline}
For the spinor of interest we should have $\partial_{x_2}\epsilon=0$.\footnote{Here we assume that the metric on the Riemann surface is the constant curvature one and thus the spinors do not depend on the coordinates on the Riemann surface. This assumptions could in principle be relaxed but the general analysis is more involved. The results of  \cite{Anderson:2011cz} however suggest that the constant curvature metric is capturing all interesting physics.} Then the equation $\delta\psi_5=0$ is of the form
\begin{equation}
i A \Gamma^{34}\epsilon + (B+iC)\Gamma^{49}\epsilon = \left[-iA + (B+iC)\Gamma^{39}\right]\epsilon= 0\;,
\end{equation}
with $\{A,B,C\}$ real and $C = - g_5/8$. This equation implies $A^2-B^2+C^2 = BC=0$, which in turn implies  $C=0$  and thus $g_5=0$ in order to have nontrivial solutions. From now on we  set $g_5=0$ in the remaining equations. The explicit form of all gravitino variations is:

\eqs{
\label{del1}
\delta\psi_1=\,&\dfrac{r}{f_1}\partial_{t}\epsilon + \frac{1}{8}\left[ \dfrac{4}{f_1}-(g_1+g_2+g_3 ) \right]\Gamma^{13}\epsilon - \frac{1}{8}\left[ \dfrac{4f_1'}{f_1f_4} +  g_4\right]\Gamma^{18}\epsilon \;,
}
\eqs{
\label{del2}
\delta\psi_2=\,&\dfrac{r}{f_1}\partial_{z}\epsilon - \frac{1}{8}\left[ \dfrac{4}{f_1}-(g_1+g_2+g_3 ) \right]\Gamma^{13}\epsilon + \frac{1}{8}\left[ \dfrac{4f_1'}{f_1f_4} + g_4\right]\Gamma^{18}\epsilon \;,
}
\eqs{
\label{del3}
\delta\psi_3=\,&\dfrac{r}{f_1}\partial_{r}\epsilon +\frac{1}{8}(g_1+g_2+g_3)\epsilon + \frac{1}{8}\left[ \dfrac{4f_1'}{f_1f_4} + g_4\right]\Gamma^{38}\epsilon \;,
}
\eqs{
\nonumber
\delta\psi_4=\,&\dfrac{1}{2f_2e^{h}}\left[2\partial_{x_1} - 2A_1a_2\partial_{\beta} -  2A_1a_3 \partial_{\psi} + i\partial_{x_2}h\right]\epsilon + \frac{1}{8}\left[ -4\dfrac{f_2'}{f_2f_4} +2 a_2 \dfrac{f_5}{f_2^2}+g_4\right]\Gamma^{59}\epsilon\\ \label{del4}
\,&+ \frac{1}{8}\left[\dfrac{2f_6(a_3+a_2f_7)}{f_2^2}+  g_1 - g_2-g_3\right]\Gamma^{510}\epsilon  \;,
}
\eqs{ \nonumber
\delta\psi_5 =\,& \dfrac{1}{2f_2e^{h}}\left[2\partial_{x_2} - 2A_2a_2\partial_{\beta} -  2A_2a_3 \partial_{\psi} - i\partial_{x_1}h\right]\epsilon + \frac{1}{8}\left[ 4\dfrac{f_2'}{f_2f_4} -2a_2\frac{ f_5}{f_2^2} - g_4\right]\Gamma^{49}\epsilon \\ \label{del5}
\,&- \frac{i}{8}\left[ 2 \dfrac{f_6(a_3+a_2f_7)}{f_2^2} +g_1 - g_2-g_3\right]\Gamma^{34}\epsilon \;,
}
\eqs{
\label{del6}
\delta\psi_6 =\,& \dfrac{1}{f_3}\partial_{\theta}\epsilon + \frac{1}{8}\left[ -4\dfrac{f_3'}{f_3f_4} -2c\frac{ f_5}{f_3^2} + g_4\right]\Gamma^{79}\epsilon + \frac{1}{8}\left[ -2 \dfrac{f_6(cf_7+b)}{f_3^2} -g_1 + g_2-g_3\right]\Gamma^{710}\epsilon \;,
}
\eqs{ 
\nonumber
\delta\psi_7 =\,& \dfrac{1}{f_3\sin\theta}\left[\partial_{\phi}+\cos\theta\left(-b\partial_{\psi}-c\partial_{\beta}-\coeff{i}{2}\right)\right]\epsilon + \frac{1}{8}\left[ 4\dfrac{f_3'}{f_3f_4} +2c\frac{ f_5}{f_3^2} - g_4\right]\Gamma^{69}\epsilon \\ \label{del7}
\,&+ \frac{1}{8}\left[ 2 \dfrac{f_6(cf_7+b)}{f_3^2} +g_1 - g_2+g_3\right]\Gamma^{610}\epsilon \;,
}
\eqs{
\label{del8}
\delta\psi_8 =\,& \left[\dfrac{1}{f_4}\partial_{y}\epsilon +\coeff{1}{8} g_4\epsilon\right] + \frac{1}{8}\left[ 2 \dfrac{f_6f_7'}{f_4f_5} -g_1 - g_2+g_3\right]\Gamma^{910}\epsilon \;,
}
\eqs{ 
\nonumber
\delta\psi_9 =\,& \dfrac{1}{f_5}\left[\partial_{\beta}-f_7\partial_{\psi} - \coeff{i}{2}\dfrac{f_5'}{f_4} -\coeff{i}{4}f_5^2\left( \dfrac{a_2}{f_2^2}-\dfrac{c}{f_3^2} \right)+\coeff{i}{8} f_5g_4\right]\epsilon \\  \label{del9}
\,&- \frac{1}{8}\left[ 2 \dfrac{f_6f_7'}{f_4f_5} -g_1 - g_2+g_3\right]\Gamma^{810}\epsilon \;,
}
\eqs{ 
\nonumber
\delta\psi_{10} =\,& \left[\dfrac{1}{f_6}\partial_{\psi}-\coeff{i}{4}f_6\left( \dfrac{(a_3+a_2f_7)}{f_2^2}-\dfrac{(cf_7+b)}{f_3^2} \right)-\dfrac{i}{8} \left(2 \dfrac{f_6f_7'}{f_4f_5} +g_1 + g_2+g_3\right)\right]\epsilon  \\ \label{del10}
\,&- \frac{1}{8}\left[ 4\dfrac{f_6'}{f_4f_6} +g_4\right]\Gamma^{810}\epsilon \;.
}

The gravitino variations are of the form $(A  +B \, \Gamma^{c_1c_2})\, \epsilon =0$ for some real $A$ and $B$, $c_1\neq c_2$. If $\epsilon$ and $\Gamma^{c_1c_2}\epsilon$ are independent spinors (\ie{} $\Gamma^{c_1c_2}$ is none of the projectors appearing in \ref{projectors}), then $A=B=0$.  Equipped with this fact we are ready to analyze the gravitino variations in  detail. 

We first focus  on the terms proportional to gamma matrices in $\delta \psi_{\mu}$; this leads to a total of eight independent equations:
\eqs{ \label{eqs gamma 1}
4-f_1 (g_1+g_2+g_3)=0\,,\\  \label{eqs gamma 2}
-g_1+g_2+g_3 - \frac{2 f_6(a_2 f_7+a_3)}{f_2^2}=0\,,\\  \label{eqs gamma 3}
g_1-g_2+g_3 + \frac{2 f_6(c f_7+b)}{f_3^2}=0\,,\\  \label{eqs gamma 4}
g_4+\frac{4 f_1'}{f_1 f_4}=0\,, \\  \label{eqs gamma 5}
g_4+\frac{4 f_6'}{f_4 f_6}=0\,,\\ \label{eqs gamma 6} 
-\frac{2 f_6 f_7'}{f_4 f_5}+g_1+g_2-g_3=0\,, \\ \label{eqs gamma 7}
-\frac{2 a_2 f_5}{f_2^2}+\frac{4 f_2'}{f_2 f_4}-g_4=0\,,\\ \label{eqs gamma 8}
\frac{2 c f_5}{f_3^2}+\frac{4 f_3'}{f_3 f_4}-g_4=0\,.
}
From (\ref{eqs gamma 1}-\ref{eqs gamma 5}) we can algebraically solve for the functions $g_{i}$:
\eqs{\nonumber
g_1=&\, \frac{2}{ f_1}-\frac{f_6(a_3+a_2 f_7)}{f_2^2}\,,\\ \nonumber
g_2 =&\,\frac{2}{f_1}+\frac{f_6(b+c f_7)}{f_3^2}\,,\\ \label{solutions g}
g_3=&\, f_6\(\frac{a_2 f_7+a_3}{f_2^2}-\frac{b+cf_7}{f_3^2}\)\,, \\ \nonumber
g_4=&\, -\frac{4f_6'}{f_4 f_6} = -\frac{4f_1'}{f_4 f_1} \,,\nonumber
}
and we also find
\equ{\label{relation f6 f1}
f_6= \alpha_{2} \, f_1\,,
}
with $\alpha_{2}$ a constant.  From now on we consider the functions $g_{i}$ determined in terms of the $f_{i}$ by (\ref{solutions g}). The remaining equations (\ref{eqs gamma 6}-\ref{eqs gamma 8}) read
\eqs{ \label{eq H 1}
\frac{2}{f_1}+\alpha_{2} f_1 \( \frac{-a_3- a_2 f_7}{f_2^2}+\frac{b+ c f_7}{f_3^2}-\frac{f_7'}{f_4 f_5}\)=0\,,\\
\frac{2 f_1'}{f_1 f_4} +\frac{2 f_2'}{f_2 f_4}-  \frac{a_2 f_5}{f_2^2}=0\,,\\ 
\frac{2 f_1'}{f_1 f_4} +\frac{2 f_3'}{f_3 f_4} +   \frac{c f_5}{f_3^2}=0\,.
}
Now we turn to the equations arising from terms proportional to the identity in $\delta \psi_{\mu}$. There are ten such equations in total. From  (\ref{del1}),  (\ref{del2}),  (\ref{del3}),  (\ref{del5}),  part of (\ref{del7}), and  (\ref{del8}) one immediately concludes that
\equ{
\partial_t \epsilon= \partial_z \epsilon=\partial_{\theta} \epsilon = \partial_{\phi} \epsilon=0\,.
}
In addition, we assume that $\partial_{x_{1}}\epsilon= \partial_{x_{2}}\epsilon=0$. Then 
\equ{
\epsilon=\(\frac{f_1(y)}{r}\)^{1/2} \, \tilde \epsilon(\beta, \psi)\,,
}  
where $\tilde \epsilon(\beta, \psi)$ is a spinor that will be fixed shortly. In addition, we have the equations
\eqs{ 
 \label{eq der H 1}
\( i \partial_{x_2} h -2 a_2 A_1 \partial_{\beta}-2 a_3 A_1\, \partial_{\psi}\) \epsilon=0\,,\\   \label{eq der H 2}
\( -i \partial_{x_1} h -2 a_2 A_2 \partial_{\beta}-2 a_3 A_2\, \partial_{\psi}\)\epsilon=0\,, \\ \label{eq b c}
\left[- i - 2 c \partial_{\beta}- 2 b \partial_{\psi}\right]\epsilon=0\,,\\ \label{eq 7 H A}
\left[2\(\frac{a_2}{f_2^2}-\frac{c}{f_3^2}\)f_5 -g_4+\frac{4 f_5'}{f_4 f_5} +\frac{8 i (\partial_{\beta}-f_7 \partial_{\psi})}{f_5}\right]\epsilon=0\,, \\ \label{eq 10 H A}
\left[\frac{4}{f_1}+\frac{8 i \partial_{\psi}}{f_6}-2 f_6\(\frac{-a_3-a_2 f_7}{f_2^2}+\frac{b+c f_7}{f_3^2}-\frac{f_7'}{f_4 f_5}\)\right]\epsilon=0\,.
}
Combining (\ref{eq H 1}) and (\ref{eq 10 H A}) implies $\partial_{\psi} \epsilon=i \alpha_{2}\,\epsilon\,$ and from (\ref{eq b c})  we have
\equ{ \label{spinor dependence}
\epsilon=e^{i (\alpha_1 \beta +\alpha_2 \psi)}\,  \(\frac{f_1(y)}{r}\)^{1/2} \,  \epsilon_{0}\,,
}
where $\epsilon_{0}$ is a constant spinor obeying the projection conditions  \eqref{chirality}, \eqref{projectors} and 
\equ{ \label{relations alphas app}
 c\, \alpha_1 + b \, \alpha_2=-\frac{1}{2}\,.
}
Thus,  we are left with the set of BPS equations
\eqs{\label{eq 1}
\frac{2 f_1'}{f_1 f_4} +\frac{2 f_2'}{f_2 f_4}-  \frac{a_2 f_5}{f_2^2}=0\,,\\ \label{eq 2 H}
\frac{2 f_1'}{f_1 f_4} +\frac{2 f_3'}{f_3 f_4} +   \frac{c f_5}{f_3^2}=0\,,\\ \label{eq 7 H}
\frac{2}{f_1}+\alpha_{2} f_1 \( \frac{-a_3- a_2 f_7}{f_2^2}+\frac{b+ c f_7}{f_3^2}-\frac{f_7'}{f_4 f_5}\)=0\,,\\ \label{eq 8 H}
\left[2\(\frac{a_2}{f_2^2}-\frac{c}{f_3^2}\)f_5 -g_4+\frac{4 f_5'}{f_4 f_5} -\frac{8  (\alpha_1-f_7 \alpha_2)}{f_5}\right]\epsilon=0\,,\\
\label{Eq Riemann surface 1}
\( i \partial_{x_2} h -2 i A_1( \alpha_1 a_2 + a_3 \alpha_2) \) \epsilon=0\,,\\  \label{Eq Riemann surface 2} 
\( -i \partial_{x_1} h -2 i A_2 (\alpha_1a_2 + a_3 \alpha_2)\)\epsilon=0 \,.
}
It is convenient to choose the coordinate $y$ such that
\equ{ \label{equation f4}
f_1^2 f_ 4f_5= D\,,
}
where $D$ is a constant. With this coordinate choice (\ref{eq 1}) and (\ref{eq 2 H}) simplify to
\eqs{\label{equations P2 and P3}
(f_1^2 f_2^2)' = a_2 \, D\, , \qquad  (f_1^2 f_3^2)' = - c \,  D\,. 
}

Let us analyze equations (\ref{Eq Riemann surface 1}), (\ref{Eq Riemann surface 2}). Taking the derivative of these equations one has 
\equ{\label{A cons eqs}
2(a_2\alpha_1 + a_3\alpha_2)dA_{\fg}=-(\partial_{x_{1}}^{2}+\partial_{x_{2}}^{2})h \, dx_{1}\wedge dx_{2}\,.
}
For $\mathfrak g\neq 1$ (\ie{} $\kappa\neq 0$) the background R-symmetry flux is set to $dA= \dvol{\Sigma_{\mathfrak g}}=e^{2h}dx_{1}\wedge dx_{2}$ in order to preserve supersymmetry. From  \eqref{hdef}, \eqref{A cons eqs} implies the consistency condition\begin{equation} \label{constraint a2 a3 kappa app}
a_2 \, \alpha_1 + a_3\, \alpha_2 = \frac{\kappa}{2}\;.
\end{equation}
Thus equations (\ref{Eq Riemann surface 1}), (\ref{Eq Riemann surface 2}) imply
\equ{\label{solution A}
A_{\fg\neq 1}=\kappa\,(\partial_{x_2} h \, dx^{1}-\partial_{x_1}h \,dx^{2})\,,
}
which is compatible with $dA_{\fg}=\dvol{\Sigma_{\fg}}$.  Of course, the connection is defined up to gauge transformations $A_{\fg}\to A_{\fg}+d\lambda$. In the case $\fg=1$, we can choose a gauge in which:
\equ{\label{A torus}
A_{\fg=1}=\frac{1}{2\pi}\,x_{1}dx_{2}\,.
}
Up to this point the only assumptions we have made are the Ansatz (\ref{MetricAnsatz app}), the projectors \eqref{projectors} and that the spinor $\epsilon$ is independent of $\{x_1,x_2\}$.

%%%%%%%%%%%%%%%%%%%%
\subsection*{Bianchi}
%%%%%%%%%%%%%%%%%%%%

In addition to the BPS equations there are a total of four Bianchi identities from (\ref{Bianchi FG}) which read:
\eqs{\label{Bianchi 1}
\partial_y(g_1 f_1^3 f_2^2)- a_2 g_3 f_1^3 f_4 f_5 - g_4 f_1^3 f_4 f_6 (a_2 f_7+a_3)=0\,, \\  \label{Bianchi 2}
\partial_y(g_2 f_1^3 f_3^2)+c g_3 f_1^3 f_4 f_5+g_4 f_1^3 f_4 f_6(c f_7+b)=0\,,\\ \label{Bianchi 3}
\partial_y (f_2^2 f_3^2 g_3 f_6)+f_4f_5 f_6( c g_2 f_2^2-a_2 g_1f_3^2 )=0\,, \\  \label{Bianchi 4}
\partial_y (g_3 f_2 ^2 f_3^2 f_6 f_7) -\partial_y (g_4 f_2^2f_3^2 f_5)+f_4 f_5 f_6(a_3 g_1 f_3^2- b g_2 f_2^2)=0\,.
}
Here we have used  $dA_{\fg}=\dvol{\Sigma_{\fg}}$ to simplify the equations. 

%%%%%%%%%%%%%%%%%%%%%%%%%%%%%%
\subsection{General solution}
\label{General solution app}
%%%%%%%%%%%%%%%%%%%%%%%%%%%%%%

To solve the BPS equations and Bianchi it is convenient to trade the functions $f_{1},f_{2}, f_{3}, f_{4},f_{7}$ by the functions $\mathcal P_{1},\mathcal P_{2},\mathcal P_{3},\mathcal Q,\mathcal P_{7}$, defined by:
\eqs{\nonumber
f_1^2(y)=\,& \sqrt{\frac{\mathcal P_2(y) \mathcal P_3(y)}{\mathcal P_1(y)}}\,, \quad f_2^2(y)= \sqrt{\frac{\mathcal P_2(y) \mathcal P_1(y)}{\mathcal P_3(y)}}\,, \quad f_3^2(y)= \sqrt{\frac{\mathcal P_3(y) \mathcal P_1(y)}{\mathcal P_2(y)}}\,,\\\label{nice form ansatz app}
f_4^{2}(y)=\,& D^{2} \frac{\sqrt{\mathcal P_{1}(y)\mathcal P_{2}(y) \mathcal P_{3}(y)}}{\mathcal Q(y)}\,, \qquad f_{7}(y)=\frac{\mathcal P_{7}(y)}{\mathcal P_{2}(y)\mathcal P_{3}(y)}\,.
}
Recall that $f_{5},f_{6}$ are given by (\ref{equation f4}), (\ref{relation f6 f1}), respectively. This form of the Ansatz requires that
\equ{
\text{sign}(\mathcal P_{1}(y)\mathcal P_{2}(y)\mathcal P_{3}(y))=+\,, \qquad \mathcal Q(y)>0\,.
}
We  assume that the range of interest for $y\in [y_{1},y_{2}]$ is finite and that in this range
\equ{
\mathcal P_{2}(y),\, \mathcal P_{3}(y)>0\,, \qquad \qquad  y \in [y_{1},y_{2}]\,,
}
which also implies that $ \mathcal P_{1}(y)>0$ in this range.  From now on we omit the argument in the functions $\mathcal P,\mathcal Q$. From (\ref{equations P2 and P3}) it follows  that 
\equ{\label{solution P2 P3 app}
\mathcal P_{2} =a_{2} D \, y+C_{2}\,,\qquad \mathcal P_{3}= - c D \, y +C_{3}\,.
}
The remaining BPS equations are (\ref{eq 7 H}) and (\ref{eq 8 H}), which using (\ref{nice form ansatz app}) can be used to write $\mathcal P_{1,7}$ in terms of the known functions $\mathcal P_{2,3}$ and the (yet undermined) function $\mathcal Q$ as: 
\equ{\label{solutions P1, P7 app}
\mathcal P_{1}= \frac{\alpha_{2}\left( a_{3} D \mathcal P_{3}-b D\mathcal P_{2}+\mathcal P_{7}'\right)}{2 D}\,,\qquad 
\mathcal P_{7}= \frac{4 D \alpha_{1}\mathcal P_{2}\mathcal P_{3}-\mathcal Q'}{4 D \alpha_{2}}\,.
}
Thus, given a function $\mathcal Q$ all local solutions to the BPS equations are given by (\ref{nice form ansatz app},\ref{solution P2 P3 app},\ref{solutions P1, P7 app}) and are characterized by the parameters $\{a_{2},a_{3},b,c,C_{2},C_{3},D,\alpha_{1},\alpha_{2}, \kappa\}$, subject to the constraints (\ref{relations alphas app}, \ref{constraint a2 a3 kappa app}). Not all these parameters are physical, $e.g$, by a rescaling of $y$ one can set $D=1$, and other parameters may also be absorbed by coordinate transformations. We will analyze this in more detail below. 

The only equations that remain to be solved are the Bianchi identities (\ref{Bianchi 1}-\ref{Bianchi 4}). The first three of these equations are automatically satisfied, assuming the  BPS equations. Thus, the only remaining equation is  (\ref{Bianchi 4}), which using the BPS equations can be written as a fourth order differential equation for the function $\mathcal Q$ and has the form
\equ{
\text{Bianchi}=\mathcal B_{2}\times \mathcal B_{4}=0\,,
}
with $\mathcal B_{2}=\mathcal Q''- 2 D^{2}(\mathcal P_{2}+\kappa \mathcal P_{3})$ and $\mathcal B_{4}$ depends on up to fourth order derivatives of $\mathcal Q$. There are two branches: $\mathcal B_{2}=0$ and $\mathcal B_{4}=0$. The former  leads to
 \equ{
\mathcal Q=\frac{D^{3}}{3}y^{3}(a_{2}-c\kappa)+D^{2}y^{2}(C_{2}+C_{3}\kappa)+\gamma_{2} y+\gamma_{1}\,,
}
where $\gamma_{1,2}$ are constants. However, using this solution in (\ref{solutions P1, P7 app}) implies $\mathcal P_{1}=0\,$ which is clearly singular.
Thus, the solutions of interest arise from the branch $\mathcal B_{4}=0$. This is a fourth order differential equation for $\mathcal Q$, which is rather complicated. However, it has the remarkable property that it can be integrated twice into the second order  equation:\footnote{We note that the differential equation below has the form $\mathcal Q(y)'^{2}-2\mathcal  Q(y) \mathcal Q(y)''+2 \mathcal Q(y) F(y)+G(y)=0$. Defining $q(y)=\mathcal Q(y)^{1/2}$, it becomes $q(y)''-\frac{G(y)}{4q(y)^{3}}-\frac{F(y)}{2 q(y)}=0$.} 
\equ{\label{equation V''=0}
\frac{1}{\mathcal P_{2}\mathcal P_{3}}\(\mathcal Q'^{2}-2 \mathcal Q(\mathcal Q'' -2 D^{2}(\mathcal P_{2}+\kappa \mathcal P_{3}))\)- 4 D^{4} y^{2}\kappa+\delta_{1}+\delta_{2} y=0\,,
}
where $\delta_{1,2}$ are integration constants.\footnote{It is worth pointing out that a similar differential equation controls a class of supersymmetric $AdS_3$ solutions of eleven-dimensional supergravity studied in \cite{Gauntlett:2006qw}.} Thus, general solutions to the BPS equations and Bianchi are characterized completely by solutions to this differential equation.  Although we have not found the most general solution to \eqref{equation V''=0},  it is straightforward to show that the most general \textit{polynomial} solution can be at most cubic, \ie
\equ{\label{Q cubic polynomial ansatz}
\mathcal Q=\sum_{i=0}^{3} q_{i}\,y^{i}\,.
}
Plugging (\ref{Q cubic polynomial ansatz}) into (\ref{equation V''=0}) leads to a system of five algebraic equations, which determine the four coefficients $q_{i}$ plus one constraint among the remaining parameters. The generic solution can be written in the iterative form\footnote{We have used a rescaling of the $y$ coordinate to set $D=1$ for convenience.}  
\eqs{ \nonumber
q_{3}=\,&-\frac{2}{3}\(-a_{2}+c\kappa \pm \sqrt{a_{2}^{2}+a_{2}\kappa c+c^{2}\kappa^{2} }\)\,,\\ \nonumber
q_{2}=\,&C_{2}+\kappa C_{3}+\frac{-4 a_{2}C_{2}+a_{2}c \delta_{2}+4 C_{3}c \kappa^{2}}{4(a_{2}-c \kappa-q_{3})}\,, \\ \label{solution polynomial}
q_{1}=\,&\frac{a_{2}c \delta_{1}+\delta_{2}(c C_{2}-a_{2}C_{3})+4 C_{2}C_{3}\kappa-4(C_{2}+\kappa C_{3})q_{2}}{4a_{2}-4 c \kappa-6q_{3}}\,,\\ \nonumber
q_{0}=\,&-\frac{C_{2} C_{3}\delta_{1}+q_{1}^2}{4 (C_{2}+C_{3} \kappa -q_{2})}\,,
}
together with the constraint
\equ{\label{constraint app}
\frac{a_{2}c (a_{2}+c \kappa)\delta_{1}+ (a_{2}^{2}C_{3}-C_{2}c^{2}\kappa)\delta_{2}+4 \kappa\(2 C_{2}C_{3}(a_{2}+\kappa c) + c C_{2}^{2}+a_{2}\kappa C_{3}^{2}\)}{16 (a_{2}+c \kappa )^2 \left(a_{2}^2+a_{2} c \kappa +c^2 \kappa ^2\right)}=0\,.
}
Since the denominator in (\ref{constraint app}) vanishes for the special values $a_{2}=- \kappa c$ or $a_{2}=0=c$ or $a_{2}=0=\kappa$, these cases must be analyzed separately. 

We note that when $\mathcal Q$ is a cubic polynomial, the function $\mathcal P_{1}$ given in (\ref{solutions P1, P7 app})  becomes linear:
\equ{ \label{P1 linear app}
\mathcal P_{1}=\frac{1}{4} (C_{2}+C_{3}\kappa -q_{2}) +\frac{1}{4}(a_{2}-c\kappa-3q_{3 })y\,,
}
and $\mathcal P_{7}$ becomes a quadratic function of $y$.

Finally, we give the expression for the 5-form $G_{(5)}$. Using (\ref{solutions g}) and the expression for $\cP_{7}$ in (\ref{solutions P1, P7 app})  and the relations (\ref{relations alphas app},\ref{constraint a2 a3 kappa app}), we find:
\equ{\label{flux G's}
G_{(5)} =  \dvol{AdS_{3}}\wedge [G_1(y) \, \dvol{\Sigma_{\mathfrak g}} +G_2(y) \, \dvol{S^{2}}+G_3(y) dy\wedge D\beta+G_4(y) dy\wedge \left(D\psi +f_7(y)\,D\beta \right) ]\
}
where
\eqs{\nonumber
G_{1}(y)&=\frac{1}{4 D \, \cP_{1}} (8 D \, \cP_{1}\cP_{2}-2 D \kappa \, \cP_{2} \cP_{3}+a_{2}\mathcal Q')\,, \qquad G_{2}(y)=\frac{1}{4 D \, \cP_{1}} (8 D \, \cP_{1}\cP_{3}-2 D \,  \cP_{2} \cP_{3}-c\, \mathcal Q')\,, \\ \label{flux general nice app}
G_{3}(y)&=\frac{2D \, \cP_{2}\cP_{3}(\cP_{2}+\kappa \, \cP_{3})+(c\, \cP_{2}-a_{2}\, \cP_{3})\mathcal Q'}{4 \cP_{1}\cP_{2}\cP_{3}}\,, \qquad G_{4}(y)=-\alpha_{2}\(\frac{\cP_{2}\cP_{3}}{\cP_{1}}\)'\,.
}
This completes our analysis of the local solutions to the BPS equations and Bianchi identities. It is worth emphasizing that despite the fact that our backgrounds preserve only 4 out of the 32 supercharges of eleven-dimensional supergravity we have managed to solve the BPS equations in full detail analytically.

%%%%%%%%%%%%%%%%%%%%%%%%%%%%%%%%%%%%%%%%%%%%
\subsection{Linear transformations on $(\beta,\psi)$}
\label{Linear transformations on beta psi}
%%%%%%%%%%%%%%%%%%%%%%%%%%%%%%%%%%%%%%%%%%%%

Let us focus on the two-dimensional part of the metric \eqref{MetricAnsatz app} corresponding to the $(\beta,\psi)$ plain:
\begin{multline}\label{metpsibeta}
ds^2_{2} =f_5(y)^2\left(d\beta + c \cos\theta d\phi+ a_2A_{\mathfrak{g}}\right)^2\\+ f_6(y)^2\left(d\psi +b\cos\theta d\phi+f_7(y)\left(d\beta + c \cos\theta d\phi+ \frac{a_2}{x_2}dx_1\right)+a_3 A_{\mathfrak{g}}\right)^2\;,
\end{multline}
and perform the linear transformation
\equ{
(d\beta,d\psi)^{\top}=\mathcal W\,(\tilde{d\beta},\tilde{d\psi})^{\top}\,,\qquad  \qquad \mathcal W\equiv \left(
\begin{array}{cc}
w_{1} &  w_{2}\\ 
w_{3} &  w_{4}
\end{array}\right)\,.
}
The Killing vectors transform as $(\partial_{\beta}, \partial_{\psi})=(\tilde{\partial_{\beta}}, \tilde{\partial_{\psi}})\mathcal W^{-1}$. It is easy to see that the metric takes the same form as in (\ref{metpsibeta}), \ie
\begin{multline}\label{metpsibetatilde}
\tilde{ds}^2_{2} =\tilde{f}_5(y)^2\left(d\tilde{\beta} + \tilde{c} \cos\theta d\phi+ \tilde{a}_2A_{\mathfrak{g}}\right)^2\\+ \tilde{f}_6(y)^2\left(d\tilde{\psi} +\tilde{b}\cos\theta d\phi+\tilde{f}_7(y)\left(d\tilde{\beta} + \tilde{c} \cos\theta d\phi+ \frac{\tilde{a}_2}{x_2}dx_1\right)+\tilde{a}_3A_{\mathfrak{g}}\right)^2\;,
\end{multline}
where the new parameters are given by
\equ{
(\tilde{a}_{2},\tilde{a}_{3})^{\top}=\mathcal W^{-1}\,(a_{2},a_{3})^{\top}\,,  \qquad (\tilde c,\tilde b )^{\top}=\mathcal W^{-1}\,(c , b)^{\top}\,,
}
and the functions by
\begin{equation}
\begin{split}
\tilde{f}_5^2 &= (\text{det}\mathcal W)^2\, \dfrac{ f_5^2f_6^2}{w_2^2f_5^2+f_6^2(w_4+w_2 f_7)^2} \;, \\
\tilde{f}_6^2 &= w_2^2f_5^2+f_6^2(w_4+w_2 f_7)^2\;, \\
\tilde{f}_7 &= \dfrac{w_1w_2 f_5^2 + f_6^2(w_3+w_1 f_7)(w_4+w_2 f_7)}{w_2^2f_5^2+f_6^2(w_4+w_2 f_7)^2}\;.
\end{split}
\end{equation}
These expressions are  useful in proving the absence of conical singularities, as shown in Section \ref{Absence of conical singularities}.

%%%%%%%%%%%%%%%%%%%
\subsubsection*{Setting $\alpha_1=0$}
%%%%%%%%%%%%%%%%%%%

Let us perform a linear coordinate transformation as above with
\begin{equation}
w_1=w_4=1\;, \qquad\qquad w_2=0\;,
\end{equation}
while keeping $w_3$ arbitrary for now. One can see that  
\begin{equation}
\tilde{a}_2=a_2\;, \qquad \tilde{c}=c\;, \qquad \tilde{f}_5^2=f_5^2\;, \qquad \tilde{f}_6^2=f_6^2\;.
\end{equation}
Moreover one has 
\begin{equation}
\tilde{a}_3=a_3-w_3a_2\;, \qquad \tilde{b}=b-w_3c\;, \qquad \tilde{f}_7=f_7+w_3 \;.
\end{equation}
Note that $a_3$ and $b$ appear in the differential BPS equations and Bianchi identity only through the combinations
\begin{equation}
a_3+a_2f_7=\tilde{a}_3+\tilde{a}_2 \tilde{f}_7\;, \qquad b+cf_7=\tilde{b}+\tilde{c} \tilde{f}_7\;.
\end{equation}
Thus the only effect of the arbitrary constant $w_3$ on the system of differential equations is in the expression
\begin{equation}
\alpha_1-\alpha_2 f_7 = (\alpha_1+\alpha_2w_3) -\alpha_2\tilde{f}_7 \;.
\end{equation}
The phase of the spinor is also modified:
\begin{equation}
\alpha_1\beta+\alpha_2\psi = (\alpha_1+\alpha_2w_3)\tilde{\beta} +\alpha_2\tilde{\psi}\;.
\end{equation}
Finally, the algebraic constraints from the BPS equations read
\begin{equation}
a_3\alpha_2+a_2\alpha_1= \tilde{a_3}\alpha_2+\tilde{a}_2(\alpha_1+\alpha_2w_3)= \dfrac{\kappa}{2}\;, \qquad b \alpha_2+c\alpha_1=\tilde{b} \alpha_2+\tilde{c}(\alpha_1+\alpha_2w_3) = -\dfrac{1}{2}\;.
\end{equation}
Since we assume $\alpha_2\neq 0$ (as we must, since  $f_6=\alpha_2 f_1$) we can always choose the arbitrary constant $w_3=-\alpha_1/\alpha_2$, thus eliminating the constant $\alpha_1$. We conclude from this analysis that we can safely set $\alpha_1=0$ from the beginning, which is often convenient. Finally, by a rescaling of the coordinate $\psi$ it is possible to set $\alpha_{2}$ to any nonzero value. Note that in this argument we have not made any assumptions on the solutions to the BPS equations or the values of the parameters $\{a_{2,3},b,c\}$.

\subsection{Solutions with no flavor flux: $a_{2}=0$}
\label{Solutions with no flavor flux: a2=0}

Setting $a_{2}=0$ leads to many simplifications. Nonetheless, the system is still quite rich and there are four cases that must be analyzed separately: (i) $\kappa \neq 0, c=0$; (ii) $\kappa\neq 0, c\neq 0$; (iii) $\kappa=0,c=0$ and; (iv) $\kappa=0,c\neq 0$. 

\subsubsection{Case (i): $\kappa\neq 0, c= 0$} 

One finds that only $\kappa=-1$ is allowed.  After some coordinate redefinitions and an overall rescaling of the metric, the solution reads
 \eqs{\nonumber
 ds^{2}_{10}=\,&ds^{2}_{AdS_{3}}+\frac{v^{2}+v+1}{4v}ds^{2}_{\Sigma_{\mathfrak g > 1}}+\frac{v^{2}+v+1}{4(v+1)}\(d \theta^2+\sin^2 \theta d \phi^2+\frac{1}{v}(dw^{2}+ \sin^{2}w d\nu ^{2})\)\\ \nonumber
 \,&\qquad \qquad \qquad \qquad \qquad \qquad  + \frac{1}{4} \(d \psi-\cos \theta d \phi-\cos w d \nu- \frac{dx_{1}}{x_{2}}\)^2\,,\\ \nonumber \\ 
 G_{(5)}=\,&\dvol{AdS_{3}}\wedge\( \frac{(v+1)^{2}}{2v} \dvol{\Sigma_{\mathfrak g>1}}+\frac{1}{2(v+1)}\(v^{2}\,\dvol{S^{2}_{\theta \phi}}+\frac{1}{v}\dvol{S^{2}_{w \nu}}\)\)\,,
 }
 where $v>0$ is a real parameter. This is the solution discussed in some detail in Section \ref{Conifold on Sigma with baryonic flux}. For the special value $v=1$ the metric of the solution can be written as
\equ{\label{eq:metT11noflavor}
ds^2_{10} = ds^2_{AdS_3} + \frac{3}{4} ds^2_{\Sigma_{\mathfrak g > 1}} + \frac{9}{4}ds^2_{T^{1,1}} \;,
}
where $ds^2_{T^{1,1}}$ is the metric for the conifold with a Reeb vector fibered over $\Sigma_{\mathfrak g}$. This solution   is an example of the solutions discussed in Section 6.1 of \cite{Gauntlett:2006qw} for $\mathbb{P}^1\times \mathbb{P}^1$ as the K\"ahler-Einstein base.

\subsubsection{Case (ii): $\kappa\neq 0, c\neq 0$}

There are two branches, corresponding to  $q_{3}\neq 0$ and $q_{3}=0$.  

\subsubsection*{Branch $q_{3}\neq 0$: Fibered $Y^{p,q}$}

In this branch, only $\kappa =-1$ is allowed. We set $\alpha_{1}=0$ and $\alpha_{2}=1/2$. After an appropriate coordinate redefinition and overall rescaling the background takes the form:
\begin{multline}\label{Metriccneq0kneq0simp text}
ds^2_{10} =ds^{2}_{AdS_{3}} + \frac{3}{4} ds^{2}_{\Sigma_{\mathfrak g>1}} + \frac{3|1-c{y}|}{8}  (d\theta^2+\sin^2\theta d\phi^2) \\+ \frac{9}{8}\frac{|1-c{y}|}{(2c{y}^3-3{y}^2+{a})}d{y}^2 +\frac{1}{8} \frac{(2c{y}^3-3{y}^2+{a})}{|1-c{y}|}\left(d\beta + c \cos\theta d\phi\right)^2\\+ \frac{1}{4}\left(d\psi -\cos\theta d\phi+{y}\left(d\beta + c \cos\theta d\phi\right)-\frac{dx_1}{x_2}\right)^2 \;,
\end{multline}
\equ{\label{Fluxcneq0kneq0simp text}
G_{(5)}=\dvol{AdS_{3}}\wedge\( 2 \dvol{\Sigma_{\mathfrak g>1}}+ \frac{|1-c y|}{4}\dvol{S^{2}}+\frac{1}{4}dy\wedge (d\beta+c \cos \theta d\phi)\)\,.
}
This solution is again of the form presented in Section 6.1 of \cite{Gauntlett:2006qw}, \ie 
\begin{equation}\label{Metriccneq0kneq061App}
ds^2_{10} = ds^2_{AdS_3} + \frac{3}{4} ds^2_{\Sigma_{\mathfrak g>1}} + \frac{9}{4}ds^2_{Y^{p,q}} \;,
\end{equation}
where $ds^2_{Y^{p,q}}$ is the local from of the metric on $Y^{p,q}$ (fibered over $\Sigma_{\mathfrak g>1}$), written in canonical form \cite{Gauntlett:2004yd}. We discuss this solution in Section \ref{subsec:univsugra}.

\subsubsection*{Branch $q_{3}=0$:  $AdS_{3}\times S^{3}\times T^{4}$}

In this case, we find that only $\kappa=1$ is allowed and we denote this sphere by $\tilde S^{2}$, with coordinates $(\tilde \theta, \tilde \phi)$. After an appropriate change of coordinates we find
\eqs{\nonumber
ds^{2}_{10}=\,&ds^{2}_{AdS_{3}}+\frac{1}{4}ds^{2}_{\tilde S^{2}}+\frac{y}{4}ds^{2}_{S^{2}}+\frac{y}{4(a+y^{2})}\,dy^{2}+\frac{(a+y^{2})}{4y}(d\beta+ \cos \theta d\phi)^{2}+\frac{1}{4}(d\tilde \psi-\cos{\tilde \theta} d{\tilde \phi})^{2}\,,\\ \label{branch kappa=-1 with q3=0}
G_{(5)}=\,&\dvol{AdS_{3}}\wedge\(\frac{y}{2}\dvol{S^{2}} -\frac{1}{2}dy\wedge(d\beta+ \cos \theta d\phi)\)\,,
}
where we defined $\tilde \psi=\psi+\beta$. This solution corresponds to  $AdS_{3}\times S^{3}\times T^{4}$ (with only 5-form flux turned on), for any value of $a$. This is easy to see in the case $a=0$;  defining $y=\rho^{2}$ the terms $ds^{2}_{S^{2}}$, $d\rho^{2}$ and the $(D\beta)^{2}$ combine into the metric on $T^{4}$ while the Riemann surface and the $(D\psi)^{2}$ term combine into an $S^{3}$, and the solution is $AdS_{3}\times S^{3}\times T^{4}$. One can check that for generic $a$ the four-dimensional metric has vanishing Riemann tensor and is therefore always $T^{4}$. Usually the $AdS_{3}\times S^{3}\times T^{4}$ background of type IIB supergravity is associate with the D1-D5 system and thus it has only 3-form flux. Here we see the same solution sourced only by 5-form flux. The two backgrounds should be related by two T-duality transformations.

\subsubsection{Case (iii): $ c=\kappa= 0$}

After appropriate coordinate redefinitions, we find
\eqs{\nonumber
ds^{2}_{10}=\,&ds^{2}_{AdS_{3}}+\frac{1}{4}ds^{2}_{\Sigma_{\mathfrak g=1}}+\frac{1}{4}ds^{2}_{S^{2}}+\frac{dy^{2}}{4 y}
+\frac{y}{4}d\beta^{2}+\frac{1}{4}(d\psi-\cos \theta d\phi-\frac{1}{2}d \beta)^{2}\,,\\
G_{(5)}=\,&\dvol{AdS_{3}}\wedge\(\frac{1}{2}\, \dvol{\Sigma_{\mathfrak g=1}} +\frac{1}{2} d y\wedge d\beta\)\,.
}
Again, by the simple change of coordinates $y=\rho^{2}$ one sees that the genus-one Riemann surface combines with $d\rho^{2}$ and $d\beta^{2}$ into a $T^{4}$ and the $S^{2}$ combines with the $(D\psi)^{2}$ part to give an $S^{3}$. Thus, the solution is again $AdS_{3}\times S^{3}\times T^{4}$ with 5-form flux.

\subsubsection{Case (iv):  $\kappa=0,\, c\neq 0$}

After appropriate redefinitions and simple coordinate transformations\footnote{For the $y$ coordinate the change of variables is  of the form $\tilde y = \gamma/y+\delta$, with $\delta, \gamma$ some specific constants.} the metric and 5-form flux read:
\eqs{\nonumber
ds^{2}_{10}=\,&\frac{b}{\sqrt{\tilde y}}ds^{2}_{AdS_{3}}+\frac{\sqrt{\tilde y}}{ b} ds^{2}_{\Sigma_{\mathfrak g=1}}+\frac{1}{4 b \sqrt{\tilde y}}ds^{2}_{S^{2}}
+\frac{b}{4\tilde{y}^{5/2}(b^{2}-(1-\tilde y)^{2})}d\tilde y^{2} \\ \nonumber
\,&  +\frac{(b^{2}-(1-\tilde y)^{2})}{ 4\sqrt{\tilde y}b^{3}}(d\beta+\cos \theta d\phi)^{2}+\frac{b}{4 \sqrt{\tilde y}}\( d\psi-\cos \theta d\phi+\frac{b^{2}+\tilde y -1}{b^{2}}(d\beta +\cos \theta d \phi)\)^{2}\,,\\ \label{GDK App} 
G_{(5)}=\,&\dvol{AdS_{3}}\wedge\Big( 2 \dvol{\Sigma_{\mathfrak g=1}}+\frac12 \dvol{S^{2}}-\frac{\tilde y-1}{2\tilde y^{2}}d\tilde y \wedge (d\beta +\cos \theta d \phi)\\  \nonumber
\qquad \qquad &+\frac{b^{2}}{2\tilde y^{2}}d \tilde y \wedge \(d \psi-\cos \theta d\phi+ \frac{b^{2}+\tilde y-1}{b^{2}}(d\beta+\cos \theta d\phi)\)\Big)\,.
}
This coincides with the solution in section 4.1 of  \cite{Donos:2008ug}. To see this one must make the linear change of coordinates:
\equ{
\beta=-\tilde \beta +\tilde \psi\,, \qquad \psi=\tilde \beta\,.
}
Using the formulae in Appendix~\ref{Linear transformations on beta psi} it is easy to see that the metric reads:
\eqs{\nonumber
ds^{2}_{10}=\,&\frac{b}{\sqrt{\tilde y}}ds^{2}_{AdS_{3}}+\frac{\sqrt{\tilde y}}{ b} ds^{2}_{\Sigma_{\mathfrak g=1}}+\frac{1}{4 b \sqrt{\tilde y}}ds^{2}_{S^{2}}
+\frac{b}{4\tilde{y}^{5/2}(b^{2}-(1-\tilde y)^{2})}d\tilde y^{2} \\ 
\,&  +\frac{(b^{2}-(1-\tilde y)^{2})}{ 4b\sqrt{\tilde y} (b^{2}-1+2\tilde y)}(d\tilde \beta+\cos \theta d\phi)^{2}+\frac{b^{2}-1+2\tilde y}{4b\sqrt{\tilde y}}\( d\tilde \psi-\frac{\tilde y}{b^{2}-1+2\tilde y}(d\tilde \beta +\cos \theta d \phi)\)^{2}\,,
}
which coincides exactly with the solution found in \cite{Donos:2008ug}, where its central charge was also computed. We discuss this solution further in Section \ref{subsec:YpqT2sugra}.

\subsection{Solutions with flavor flux: $a_{2} \neq 0$}
\label{Solutions with flavor flux: a2 not zero}

Here there are also a few special cases: 1) $a_{2}=- \kappa c \neq 0$ and; 2) $c=0$. 

\subsubsection{Case $c=0$}

When solving the Bianchi identity there are two branches: $q_{3}=0$ and $q_{3}\neq 0$. The latter, however is inconsistent as it leads to $\cP_{1}/\cP_{2}<0$, in contradiction with the assumption that all $\cP_{i}>0$. The former branch leads to
\eqs{\nonumber
ds^{2}_{10}=\,&ds^{2}_{AdS_{3}}+y\,ds^{2}_{\Sigma_{\mathfrak g}}+\frac{1}{4}ds^{2}_{S^{2}}+\frac{y}{\kappa y^{2}+\lambda}dy^{2}
+\frac{\kappa y^{2}+\lambda}{y}(d\beta+A_{\mathfrak{g}})^{2}+\frac{1}{4}(d\psi-\cos \theta d \phi)^{2}\,,\\
G_{(5)}=\,&\dvol{AdS_{3}}\wedge\(2y\, \dvol{\Sigma_{\mathfrak g}}+2 dy \wedge (d\beta+A_{\mathfrak{g}})\)\,.
}
Here $\kappa=\{0,\pm1\}$ and $\lambda$ is a real parameter and  $y>0$. For $\kappa=\{1,0,-1\}$, positivity of the internal manifold requires for $y^{2}>-\lambda$,  $\lambda>0$, and $y^{2}<\lambda$, respectively.  The connection $A_{\mathfrak{g}}$ is given in (\ref{solution A}) for $\kappa=\pm1$ and  for $\kappa=0$ in (\ref{A torus}). The case $\kappa=-1$ is not topologically allowed since the Riemann surface $\Sigma_{\mathfrak{g}}$ shrinks to zero size at $y=0$. One can see that the cases $\kappa=\{1,0\}$ lead again to $AdS_{3}\times S^{3}\times T^{4}$ for any value of $\lambda$.

\subsubsection{Case $a_{2}=-\kappa c  \neq 0$}

There are two branches: branch A with $q_{3}=-\frac{2}{3} c\kappa$; and branch B with $q_{3}=-2 c\kappa$. 

\subsubsection*{Branch A}

For this branch, we find that only $\kappa=1$ is allowed and after appropriate coordinate redefinitions
\eqs{\nonumber
ds^2_{10} =\,& y \, ds^{2}_{AdS_{3}} + \frac{3}{8}\(ds^{2}_{\Sigma_{\mathfrak g=0}}+  ds^{2}_{S^{2}}\) + \frac{9y}{4(4y^{3}-9y^{2}+6 ay-a^{2})}dy^2\\ \nonumber
\,&\qquad    +\frac{4y^{3}-9y^{2}+6 ay-a^{2}}{16y^{3}}\left(d\beta+\cos \theta d\phi-A_{\mathfrak{g}}\right)^2+ y \left(d\psi+\frac{a-3y}{4y^{2}}(d\beta+\cos \theta d\phi-A_{\mathfrak{g}}) \right)^2\;,\\ \label{metric case a2 = k c}
G_{(5)} =\,& \dvol{AdS_{3}}\wedge \Big(\frac{a}{4}\dvol{\Sigma_{\mathfrak g=0}}+\frac{a}{4}\dvol{S^{2}}+\frac{(a-3y)}{2y}dy\wedge(d\beta+\cos \theta d\phi-A_{\mathfrak{g}})  \\ \nonumber
\,&\qquad \qquad \qquad - 2 y dy \wedge \left(d\psi+\frac{a-3y}{4y^{2}}(d\beta+\cos \theta d\phi-A_{\mathfrak{g}}) \right)  \Big)\,.
}
This matches the solution presented in \cite{Gauntlett:2006af}, with $KE_{4}=S^{2}\times S^{2}$. We were not able to identify a candidate dual field theory to this solution.

\subsubsection*{Branch B}

After appropriate coordinate redefinitions we find that in this branch the metric reads:
\eqs{\nonumber
ds^{2}_{10}=\,&\sqrt{\frac{y(y-1) \kappa}{a-y}}ds^{2}_{AdS_{3}}+\sqrt{\frac{y (a-y)\kappa}{y-1}}ds^{2}_{\Sigma_{\mathfrak g\neq1}}+\sqrt{\frac{(y-1)(a-y)\kappa}{y}}ds^{2}_{S^{2}}\\\nonumber
&+\frac{\sqrt{y(y-1) (a-y)\kappa}}{(2y-1)(a-2ay+y^{2})}dy^{2}\\ \nonumber
&+\frac{(2y-1) (a-2ay+y^{2})\sqrt{y(y-1)(a-y)\kappa}}{y^{2}(y-1)^2}\(d\beta+\cos \theta d\phi-\kappa A_{\mathfrak{g}}\)^{2}\\  \label{sol43}
&+\frac{1}{4}\sqrt{\frac{y(y-1) \kappa}{a-y}}\(d\psi'+\frac{2(a-2ay+y^{2})}{y(y-1)}\(d\beta+\cos \theta d\phi-\kappa A_{\mathfrak{g}}\)\)^{2}\,,
}
where $a$ is a real parameter and the flux reads:
\eqs{\nonumber
G_{(5)}=\,&\dvol{AdS_{3}}\wedge\Big( \frac{a-y^{2}}{a-y}\dvol{\Sigma_{\mathfrak g \neq 1}}-\frac{(a-2y +y^{2})\kappa}{a-y}\dvol{S^{2}} \\ 
\,&\qquad \qquad \qquad +\frac{(2y-1)(a-2ay+y^{2})\kappa}{y(y-1)(a-y)}dy \wedge (d\beta+\cos \theta d\phi-\kappa A_{\mathfrak{g}})\\ \nonumber
\,&\qquad\qquad \qquad +\frac{(a-2ay+y^{2})\kappa}{2(a-y)^{2}}dy \wedge \(d\psi'+\frac{2(a-2ay+y^{2})}{y(y-1)}\(d\beta+\cos \theta d\phi-\kappa A_{\mathfrak{g}}\)\) \Big)\,.
}
The connection $A_{\mathfrak{g}}$ is given in (\ref{solution A}). For the metric to be positive definite one needs\footnote{For $\kappa=-1$ one also finds the possibilities $(a+\sqrt{a(a-1)}<y,a>1)$ or  $\(y>1,a\leq1\)$. However, we exclude these since the warp factor is not bounded.}
\eqsn{
\kappa&=1:\, \(0<y<a+\sqrt{a(a-1)},a\leq0\) \quad  \text{or} \quad \(a<y<1, \frac{1}{2}<a<1\)\quad \text{or}\quad \(\frac{1}{2}<y<1,a\leq \frac{1}{2}\)\\
\kappa&=-1:\, \(\frac{1}{2}<y<a, \frac{1}{2}<a\leq1\) \quad  \text{or} \quad \(\frac{1}{2}<y<a-\sqrt{a(a-1)},a>1\)\,.
}
We note that for $\kappa=1$ there are no real zeroes for the metric functions; thus the solutions for $\kappa=1$ are not compact and we must take $\kappa=-1$.  In the case $\kappa=-1$ and $a=1$ the space is topologically $AdS_{3}\times \Sigma_{\mathfrak g>1}\times S^{5}$. This is easy to see; setting  $a=1$ the metric can be written as
\equ{
ds^{2}_{10}=\sqrt{y}\Big[ds^{2}_{AdS_{3}}+ds^{2}_{\Sigma_{\mathfrak g>1}}+ds^{2}_{5} \Big]\,,
}
and with the change of variables $\sin^{2}\mu=\frac{1-y}{y}$ the five-dimensional metric reads
\eqs{\nonumber
\frac{1}{4}ds^{2}_{5} =\frac{ d\mu^{2}}{(1+\sin^{2}\mu)^{2}}&+\frac14\sin^{2}\mu\(ds^{2}_{S^{2}}+\cos^{2}{\mu}\,(d\beta +\cos \theta d\phi+ A_{\mathfrak{g}})^{2}\)\\ \label{metric squashed S5 app}
&+\frac{1}{16}\(d\psi'-2 \sin^{2} \mu\,(d\beta +\cos \theta d\phi+ A_{\mathfrak{g}})\)^{2}\,. 
}
This metric looks like the metric on $S^5$ written as a $U(1)$ bundle over $\mathbb{CP}^2$. Due to the denominator in the $d\mu^2$ term this is not the Einstein metric on $S^5$. Since we have $a_2\neq 0$ we believe that this supergravity solution is dual to the $Y^{p,p}$ theory compactified on $\Sigma_{\mathfrak{g}}$ with some particular value of the background flavor flux $b_2$ and possibly non-zero baryonic flux. It will be of course interesting to understand better this supergravity background, compute the supergravity value of the central charge and compare to the expressions in Section \ref{subsec:YppQFT}.

\subsubsection{Case $\kappa=0$, $c\neq 0$} 
\label{Case kappa=0, c not 0} 

There are two branches:
\subsubsection*{Branch A: $q_{3}=0$}
In this branch the metric reads
\eqs{\nonumber
ds^{2}_{10}=&\sqrt{\frac{y(b-a_{2}y)}{b-1-a_{2}y}}\,ds^{2}_{AdS_{3}}+\frac{1}{4}\sqrt{\frac{
(b-a_{2}y)(b-1-a_{2}y)}{y}}\,ds^{2}_{T^{2}}+\frac14 \sqrt{\frac{y(b-1-a_{2}y)}{b-a_{2}y}}\,ds^{2}_{S^{2}}
\\
&+\frac{dy^{2}\sqrt{(b-a_{2}y)(b-1-a_{2}y)}}{ 4y^{3/2} }
+\frac{\sqrt{y(b-1-a_{2}y)}}{4(b-a_{2}y)^{3/2}}D\beta^{2}\\ \nonumber
&+\frac14 \sqrt{\frac{y(b-a_{2}y)}{b-1-a_{2}y}}\(D\psi+\frac{1}{b-a_{2}y}D\beta\)^{2}\,,
}
where $D\psi=d\psi-\cos \theta d\phi, D\beta= d\beta +\cos \theta d\phi+a_{2}A$ and the 5-form flux:
\eqs{\nonumber
G_{(5)}=\frac12\dvol{AdS_{3}}\wedge\Big(& \(1+b-a_{2}y-\frac{b-1}{b-1-a_{2}y}\) \dvol{T^{2}}\\ 
&-\(1+ \frac{a_{2}y}{(b-a_{2}y)(b-1-a_{2}y)}\)dy\wedge D\beta\\ \nonumber
&-\frac12 \(1-\frac{b-1}{b-1-a_{2}y}\)dy\wedge(D\psi+\frac{1}{b-a_{2}y}D\beta)\Big)\,.
}

\subsubsection*{Branch B: $q_{3}\neq 0$}
We find:
\eqs{\nonumber
ds^{2}&=\sqrt{\frac{y(y-1)}{q_{0}-a_{2}y}}ds^{2}_{AdS_{3}}+\frac34 \sqrt{\frac{(q_{0}-a_{2}y)(y-1)}{y}} \,ds^{2}_{T^{2}}+\frac34 \sqrt{\frac{y(q_{0}-a_{2}y)}{a_{2}^{2}(y-1)}}\,ds^{2}_{S^{2}}\\ \
&+\frac{9dy^{2}\sqrt{y(y-1)(q_{0}-a_{2}y)}}{4 w(y)}+\frac{w(y)\sqrt{y(y-1)(q_{0}-a_{2}y)}}{4a_{2}^{2}y^{2}(y-1)^{2}} \, D\beta^{2}\\\nonumber
&+\frac14 \sqrt{\frac{y(y-1)}{q_{0}-a_{2}y}}\(D\psi-\frac{w'(y)}{6a_{2}y(y-1)} D\beta\)^{2}\,,
}
where $q_{0}$ is a parameter and we assumed $a_{2}>0$, $D\psi=d\psi-\cos \theta d\phi, D\beta= d\beta +\cos \theta d\phi+a_{2}A$ and defined
\equ{
w(y)=q_{0}(2-3y)^{2}+a_{2}y^{2}(3-4y)\,,
}
and the flux is given by
\eqs{\nonumber
G_{(5)}=\dvol{AdS_{3}}\wedge\Big(& \frac{q_{0}+a_{2}y(y-2)}{2(q_{0}-a_{2}y)}\,\dvol{T^{2}}
+ \frac{q_{0}-a_{2}y^{2}}{a_{2}(q_{0}-a_{2}y)}\,\dvol{S^{2}} \\ 
&+\frac{(q_{0}-2q_{0}y+a_{2}y^{2})}{2(q_{0}-a_{2}y)^{2}}(-1+\frac{w'(y)}{6a_{2}y(y-1)})\,dy \wedge D\beta\,\\ \nonumber
&+\frac{(q_{0}-2q_{0}y+a_{2}y^{2})}{2(q_{0}-a_{2}y)^{2}}\,dy \wedge \(D\psi-\frac{w'(y)}{6a_{2}y(y-1)} D\beta\)  \Big)\,,
}
which can also be written as
\eqs{\nonumber
G_{(5)}=\dvol{AdS_{3}}\wedge&\Big( \frac{q_{0}+a_{2}y(y-2)}{2(q_{0}-a_{2}y)}\,\dvol{T^{2}}
+ \frac{q_{0}-a_{2}y^{2}}{a_{2}(q_{0}-a_{2}y)}\,\dvol{S^{2}} \\
&+\frac{(q_{0}-2q_{0}y+a_{2}y^{2})}{2(q_{0}-a_{2}y)^{2}}\,dy \wedge \(D\psi-D\beta\)  \Big)\,.
}
We leave the analysis of the global properties of the solutions in branches A and B above for future work.

%%%%%%%%%%%%%
\subsubsection{Case $\kappa\neq 0$, $c\neq 0$} 
\label{Case kappa not 0, c not 0} 
%%%%%%%%%%%%%

Here we assume $\kappa=-1$ and  $c\neq 0$, which we rescale to 1.  As discussed in Appendix \ref{Linear transformations on beta psi} it is always possible to set $\alpha_{1}=0$ and $\alpha_{2}=\frac12$ by a coordinate transformation, which we do.  Since  $c\neq0$ we may shift $y$ to set $C_{3}=0$.  The parameter $C_{2}$ can be similarly rescaled away, thus we can set $C_{2}$ to any nonzero numerical value (we set $C_{2}=\frac{3}{4}$). The Bianchi identity leads to a number of constraints among the remaining parameters. Finally,  the solution is given by
\eqs{
\cP_{1}&=\frac34(a-1) a_{2}^{2}+C_{1}y\,, \qquad \cP_{2}=\frac34+ a_{2}y\,, \qquad \cP_{3}=-y\,, \qquad \cP_{7}=-\frac12 Q'\,,\\ \nonumber \\
\mathcal Q&=\frac{-3(a-1)(7+a_{2}^{2}+8C_{1}-4a_{2}(1+C_{1}))}{16(a_{2}-1)^{2}}-\frac{3a_{2}(a-1)(a_{2}-4C_{1}-5)}{4(a_{2}-1)}y\\\nonumber
&\qquad \qquad +\frac34\(1-4(a-1)a_{2}^{2}\)y^{2}+\frac13(1+a_{2}-4C_{1})y^{3}\,,
}
with $C_{1}=-\frac{1}{4}(1+a_{2}+2\sqrt{1-a_{2}+a_{2}^{2}})$.
Explicitly, the metric reads
\eqs{\nonumber
ds^{2}_{10}=\,&\sqrt{\frac{y(3+4a_{2}y)}{3(1-a)a_{2}^{2}-4C_{1}y}}ds^{2}_{AdS_{3}}+\frac14\sqrt{\frac{(3+4a_{2}y)(3(1-a)a_{2}^{2}-4C_{1}y)}{y}}\,  ds^{2}_{\Sigma_{\mathfrak g>1}}\\ \nonumber
&+\sqrt{\frac{y(3(1-a)a_{2}^{2}-4C_{1}y)}{3+4a_{2}y}}\, \,  ds^{2}_{S^{2}}+\frac14\frac{\sqrt{y(3+4a_{2}y)(3(1-a)a_{2}^{2}-4C_{1}y)}}{\mathcal Q}dy^{2}\\ \label{metric general nice app}
&+\frac{4\mathcal{Q} \sqrt{y(3(1-a)a_{2}^{2}-4C_{1}y)}}{y^{2}(3+4a_{2}y)^{3/2}} D\beta^{2}+\frac{1}{4}\sqrt{\frac{y(3+4a_{2}y)}{3(1-a)a_{2}^{2}-4C_{1}y}}\(D\psi+\frac{2\mathcal{Q}'}{y(3+4a_{2}y)}D\beta\)^{2}\,,
}
where $D\beta=d\beta+\cos \theta +a_{2}A_{\fg}$, $D\psi=d\psi-\cos \theta-A_{\fg}$. The 5-form, which we do not write explicitly here, is given by (\ref{flux G's}) and (\ref{flux general nice app}).
Setting $a_{2}=0$ and  sending $y\to -\frac{3}{8}(y+1)$ and redefining the parameter $a$, the metric coincides with the solution with no flavor flux (\ref{Metriccneq0kneq0simp text}), with $c$ scaled to 1. It is also possible to take the limit $a_{2}\to 1$. Taking this limit and performing the change of  variables $y\to 3/4 (y - 1)$ and changing  $\psi \to \psi-\beta$ the metric coincides with \eqref{sol43} with $\kappa=-1$. Thus, this solution contains all previous cases.

\paragraph{Case $a=1$.}

Let us consider the local form of the metric more carefully for $a=1$. In this case the cubic $\mathcal Q$ becomes
\equ{
\mathcal Q=\frac{3}{4y_{1}}y^{2}(y_{1}-y)\,, \qquad y_{1}\equiv -\frac{9}{8(1+a_{2}+\sqrt{1+a_{2}^{2}-a_{2}})}\,.
}
There is a negative root at $y=y_{1}$ and a doubly-degenerate root at $y=0$. It is convenient to go to a basis $(\tilde \beta, \tilde \psi)$ in which it is manifest that there are no conical singularities, provided $\tilde \beta$ period $2\pi$. As discussed in Section~\ref{Absence of conical singularities} this is accomplished  by a coordinate transformation mixing $(\psi,\beta)$. Setting $\alpha_{1}=0, \, \alpha_{2}=1/2$ and $c=-b=1$ we may choose $w_{1}=-1, w_{2}=\frac{1}{2},w_{3}=1, w_{4}=0$, \ie
\equ{
\beta=\frac{1}{2}\tilde \psi-\tilde \beta\,, \qquad \psi=\tilde \beta\,.
}
It is also convenient to make the coordinate transformation
\equ{
y=\frac{2y_{1}\sin^{2}\mu}{1+\sin^{2}\mu}\, , \qquad \mu\in[0,\pi/2]\,,
}
in terms of which  $\mathcal Q=\frac{3y_{1}^{2}\cot^{2} \mu}{(1+\csc^{2}\mu)^{3}}$, with zeros at   $\mu=\{0,\pi/2\}$. In these coordinates the metric reads
\equ{
ds^{2}_{10}=f_{1}
^{2}\Big[ds^{2}_{AdS_{3}}+|C_{1}|ds^{2}_{\Sigma_{\mathfrak g>1}}+\tfrac{32|C_{1}y_{1}|}{3}ds^{2}_{5} \Big]\,,
}
with
\eqs{\nonumber
ds^{2}_{5} =\frac{ d\mu^{2}}{(1+\sin^{2}\mu)^{2}}&+\frac{1}{4g(\mu)}\sin^{2}\mu\(ds^{2}_{S^{2}}+\frac{1}{f(\mu)}\cos^{2}{\mu}\,(d\tilde \beta +\cos \theta d\phi+ A_{\mathfrak{g}})^{2}\)\\ \label{metric squashed S5 app flux}
&+\frac{3}{8|C_{1}|}\frac{f(\mu)}{16\,g(\mu)}\(\frac{d\tilde \psi}{2} +(a_{2}-1)A_{\mathfrak{g}}-\(\frac{9+8y_{1}}{3}\)\frac{\sin^{2} \mu}{f(\mu)}\,(d\tilde \beta +\cos \theta d\phi+ A_{\mathfrak{g}})\)^{2}\,,
}
where we defined the functions
\equ{
f(\mu)\equiv  1+\(1+\frac{8y_{1}}{3}\)\sin^{2}\mu\,,\qquad g(\mu)\equiv 1+\(1+\frac{8a_{2}y_{1}}{3}\)\sin^{2}\mu\,, 
}
and the overall warp factor is given by $f_{1}^{2}=\(\frac{3g(\mu)}{4|C_{1}|(1+\sin^{2}\mu)}\)^{1/2}$. We note that for the special value $a_{2}=1$ then $ y_{1}=-\frac{3}{8},C_{1}=-1$ and $g(\mu)=f(\mu)=1$ and the metric coincides with \eqref{metric squashed S5 app} with $\tilde \psi =2\psi'$. Similalry to the metric in \eqref{metric squashed S5 app} the metric in \eqref{metric squashed S5 app flux} looks like a squashed metric on $S^5$ written as a $U(1)$ bundle over $\mathbb{CP}^2$. We believe that this background is dual to the $Y^{p,p}$ theory on $\Sigma_{\mathfrak{g}}$ with general value of the flavor flux $b_2$ which is related to the value of the supergravity parameter $a_2$. However we leave a global analysis of this background and a supergravity calculation of the central charge for future work.

%%%%%%%%%%%%%%%%%%%%%%%%%%%%%%%%%%%%%%%%%  
\section{General formulas for the central charge}
\label{General formulas for the central charge}
\renewcommand{\theequation}{B.\arabic{equation}}
\renewcommand{\thetable}{B.\arabic{table}}
\setcounter{equation}{0}

%%%%%%%%%%%%%%%%%%%%%%%%%%%%%%

Here we provide some useful normalizations and formulas for computing the gravitational central charge for the class of solutions considered in the paper. 

Consider a metric and flux of the form 
\begin{equation}\label{Lmetric}
ds^2 = L^2 e^{2\lambda}  ds^2_{\text{AdS}_3} + L^2 ds^2_{\cM_7} \;,\qquad\qquad g_s F_{(5)} = L^4 (1+\ast_{10}) G_{(5)} \;.
\end{equation}
 The quantization condition for $F_{(5)}$ is (here we follow the conventions of \cite{Gauntlett:2006af})
\begin{equation}\label{T11quant}
N(D)=\frac1{(2\pi \ell_s)^4} \int_{D} F_{(5)} \;,
\end{equation}
where $D$ is any five-cycle in $\cM_{7}$ and $N(D)$ is an integer. The type IIB supergravity action is
\begin{equation}\label{IIBaction}
S_{\text{IIB}} = \dfrac{1}{16\pi G_{N}^{(10)}} \int d^{10} x\sqrt{-g^{(10)}} R^{(10)}+\ldots\;,
\end{equation}
where the dots stand for other terms in the action that will not be important for our discussion. The normalization we use (see for example Appendix D in \cite{Bena:2008dw}) is such that
\begin{equation}\label{G10def}
16\pi G_{N}^{(10)} = (2\pi)^7 g_s^2\ell_s^8\;,
\end{equation}
where $g_s$ is the string coupling constant and $\ell_s$ is the string length.  The central charge of the dual CFT is given by the Brown-Henneaux formula \cite{Brown:1986nw}
\begin{equation}
c_{\text{sugra}} = \dfrac{3L}{2G_{N}^{(3)}} \;,
\end{equation}
where $L$ is the same as in \eqref{Lmetric} and $G_{N}^{(3)}$ is the 3d Newton constant which can be read off from the 3d effective gravitational action
\begin{equation}
S_{3d} = \dfrac{1}{16\pi G_{N}^{(3)}} \int d^{3} x\sqrt{-g^{(3)}} R^{(3)} + \ldots \;.
\end{equation}
The goal now is to find $G_{N}^{(3)}$ by reducing the type IIB action on the manifold $\mathcal{M}_7$.  To do this one has to plug the metric  \eqref{Lmetric} in the type IIB action \eqref{IIBaction} which leads to\footnote{Under a conformal transformation $\tilde g= e^{2\lambda}g$,  $\sqrt{\tilde g}\tilde R=e^{(D-2)\lambda} R\sqrt{g}+...$, where $D$ is the dimension; this leads to the factor $e^{\lambda}$ in the integrand. } 
\equ{
 \dfrac{1}{16\pi G_{N}^{(3)}}= \dfrac{L^{7}}{16\pi G_{N}^{(10)}}\int d^{7}x \sqrt{g_{\cM_7}} \, e^{\lambda} \;,
}
and therefore
\equ{\label{central charge general}
c_{\text{sugra}} = \dfrac{3L^{8}}{2G_{N}^{(10)}} \int d^{7}x \sqrt{g_{\cM_7}} \, e^{\lambda} \;.
}

Now we specialize these general expressions to the Ansatz in equation \eqref{MetricAnsatz}.  Using \eqref{nice form ansatz} we have
\bea
\int d^7x\, \sqrt{g_{\cM_7}} \, e^\lambda &= \int d^7x \, \sqrt{g_{\Sigma_{\mathfrak g}}}\sqrt{g_{S^{2}}}f_{1} f_{2}^{2}f_{3}^{2}f_{4}f_{5}f_{6} \\
&= \alpha_{2}  \, (4\pi)^{2} \,(\mathfrak g-1) \Delta \beta \,\Delta \psi\int_{y_{1}}^{y_{2}} dy  \, \cP_{1}(y) \;,
\eea
where $\Delta \beta=2\pi \ell_{\beta}$ and $\Delta \psi=2\pi \ell_{\psi}$ denote the periods of the corresponding coordinate and the integral over $y$ is between two roots $y_{1},y_{2}$ of $\mathcal Q$, between which the function $\mathcal Q$ is positive. 

Now we look at the quantization condition for the 5-form $F_{(5)}$. In general there can be several five-cycles in $\cM_{7}$, one of them being the manifold $\cM_{5}$, spanned by $\{\theta, \phi, y, \beta, \psi\}$, itself. The 5-form flux through $\cM_{5}$ corresponds to the number $N$ of D3-branes. The only term that contributes to this integral is%
\equ{
\int _{\cM_{5}} \big( 1+\ast_{10} G_{(5)} \big) = \int _{\cM_{5}} g_{1}f_{3}^{2} f_{4}f_{5}f_{6} \, \dvol{S^{2}}\wedge dy \wedge d\beta \wedge d\psi\,.
}
Using (\ref{solutions g}), the expression for $\cP_{7}$ in \eqref{solutions P1, P7 app}, and the relations \eqref{relations alphas app}, \eqref{constraint a2 a3 kappa app} we can write
\equ{
g_{1}f_{3}^{2} f_{4}f_{5}f_{6} =\frac{\alpha_{2}}{4} \, \partial_{y}\(2y-\frac{\mathcal Q'}{\cP_{2}}\)\,.
}
Thus, 
\equ{
\int _{\cM_{5}} \big( 1+\ast_{10} G_{(5)} \big) = \pi \alpha_{2}\,\Delta \beta \Delta \psi \(2y-\frac{\mathcal Q'}{\cP_{2}}\)\Big|^{y_{2}}_{y_{1}}\,.
}
Thus, the quantization condition reads
\equ{\label{quant condition general}
N=\frac1{(2\pi \ell_s)^4} \frac{L^{4}}{g_{s}} \,\pi\alpha_{2} \, \Delta \beta \Delta \psi \, S(y_{1},y_{2})\,, \qquad S(y_{1},y_{2})\equiv \(2y-\frac{\mathcal Q'}{\cP_{2}}\)\Big|^{y_{2}}_{y_{1}}\,.
}
Putting everything together, the central charge is given by:
\equ{
c_{\text{sugra}}=\frac{192(\mathfrak g-1)N^{2}}{\alpha_{2}\, \ell_{\beta} \ell_{\psi}}\frac{1}{S(y_{1},y_{2})^{2}}\int_{y_{1}}^{y_{2}} dy \,\cP_{1}(y)\,.
}

%%%%%%%%%%%%%%%%%%%%%%%%%%%%%%%%%%%%%
\end{appendices}
%%%%%%%%%%%%%%%%%%%%%%%%%%%%%%%%%%%%%

%%%%%%%%%%  Bibliography  %%%%%%%%%%%%
{\small
\bibliographystyle{utphys}
\bibliography{YpqSigma}
}

\end{document}